\documentclass[a4paper,11pt]{article}
\pdfoutput=1 
\usepackage{mathtools}
\usepackage{enumitem}  
\usepackage{jinstpub} 
\usepackage{lineno}

\title{\boldmath Monte Carlo simulation method of polarization effects in Laser Compton Scattering on relativistic electrons}

\author[a]{Dan Filipescu}

\affiliation[a]{Horia Hulubei National Institute for R\&D in Physics and Nuclear Engineering (IFIN-HH),\\ 30 Reactorului, 077125 Bucharest, Romania }

\emailAdd{dan.filipescu@nipne.ro}

\abstract{

Quasi-monochromatic, high energy and highly polarized $\gamma$-ray beam sources based on Compton scattering of laser photons (LCS) on relativistic electrons have developed for the last few decades as established instruments for nuclear physics studies. Following an extensive photoneutron experimental campaign at the LCS $\gamma$-ray beam line of the NewSUBARU synchrotron radiation facility at SPring8, Japan, a dedicated simulation code was developed for characterizing the incident $\gamma$-ray beams. The \texttt{eliLaBr} code is implemented using \textsc{Geant4} and is available on the GitHub repository (\href{https://github.com/dan-mihai-filipescu/eliLaBr}{github.com/dan-mihai-filipescu/eliLaBr}). The present work describes step-by-step the Monte Carlo algorithm with focus on modeling the polarization properties of the scattered photon. The polarization is treated independently both in the Stokes parameters and in the polarization vector formalisms. An intervalidation between the two methods is given. Based on polarization state description requirements of different \textsc{Geant4} physics classes, user recommendations are given on which of the two methods to be employed.   

The spatial and energy distributions for the LCS $\gamma$-ray beam and its Stokes parameters are obtained for head-on laser~--~relativistic electron collisions, where several incident laser polarization states were considered: linear, unpolarized, circular  and mixed linear and circular polarization. Results of previous investigations on the polarization of Compton scattered photons are reproduced.

The influence of variable incident angle between photon and electron beam was also investigated. We show that the degree of polarization transfer from the incident photon to the scattered photon increases with the collision angle, where head-on is considered 0$^\circ$. However, as the polarization transfer is strongly influenced by the incident photon energy, we show that, for $\gamma$-ray sources based on Compton scattering of laser photons on relativistic electrons, the polarization degree of the incident photon is almost completely transferred to the scattered photon for any incident angle. 
  }

\keywords{Simulation methods and programs; Polarisation; Accelerator Applications; Beam-line instrumentation (beam position and profile monitors, beam-intensity monitors, bunch length monitors); }

\begin{document}
\maketitle
\flushbottom

\section{Introduction}

A series of photoneutron experimental campaigns were performed with the $\gamma$-ray beam produced at the NewSUBARU electron synchrotron radiation facility from SPring8 laboratory (Japan)~\cite{Kawano2020}. Quasi-monochromatic $\gamma$-ray beams were produced at NewSUBARU beamline BL01 long racetrack~\cite{amano09,Horikawa2010} by laser Compton scattering (LCS) on relativistic electron beams, where several lasers have been used (CO$_2$ 10500 nm, INAZUMA 1064 nm, TALON 532 nm). The electron beam is injected in the NewSUBARU synchrotron ring from SPring8 linac with fixed electron energy $\sim$1~GeV and afterwards the electron beam can be slowed down to $\sim$500~MeV or ramped up to $\sim$1.5~GeV. After Compton interaction of laser photons with relativistic electrons, the photons are back-scattered in a very narrow angle cone and their energy is shifted into $\gamma$-ray energy range with a characteristic continuous spectrum. By a tight and precise alignment collimation, only a narrow part from the highest energies of the $\gamma$-ray spectrum is selected, thus generating a quasi-monochromatic $\gamma$-ray beam. The characteristics of the LCS $\gamma$-ray beam are determined by the properties of the interacting electron and laser beams. By varying the electron beam energy, an energy tunable quasi-monochromatic $\gamma$-ray beam is obtained.

Several neutron multiplicity sorting campaigns with a flat-efficiency moderated neutron detection array~\cite{utsunomiyaNimDNM,IGheorghe_2021_MF} were dedicated to cross section measurements for photoneutron reactions ($\gamma, xn$) with $x=1,2,3,4$~\cite{Gheorghe2017,Kawano2020} and photofission for actinides~\cite{filipescu22ND} at $\gamma$-ray energies starting from the neutron separation energy and covering the entire Giant Dipole Resonance (GDR) energy region. The characterization of the incident $\gamma$-ray beam energy spectrum is crucial for photoneutron experiments, where we underline two main reasons:
\begin{itemize}[leftmargin=*]
\item[--] at energies just above a new channel opening, only a small part of the $\gamma$-ray spectrum generates that particular reaction, thus it is important to precisely know which fraction from the total $\gamma$-ray flux contributes to the reaction rate;
\item[--] especially at excitation energies above the GDR peak, where the photoabsorbtion cross sections become small, although the energy spectrum of the quasi-monochromatic $\gamma$-ray beam is concentrated in the narrow high energy range that one is interested to investigate, the low intensity tail of the $\gamma$-ray beam spectrum from the low energy region may produce important contribution over the GDR peak due to its high cross section values.
\end{itemize}

For this purpose, the \textsc{Geant4}~\cite{geant_agostinelli_2003,geant_allison_2006,geant_allison_2016} simulation code \texttt{eliLaBr}~\cite{eliLaBr_github} has been developed, in which the Compton scattering process between a relativistic electron beam described with Twiss parameters formalism and a laser beam described with Gaussian beam formalism is used to generate the primary particles, then a full simulation of particle transport through the collimator and further to the experimental setup is performed. The present work describes step-by-step the Monte Carlo algorithm for simulating the laser Compton scattering between laser photons and relativistic electrons with focus on describing the photon polarization. Details on the realistic modeling of unsynchronized laser and relativistic electron beams for flux and energy spectrum characterization of LCS $\gamma$-ray beams will be given in a follow-up paper~\cite{filipescu_NIM_LCS_22}.

Due to their high selectivity, polarized photon beams are used in nuclear resonance fluorescence (NRF) experiments for determining the parity of nuclear states~\cite{Pietralla_2002_NIM,Beller_2015}. In ($\gamma$,~n) experiments with polarized incident photons, information on the nuclear structure can be extracted from the anisotropy of the neutron emission~\cite{Kondo_2012}. 

The polarization properties of the scattered photon in the Compton interaction between laser and relativistic electron beams have been extensively discussed in the literature, with many recent works dedicated to their analytical description. It has been shown that the high energy, back-scattered photons retain the initial laser linear polarization, while a flip of the helicity configuration is observed for initial laser circular polarization states~\cite{Angelo2000}. An analytical analysis with direct application for the HI$\gamma$S~\cite{litvinenko_1997} LCS $\gamma$-ray source characterization is done by Sun in reference~\cite{Sun2011_STAB}, where integral values of polarization related variables are given. The spatial distribution of Stokes parameters in a plane perpendicular on the $\gamma$-ray beam propagation axis has been treated in an analytical formalism by Petrillo, in reference~\cite{petrillo2015}, where both Thomson and Compton sources have been considered. A classical electrodynamics treatment on the photon polarization transfer in Thomson interactions is given by Zhijun~Chi in reference~\cite{zhijun_chi2020}. The study shows that the Thomson scattered photon can retain the polarization state of the incident laser photon in arbitrary interaction geometries, which is of great interest for \emph{slant-scattering}~\cite{ohgaki_2007,xu_fan_2022} $\gamma$-ray sources such as SLEGS~\cite{wang_fan_2022,hao_fan_2022}. The nonlinear Thomson scattering of high intensity laser beams on relativistic electrons is treated in reference~\cite{zhijun_chi2022}, where analytical expressions are given for the scattered photon polarization of the fundamental and higher order harmonics. In the present work, the un-coherent Compton scattering of low and medium power lasers is treated, for which the nonlinear effects are not present. 

The main formalism used in the present LCS generator is the Stokes parameters formalism based on the theoretical work described in \cite{McMaster1,McMaster2,Landau}. However, we found that the polarization vector formalism is not only useful for intervalidation with the Stokes parameters one, but also required by some \textsc{Geant4} physics classes for describing the polarization related properties of a particle. Thus, the polarization vector formalism has also been implemented in the present algorithm mostly following the prescription given in the work of Depaola~\cite{Depaola}, but with an important difference regarding the possible orientations of the polarization vector. 

\section{Polarization and the Stokes Parameters} 

In the study of electromagnetic radiation, the formalism of the Stokes parameters offers a very convenient tool for the description of polarization. Although various publications were dedicated to this subject, the Stokes parameters theory is well organized and clearly presented in the articles \cite{McMaster1} and \cite{McMaster2} of William H.~McMaster. Here we outline the concept of density matrix and give the Stokes parameters definition as introduced in the mentioned references. 

\subsection{Polarization density matrix and Stokes parameters for electromagnetic radiation}

The wave function describing the state of polarization for electromagnetic radiation can be expanded in a complete set of orthonomal eigenfunctions with only two terms given by
\begin{equation} \label{eq_10_fromStokesArticle} \psi = a_1 \psi_1 + a_2 \psi_2,
\end{equation}
where $\psi_1$ and $\psi_2$ can either represent two orthogonal states of plane polarization or two states of circular polarization. Then $ \arrowvert a_1 \arrowvert ^2$ and $ \arrowvert a_2 \arrowvert ^2$ give the probabilities of detection of quanta by a detector sensitive only to the $\psi_1$ and $\psi_2$ states, respectively. Using this formalism, the beam polarization is completely characterized by the density matrix
\begin{equation} \label{eq_11_fromStokesArticle}
\rho =
\begin{pmatrix}
a_{1}a_{1}^* & a_{1}a_{2}^* \\
a_{2}a_{1}^* & a_{2}a_{2}^* 
\end{pmatrix},
\end{equation}
as the diagonal elements give the intensities of the two polarization states while the offdiagonal elements give the relative phase between the two states.

A complete set of measurements must be made to retrieve quantities that fully characterize the state of polarization of an arbitrary beam of photons. In the following, we give those experimental quantities, also known as Stokes parameters, along with their expressions as function of density matrix elements and their values for pure states of polarization: 
\begin{enumerate}
\item Beam intensity,  
\begin{equation}\label{eq_Stokes0_expression}
I = \rho_{11} + \rho_{22} = 
\begin{pmatrix}
a_1^* & a_2^*
\end{pmatrix} 
\begin{pmatrix}
1 & 0  \\
0 & 1
\end{pmatrix} 
\begin{pmatrix}
a_1  \\
a_2
\end{pmatrix},
\end{equation}
where we shall consider throughout this discussion the beam to be normalized to unit intensity.  
\item The degree of plane polarization with respect to two arbitrary orthogonal axes $\boldsymbol e_1$ and $\boldsymbol e_2$,
\begin{equation}  \label{eq_Stokes1_expression}
P_1 = \rho_{11} - \rho_{22} = 
\begin{pmatrix}
a_1^* & a_2^*
\end{pmatrix} 
\begin{pmatrix}
1 & 0  \\
0 & -1
\end{pmatrix} 
\begin{pmatrix}
a_1  \\
a_2
\end{pmatrix}.
\end{equation}

$P_1$ = +1 for plane polarization aligned with $\boldsymbol e_1$; 

$P_1$ = $-$1 for plane polarization aligned with $\boldsymbol e_2$. 

\item The degree of plane polarization with respect to a set of axes rotated around $\boldsymbol k_0$ axis ($\boldsymbol k_0 = \boldsymbol e_1 \times \boldsymbol e_2$), with 45$^\circ$ to the right as we are looking toward $+\boldsymbol k_0$,
\begin{equation} \label{eq_Stokes2_expression}
P_2 = \rho_{12} + \rho_{21} = 
\begin{pmatrix}
a_1^* & a_2^* =
\end{pmatrix} 
\begin{pmatrix}
0 & 1  \\
1 & 0
\end{pmatrix} 
\begin{pmatrix}
a_1  \\
a_2
\end{pmatrix}.
\end{equation}

$P_2$ = +1 for plane polarization aligned with $\boldsymbol e_1$ rotated by 45$^\circ$; 

$P_2$ = $-$1 for plane polarization aligned with $\boldsymbol e_2$ rotated by 45$^\circ$.

\item The degree of circular polarization,
\begin{equation} \label{eq_Stokes3_expression}
P_3 = i (\rho_{21} - \rho_{12}) = 
\begin{pmatrix}
a_1^* & a_2^*
\end{pmatrix} 
\begin{pmatrix}
0 & -i  \\
i & 0
\end{pmatrix} 
\begin{pmatrix}
a_1  \\
a_2
\end{pmatrix}.
\end{equation}

$P_3$ = +1 for right circular polarization; 

$P_3$ = $-$1 for left circular polarization. 

\end{enumerate}

The McMaster convention introduced in ref.~\cite{McMaster1} for indexing the Stokes parameters has been adopted here. A different convention regarding indexes is used by Landau in ref.~\cite{Landau}. In order to switch from McMaster convention to Landau convention, one has to perform the following indexes change $1 \rightarrow 3$ (plane polarization), $2 \rightarrow 1$ (plane polarization in a $45^\circ$ rotated system), $3 \rightarrow 2$ (circular polarization) in all equations when dealing with Stokes parameters. 

From equations~\eqref{eq_Stokes0_expression},\eqref{eq_Stokes1_expression},\eqref{eq_Stokes2_expression},\eqref{eq_Stokes3_expression}, the density matrix is easily expressed in terms of the Stokes parameters as:
\begin{equation} 
\rho = \cfrac{1}{2}
\begin{pmatrix}
1 + P_1 & P_2 + i P_3  \\
P_2 - i P_3 & 1 - P_1
\end{pmatrix}. 
\end{equation} 

\subsection{Properties of the Stokes parameters}

It is customary to write the Stokes parameters in the form of a four vector:
\begin{equation} 
\begin{pmatrix}
I \\
P_1 \\
P_2 \\
P_3  
\end{pmatrix} 
=
\begin{pmatrix}
I \\
{\bf P}  
\end{pmatrix}. 
\end{equation} 
One could imagine ${\bf P}$ as a vector with the origin in the center of the Poincar\'e sphere of radius one and $P_1$, $P_2$ and $P_3$ Cartesian components. The three Stokes parameters take real values within the [-1, 1] interval conditioned by the resulting polarization ${\bf P}$ to stay inside the sphere (0 $\le$ $|{\bf P}|$ $\le$ 1). Consequently, the beam polarization degree is defined as
\begin{equation}\label{Total_polarization}
P = \cfrac {\sqrt{P_1^2 + P_2^2 + P_3^2}}{I}.
\end{equation}

\begin{figure}[t]
\centering
\includegraphics [width=0.30\columnwidth, angle=0]{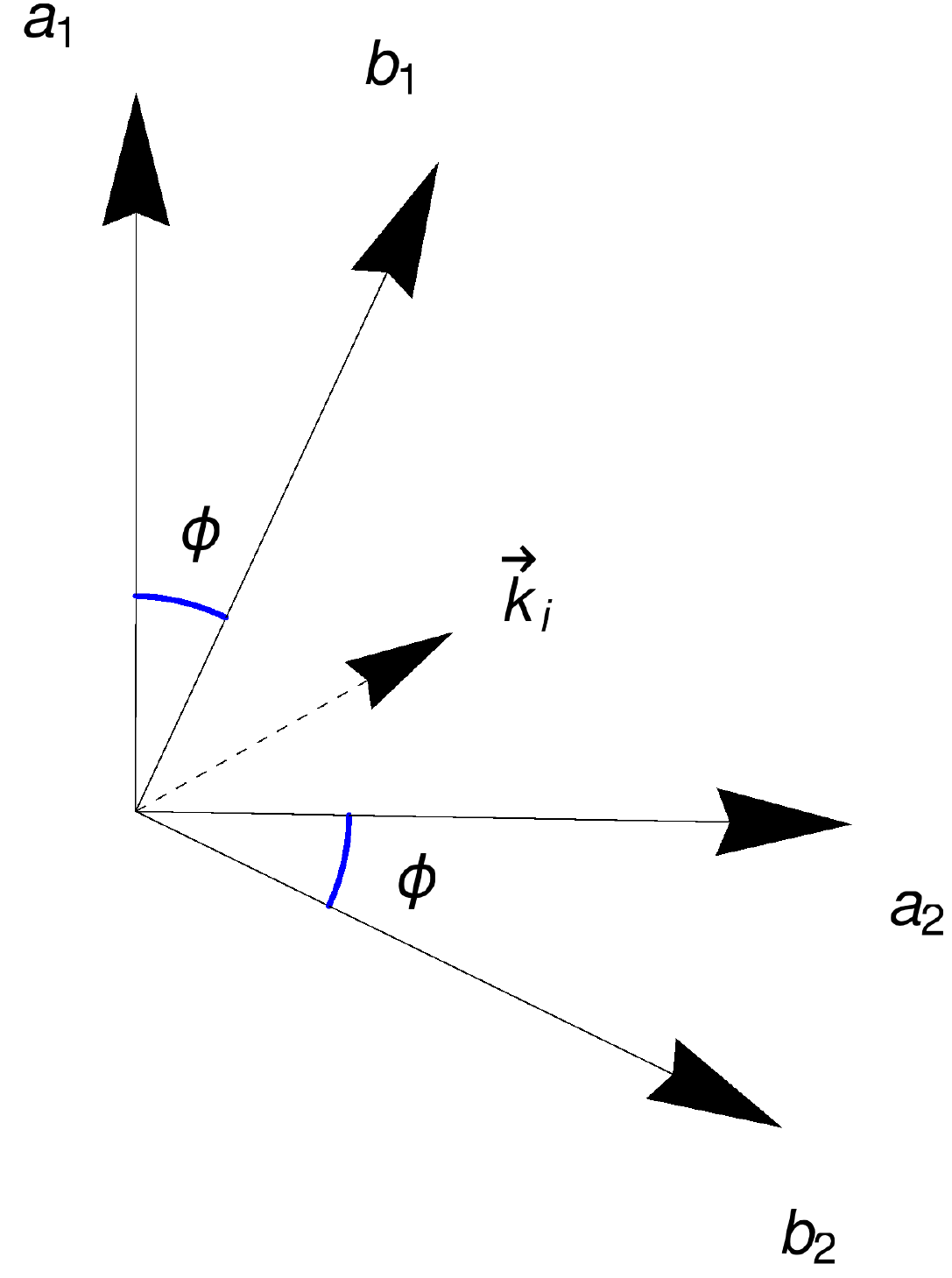} 
\caption {\label{rotated_ref_system} 
 Illustration of space rotation transformation. The initial reference system defined by $\boldsymbol k_i = \boldsymbol a_1 \times \boldsymbol a_2$ is rotated around $\vec{k}_i$ axis with angle $\phi$ to the right, looking in the direction $+\vec{k}_i$. Thus, one obtains the final reference system defined by the axes $\boldsymbol k_i = \boldsymbol b_1 \times \boldsymbol b_2$.  
}
\end{figure}

The coefficients in equation~\eqref{eq_10_fromStokesArticle} depend on the chosen reference system because the system is defining the orthogonal states. Thus, the density matrix and consequently the Stokes parameters are dependent on the reference system. As the system of reference is generally taken with one axis along the photon propagation vector, it is often useful to express the Stokes parameters in a second coordinate system rotated about the direction of propagation $\vec{k}_i$ to an angle $\phi$ to the right of the original system as we are looking along the +$\vec{k}_i$ direction. Figure~\ref{rotated_ref_system} gives a schematic illustration of the initial and the rotated system. We can define an $M$ rotation matrix which, when applied on the initial Stokes vector, generates the Stokes vector in the rotated system:
\begin{equation} 
\begin{pmatrix}
I' \\
{\bf P'}
\end{pmatrix} 
= M
\begin{pmatrix}
I \\
{\bf P}
\end{pmatrix}.
\end{equation}
One can easily demonstrate that the $M$ matrix can be expressed as
\begin{equation}\label{Rotation_Matrix} 
M = 
\begin{pmatrix}
1 & 0 & 0 & 0\\
0 & \cos 2\phi & \sin2\phi & 0 \\
0 & -\sin 2\phi & \cos2\phi & 0 \\
0 & 0 & 0 & 1
\end{pmatrix}. 
\end{equation}
Thus if in the old system we had $\begin{pmatrix}I, & P_1, & P_2, & P_3\end{pmatrix}$, then, in the new system of coordinates, the Stokes parameters of the same beam are
\begin{equation} \label{eq_StokesRotationTransformation}
\begin{pmatrix}
I \\
P_1 \cos 2\phi + P_2 \sin 2\phi \\
-P_1 \sin 2\phi + P_2 \cos 2\phi \\
P_3
\end{pmatrix}. 
\end{equation}
A rotation in the opposite direction changes the sign of the $\sin 2\phi$ terms as expected. It follows directly from equation~\eqref{eq_StokesRotationTransformation} that the polarization degree from \eqref{Total_polarization} is invariant to spatial rotations. 

The probability of detecting a photon characterized by the Stokes parameters $\begin{pmatrix}1&\bf Q\end{pmatrix}$, (or in other words, using an analyzer which will pass only states of that particular polarization $\bf Q$ without affecting the beam intensity) in an arbitrary beam characterized by the Stokes parameters $\begin{pmatrix}I&\bf P\end{pmatrix}$ is given by
\begin{equation}\label{general_stokes_transform}
W = \cfrac{1}{2} (1 + \boldsymbol{P} \cdot \boldsymbol{Q}) = \cfrac{1}{2}
\begin{pmatrix}1, & \boldsymbol Q\end{pmatrix} 
\begin{pmatrix}I \\ \boldsymbol P\end{pmatrix}. 
\end{equation}
Considering an interaction which affects the polarization state of the incident photon beam, we can define a $4 \times 4$ matrix operator $T$ characteristic to that specific interaction, which transforms the initial Stokes parameters as following
\begin{equation}\label{Tmatrix_action}
\begin{pmatrix}I \\ \boldsymbol P\end{pmatrix} = 
T \begin{pmatrix}I_0 \\ \boldsymbol P_0\end{pmatrix}.
\end{equation}
From equations~\eqref{general_stokes_transform} and \eqref{Tmatrix_action} it follows that the fractional intensity given by an analyzer $\begin{pmatrix}1, & \boldsymbol Q\end{pmatrix}$ applied to a $\begin{pmatrix}I&\boldsymbol P\end{pmatrix}$ beam which suffered an interaction characterized by operator $T$ is given by 
\begin{equation}
W = \cfrac{1}{2} 
\begin{pmatrix}1, & \bf Q\end{pmatrix} T
\begin{pmatrix}I_0 \\ \bf P_0\end{pmatrix}. 
\end{equation}

\section{Polarization of $\gamma$ rays by Compton scattering}

Using the Stokes parameters formalism, we now move on to describe the particular case of Compton scattering of polarized photons on unpolarized electrons at rest. Following energy and momentum conservation in Compton scattering, one obtains the well known expression
\begin{equation}\label{eq_compton_int_K_1} \cfrac {1}{k_f} - \cfrac {1}{k_i} = 1 - \cos \theta, \end{equation}
where $k_i = E_i/{mc^2}$ and $k_f = E_f/{mc^2}$ are the initial and final photon energies in units of $mc^2$ and $\theta$ is the photon scattering angle. 

The Compton cross section has been shown to depend on the angle between polarization vectors of incident and scattered photons, with the first well-known formulation of Klein-Nishina~\cite{nishina1929}, which treats only linear photon polarization. In the Stokes parameters formalism, the general treatment of Compton scattering with consideration of polarization for the incident and scattered photons and electrons has been given by Wightman in ref.~\cite{wightman1948} and brought into a convenient form by Fano in ref.~\cite{fano1949}. Reference~\cite{krafft2016} gives a generalization of the Klein-Nishina formula for vectorial treatment of polarized photons on unpolarized electrons for general complex polarization vectors, e.g. for scattering of elliptically or circularly polarized lasers. In the present section, we follow Fano's matrix formalism, which involves the use of a particular reference system referred to as \emph{Fano reference system} from here on. 

\begin{figure}[t]
\centering
\includegraphics [width=0.60\columnwidth, angle=0]{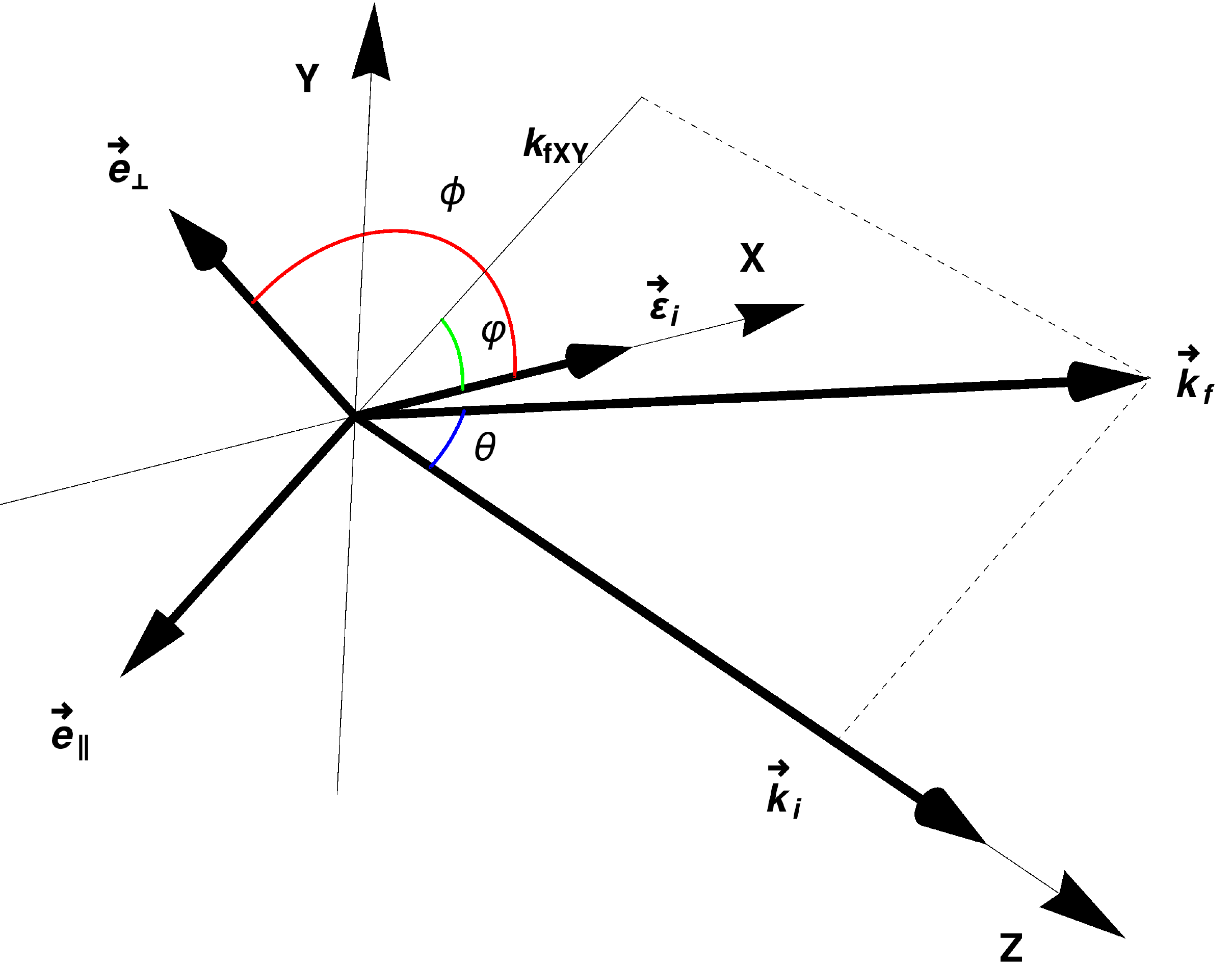}  
\caption {\label{Fano_ref_system} 
 Representation of Compton scattering process on electron at rest and the associated Fano reference system. The incident photon direction $\boldsymbol k_i = k_i \boldsymbol n_i$ is oriented along $OZ$ axis towards positive values and has the polarization vector $\boldsymbol \varepsilon _i$ aligned with $OX$ axis. After Compton scattering, photon direction becomes $\boldsymbol k_f = k_f \boldsymbol n_f$ defined by the polar $\theta$ and azimuthal $\varphi$ spherical angles with modulus $k_f$ given by equation~\eqref{eq_compton_int_K_1}. Fano reference system for the incident photon has the axes directions defined by the vectors $\boldsymbol k_i$, $\boldsymbol e_{\perp} \propto \boldsymbol k_i \times \boldsymbol k_f$ and $\boldsymbol e_{\parallel} \propto \boldsymbol k_i \times \boldsymbol e_{\perp}$. For the scattered photon, the Fano system (not represented here) is given by $\boldsymbol k_f$, $\boldsymbol e_{\perp}$ and $\boldsymbol e'_{\parallel} \propto \boldsymbol k_f \times \boldsymbol e_{\perp}$. 
}
\end{figure}

\subsection{Fano representation of Compton scattering matrix}

If we consider that the initial electron at rest is unpolarized and we express the Stokes parameters in a reference system where $P_1=+1$ refers to \emph{linear polarization perpendicular to the plane of scattering} (formed by the direction of the incident and scattered photon), it can be proved that the $T$ matrix describing Compton interaction can be expressed as
\begin{equation}\label{eq_Fano_matrix}
T = \\
\cfrac{1}{2} r^2_0 \Big (\cfrac {k_f}{k_i} \Big )^2
\begin{pmatrix}
1+\cos ^2{\theta} + (k_i - k_f)(1 - \cos {\theta}) & \sin ^2{\theta} & 0 & 0 \\
\sin ^2{\theta} & 1+\cos ^2{\theta} & 0 & 0 \\
0 & 0 & 2 \cos {\theta} & 0 \\
0 & 0 & 0 & 2 \cos {\theta} + (k_i - k_f)(1 - \cos {\theta}) \cos {\theta}
\end{pmatrix},
\end{equation}
where $r_0 = e^2/m c^2$ is the classical radius of the electron. 

This matrix was introduced by Fano and it is simplified in the present representation by eliminating all the terms related with the electron spin, as we consider that the target electron is unpolarized.
As stated above, the Stokes parameters vector on which the Fano's $T$ matrix is applied are defined relative to a coordinate system defined by the scattering plane (see Figure \ref{Fano_ref_system}). Thus, the $\boldsymbol e_{\perp}$ axis that defines $P_1 = +1$ is perpendicular to the scattering plane 
\begin{equation}
\boldsymbol e_{\perp} \parallel \boldsymbol k_i \times \boldsymbol k_f,
\end{equation}
while the $\boldsymbol e_{\parallel}$ axis that defines $P_1 = -1$ is located in the scattering plane
\begin{equation}
\boldsymbol e_{\parallel} \parallel \boldsymbol k_i \times (\boldsymbol k_i \times \boldsymbol k_f)
\end{equation}
and, of course, both of the above axes are perpendicular on the incident photon direction $\boldsymbol k_i$. The Stokes parameter $P_2=+1$ is related to a reference system rotated with 45$^\circ$ to the right of the above reference system as we are looking toward $+\boldsymbol k_i$.

In a similar way the Fano reference system is defined, to which is related the Stokes vector resulted after Compton interaction. Thus, the Fano reference system for the scattered photon is defined by the $\boldsymbol k_f$ axis, the same $\boldsymbol e_{\perp}$ axis from above and
\begin{equation}
\boldsymbol e'_{\parallel} \parallel \boldsymbol k_f \times (\boldsymbol k_i \times \boldsymbol k_f)
\end{equation}
in the scattering plane. 

Although the $T$ matrix defined in equation~\eqref{eq_Fano_matrix} doesn't contain explicit dependence on the azimuthal spherical angle, it is clear from the above definitions of Fano reference systems that the Compton scattering process depends on the relative orientations of the polarization vectors before and after scattering.

At this point it is worth to define two azimuthal angles that will be used from now on and the relation between them, in a reference system in which the incident photon direction is oriented along $OZ$ axis towards positive values and its polarization vector $\boldsymbol \varepsilon _i$ is aligned with the $OX$ axis:
\begin{enumerate}
\item $\varphi$ - azimuthal spherical angle defining the $\boldsymbol k_f$ direction of scattered photon after Compton interaction;
\item $\phi$ - the rotation angle of $\boldsymbol e_{\perp}$ vector to the right of $\boldsymbol \varepsilon _i$, as we are looking in the direction $+\boldsymbol k_i$.
\end{enumerate}
By analysing figure \ref{Fano_ref_system}, we deduce that:
\begin{equation}\label{eq_polar_angle_link}
\phi = \varphi + \cfrac{\pi}{2}
\end{equation}

\subsection{Compton scattering of polarized photons~--~general case.}

In the general case of polarized gamma beam we consider the matrix with Stokes parameters of incident gamma beam $\begin{pmatrix}1 & P_1 & P_2 & P_3\end{pmatrix}$. The probability of detecting a photon characterized by the Stokes parameters $\begin{pmatrix}1 & P_1' & P_2' & P_3' \end{pmatrix}$ after Compton scattering on unpolarized electron is
\begin{align}\label{eq_ComptonCS_PolarizedBeam_General}
\cfrac {\mathrm d \sigma}{\mathrm d \Omega} & = \cfrac{1}{2} \begin{pmatrix}1 & P_1' & P_2' & P_3'\end{pmatrix} T
\begin{pmatrix}1 \\ P_1 \\ P_2 \\ P_3\end{pmatrix} = \\
& = \nonumber \cfrac{1}{4} r_0^2 \Big ( \cfrac{k_f}{k_i} \Big )^2 \begin{pmatrix}1 & P_1' & P_2' & P_3'\end{pmatrix}
\begin{pmatrix}(k_i-k_f)(1-\cos \theta) + 1 + \cos^2 \theta + P_1 \sin^2 \theta\\ 
\sin^2 \theta + P_1 (1+\cos^2 \theta)\\
2 P_2 \cos \theta \\
P_3 \left [2\cos \theta + (k_i-k_f)(1-\cos \theta)\cos \theta \right ] \end{pmatrix} = \\
& = \nonumber \cfrac{1}{4} r_0^2 \Big ( \cfrac{k_f}{k_i} \Big )^2 \left \{ (k_i-k_f)(1-\cos \theta) + 1 + \cos^2 \theta + \sin^2 \theta P_1 + \right. \\
& + \nonumber \sin^2 \theta P_1' + (1+\cos^2 \theta) P_1 P_1' + 2\cos \theta P_2 P_2' + \\
& + \nonumber \left. \left [2\cos \theta + (k_i-k_f)(1-\cos \theta)\cos \theta \right ] P_3 P_3' \right \} = \\
& = \nonumber \cfrac{1}{4} r_0^2 \Big ( \cfrac{k_f}{k_i} \Big )^2 \left \{ F_0 + F_1 (P_1 + P_1') + F_{11} P_1 P_1' + F_{22} P_2 P_2' + F_{33} P_3 P_3' \right \}
\end{align}
where
\begin{align}
\nonumber F_0 = (k_i-k_f)(1-\cos \theta) + 1 + \cos^2 \theta = \cfrac {k_i}{k_f} + \cfrac {k_f}{k_i} - \sin^2 \theta, \\
F_1 = \sin^2 \theta , \; F_{11} = 1 + \cos^2 \theta, \; F_{22} = 2 \cos \theta, \\
\nonumber F_{33} = 2\cos \theta + (k_i-k_f)(1-\cos \theta)\cos \theta = \left ( \cfrac {k_i}{k_f} + \cfrac {k_f}{k_i} \right ) \cos \theta
\end{align}
These equations correspond to equations (88,12-13) from ref.~\cite{Landau} of Landau given in the McMaster indexing convention.

By summing the cross sections of \eqref{eq_ComptonCS_PolarizedBeam_General}, taking both signs of $P'$ Stokes parameters when detecting the outgoing photon (this involves making $P'=0$ and doubling the result), meaning that we detect the outgoing photon regardless of its polarization, we get the Compton scattering cross section of a polarized photon on an unpolarized electron
\begin{equation}\label{eq_ComptonCS_PolarizedBeam_General_Int}
\cfrac {\mathrm d \sigma}{\mathrm d \Omega} = \cfrac{1}{2} r_0^2 \Big ( \cfrac{k_f}{k_i} \Big )^2 F,
\end{equation}
where
\begin{equation}
F = F_0 + F_1 P_1 = \cfrac {k_i}{k_f} + \cfrac {k_f}{k_i} - (1-P_1) \sin^2 \theta.
\end{equation}
From this equation, one can deduce that the scattering cross section of incident photons having polarization vector perpendicular on the plane of scattering ($P_1=+1$) is much greater than the scattering cross section of photons with polarization vector in the plane of scattering ($P_1=-1$). 
We also note that, considering a detector insensitive to the polarization state of the outgoing photon, the cross section for Compton scattering of polarized photon given by \eqref{eq_ComptonCS_PolarizedBeam_General_Int} is independent on the circular polarization $P_3$ Stokes parameter and on $P_2$. 

Further on, let us consider that a photon with polarization characterized by the Stokes vector $\boldsymbol P$ undergoes a process and is recorded by a detector sensitive to $\boldsymbol P'$ polarization vector. 
It is demonstrated by Landau in chapter 66 of ref.~\cite{Landau} that, if we express the scattering amplitude as a function of the particular $\boldsymbol P'$ vector as
\begin{equation}
|M_{fi}|^2 = \alpha + \boldsymbol \beta \boldsymbol P',
\end{equation}
the polarization state of the scattered photon which can be extracted experimentally is
\begin{equation}\label{eq_PolarizationNormalization}
\boldsymbol P^{(f)} = \cfrac{\boldsymbol \beta}{\alpha}.
\end{equation}
Thus, the experimental values for the Stokes parameters of the Compton scattered photon are given by
\begin{equation}\label{eq_Stokes_Compton_transformation}
P_1^{(f)} = \cfrac{F_1+F_{11}P_1}{F}; \; P_2^{(f)} = \cfrac{F_{22}}{F} P_2; \; P_3^{(f)} = \cfrac{F_{33}}{F} P_3.
\end{equation}
It is interesting to note that, in the case of unpolarized incident photon beam, we get
\begin{equation}
P_1^{(f)} = \cfrac{\sin^2 \theta}{\cfrac{k_i}{k_f}+\cfrac{k_f}{k_i}-\sin^2 \theta}, \; P_2^{(f)} = P_3^{(f)} = 0.
\end{equation}
Thus, the forward and backward scattered photons ($\theta$~=~0 and $\pi$) are unpolarized ($\boldsymbol P^{(f)}$~=~$\boldsymbol 0$) whereas for other scattering angles they get some degree of polarization. 

Also, it follows directly from equation~\eqref{eq_Stokes_Compton_transformation} that after Compton scattering of a photon on an unpolarized electron, the scattered photon can have circular polarization only if the incident photon is circular polarized.

\subsection{Compton scattering of partially linear polarized photons}

The simplest representation of such polarized photon beam is $\begin{pmatrix} 1 & P_T & 0 & 0\end{pmatrix}$, where $P_T\leq1$ is the partial linear polarization degree. Let's recall that, when using Fano's matrix formalism, $P_1$ positive values  refer to plane polarization perpendicular to the plane of scattering (along $\boldsymbol e_{\perp}$ in figure~\ref{Fano_ref_system}). But this is in a coordinate system rotated with an angle $\phi$ to the right of $\boldsymbol e_{\perp}$ (looking in the direction $-\bf k_i$). Therefore, the coordinate system in which the polarization vector is $\begin{pmatrix} 1 & P_T & 0 & 0\end{pmatrix}$, must be rotated with an angle $\phi$ to the right when looking toward $+\bf k_i$, by using the matrix $M$ given in equation~\eqref{Rotation_Matrix}. Thus, using the plane of scattering as a reference plane, we have
\begin{equation}\label{eq_Rotate_PartialLinear_Stokes}
\begin{pmatrix}I \\ \bf P\end{pmatrix} = M \begin{pmatrix} 1 \\ P_T \\ 0 \\ 0\end{pmatrix} = \begin{pmatrix} 1 \\ P_T\cos 2\phi \\ -P_T\sin 2\phi \\ 0 \end{pmatrix} =
\begin{pmatrix} 1 \\ P_T(\cos ^2 \phi - \sin ^2 \phi) \\ -2P_T \sin \phi \cos \phi \\ 0\end{pmatrix}.
\end{equation}
Let's recall that if the Compton scattered photon is recorded by a detector insensitive to the polarization state, the cross section given in equation~\eqref{eq_ComptonCS_PolarizedBeam_General_Int} can be re-written as
\begin{equation}\label{eq_ComptonCS_PolarizedBeam_General_Int_explicit}
\cfrac {\mathrm d \sigma}{\mathrm d \Omega} = \cfrac{1}{2} r_0^2 \Big ( \cfrac{k_f}{k_i} \Big )^2 \left [ \cfrac {k_i}{k_f} + \cfrac {k_f}{k_i} - (1-P_1) \sin^2 \theta \right ].
\end{equation}
In previous equation we switch from variable $\theta$ to variable $E_f$, and by using equation~\eqref{eq_compton_int_K_1} we get the following expression for the energy and azimuthal angle differential Compton scattering cross section:
\begin{equation}\label{eq_ComptonCS_PolarizedBeam_General_DiffEne}
\cfrac{\mathrm d \sigma}{\mathrm d E_f \mathrm d \phi} = \cfrac{1}{2} r_0^2 \cfrac{mc^2}{E^2_i} \left \{ \cfrac {k_i}{k_f} + \cfrac {k_f}{k_i} + (1-P_1) \left [ 2 \Big ( \cfrac{1}{k_i} - \cfrac{1}{k_f} \Big ) + \Big ( \cfrac{1}{k_i} - \cfrac{1}{k_f} \Big )^2 \right ] \right \}
\end{equation}
and further replace $P_1$ with $P_T\cos 2\phi$ from \eqref{eq_Rotate_PartialLinear_Stokes}. We note that equation~\eqref{eq_ComptonCS_PolarizedBeam_General_DiffEne} has been expressed in terms of the $\phi$ characteristic to the Fano reference system, as shown in figure~\ref{Fano_ref_system}. However, when expressing the double differential cross section in the electron rest frame, one must change the $\phi$ variable with $\varphi$ variable using equation~\eqref{eq_polar_angle_link}. 

The energy differential Compton scattering cross section is obtained by integrating equation~\eqref{eq_ComptonCS_PolarizedBeam_General_DiffEne} over $\phi$ on its entire definition range (from $0$ to $2\pi$), after $P_1$ replacement:
\begin{equation}\label{eq_ComptonCS_General_DiffEne}
\cfrac{\mathrm d \sigma}{\mathrm d E_f} = \pi r_0^2 \cfrac{mc^2}{E^2_i} \left [ \cfrac {k_i}{k_f} + \cfrac {k_f}{k_i} + 2 \Big ( \cfrac{1}{k_i} - \cfrac{1}{k_f} \Big ) + \Big ( \cfrac{1}{k_i} - \cfrac{1}{k_f} \Big )^2 \right ]
\end{equation}
From equation~\eqref{eq_compton_int_K_1}, one can deduce that $k_f$ varies in the interval 
\begin{equation}\label{eq_k_interval}
k_f\in \left [ \cfrac{k_i}{1+2k_i}, \; k_i \right ]. 
\end{equation}
The maximum of equation~\eqref{eq_ComptonCS_General_DiffEne} is met for $\cfrac{k_i}{1+2k_i}$ and is equal with
\begin{equation}\label{eq_ComptonCS_Maximum}
{\left ( \cfrac{\mathrm d \sigma}{\mathrm d E_f}\right )}_{MAX} = \pi r_0^2 \cfrac{mc^2}{E^2_i} \left ( 1+2k_i + \cfrac{1}{1+2k_i} \right ).
\end{equation}

\subsection{Perpendicular to parallel polarization cross section ratio and polarization degree}

From the experimental point of view, the general technique of characterizing the  polarization of a Compton scattered photon beam is to measure the ratio of intensities with opposite polarizations relative to the scattering plane
\begin{equation}\label{eq_Intensity_ratio}
p = \cfrac{\mathrm d \sigma_{\perp} / \mathrm d \Omega}{\mathrm d \sigma_{\parallel} / \mathrm d \Omega},
\end{equation}
where $\mathrm d \sigma_{\perp} / \mathrm d \Omega$ and $\mathrm d \sigma_{\parallel} / \mathrm d \Omega$ give the probabilities for the polarization vector of the scattered photon to be perpendicular on the scattering plane and in the scattering plane, respectively. Using the general equation~\eqref{eq_ComptonCS_PolarizedBeam_General}, the two differential cross sections can be further expressed as
\begin{align}\label{eq_ComptonCS_PolarizedBeam_General_perpendicular}
\cfrac {\mathrm d \sigma_{\perp}}{\mathrm d \Omega} & = \cfrac{1}{2}\begin{pmatrix}1 & 1 & 0 & 0\end{pmatrix} T
\begin{pmatrix}1 \\ P_1 \\ P_2 \\ P_3\end{pmatrix} = \\
& = \nonumber \cfrac{1}{4} r_0^2 \Big ( \cfrac{k_f}{k_i} \Big )^2 \begin{pmatrix}1 & 1 & 0 & 0\end{pmatrix}
\begin{pmatrix}(k_i-k_f)(1-\cos \theta) + 1 + \cos^2 \theta + P_1 \sin^2 \theta\\ 
\sin^2 \theta + P_1 (1+\cos^2 \theta)\\
2 P_2 \cos \theta \\
P_3 \left [2\cos \theta + (k_i-k_f)(1-\cos \theta)\cos \theta \right ] \end{pmatrix} = \\
& = \nonumber \cfrac{1}{4} r_0^2 \Big ( \cfrac{k_f}{k_i} \Big )^2 \left \{ (k_i-k_f)(1-\cos \theta) + 2 (1 + P_1) \right \}; 
\end{align}
\begin{align}\label{eq_ComptonCS_PolarizedBeam_General_parallel}
\cfrac {\mathrm d \sigma_{\parallel}}{\mathrm d \Omega} & = \cfrac{1}{2}\begin{pmatrix}1 & -1 & 0 & 0\end{pmatrix} T
\begin{pmatrix}1 \\ P_1 \\ P_2 \\ P_3\end{pmatrix} = \\
& = \nonumber \cfrac{1}{4} r_0^2 \Big ( \cfrac{k_f}{k_i} \Big )^2 \begin{pmatrix}1 & -1 & 0 & 0\end{pmatrix}
\begin{pmatrix}(k_i-k_f)(1-\cos \theta) + 1 + \cos^2 \theta + P_1 \sin^2 \theta\\ 
\sin^2 \theta + P_1 (1+\cos^2 \theta)\\
2 P_2 \cos \theta \\
P_3 \left [2\cos \theta + (k_i-k_f)(1-\cos \theta)\cos \theta \right ] \end{pmatrix} = \\
& = \nonumber \cfrac{1}{4} r_0^2 \Big ( \cfrac{k_f}{k_i} \Big )^2 \left \{ (k_i-k_f)(1-\cos \theta) + 2\cos^2 \theta (1 - P_1) \right \} 
\end{align}
The sum of the two differential cross sections given by the above equations~\eqref{eq_ComptonCS_PolarizedBeam_General_perpendicular} and \eqref{eq_ComptonCS_PolarizedBeam_General_parallel} is equal to the total differential cross section given in \eqref{eq_ComptonCS_PolarizedBeam_General_Int_explicit}, proving the conservation of the total scattered photon flux. Again, equations~\eqref{eq_ComptonCS_PolarizedBeam_General_perpendicular} and \eqref{eq_ComptonCS_PolarizedBeam_General_parallel} have been given in the Fano reference system. In order to use them in the electron rest frame, where the linear polarization vector for the incident photon is set to be along the $OX$ axis, $P_1$ must be replaced with $P_T\cos 2\phi$ from \eqref{eq_Rotate_PartialLinear_Stokes}.

When taking into account \eqref{eq_ComptonCS_PolarizedBeam_General_perpendicular} and \eqref{eq_ComptonCS_PolarizedBeam_General_parallel}, the intensity ratio \eqref{eq_Intensity_ratio} becomes
\begin{equation}\label{eq_Intensity_ratio_GenCase}
p = \cfrac{(k_i-k_f)(1-\cos \theta) + 2 (1 + P_1)}{(k_i-k_f)(1-\cos \theta) + 2\cos^2 \theta (1 - P_1)}.
\end{equation}

Considering a partial linear polarized photon beam undergoing Compton scattering, the total linear polarization degree $P$ of the scattered photon is expressed as:
\begin{equation}\label{eq_LinearPolDegree_GenCase}
P = \cfrac{\sqrt{T_1 + T_2 \cdot T_3}}{\cfrac{k_i}{k_f} + \cfrac{k_f}{k_i} - T_2 - 2P_T \cdot A \cdot B}
\end{equation}
where we used the polarization degree definition \eqref{Total_polarization} and applied the Compton scattering matrix $T$ on the Stokes parameters matrix from equation~\eqref{eq_Rotate_PartialLinear_Stokes}. The following terms have been introduced:
\begin{equation}
T_1 = 4 (P_T - A \cdot B)^2, \;\; T_2 = (1-P_T) \cdot A, \;\; T_3 = T_2 + 4P_T \cdot (1+A \cdot B) - 4(P_T+1) \cdot A \cdot B^2 \nonumber
\end{equation}
in which we followed the notations $A = \sin^2 \theta$ and $B = \sin^2 \phi$.

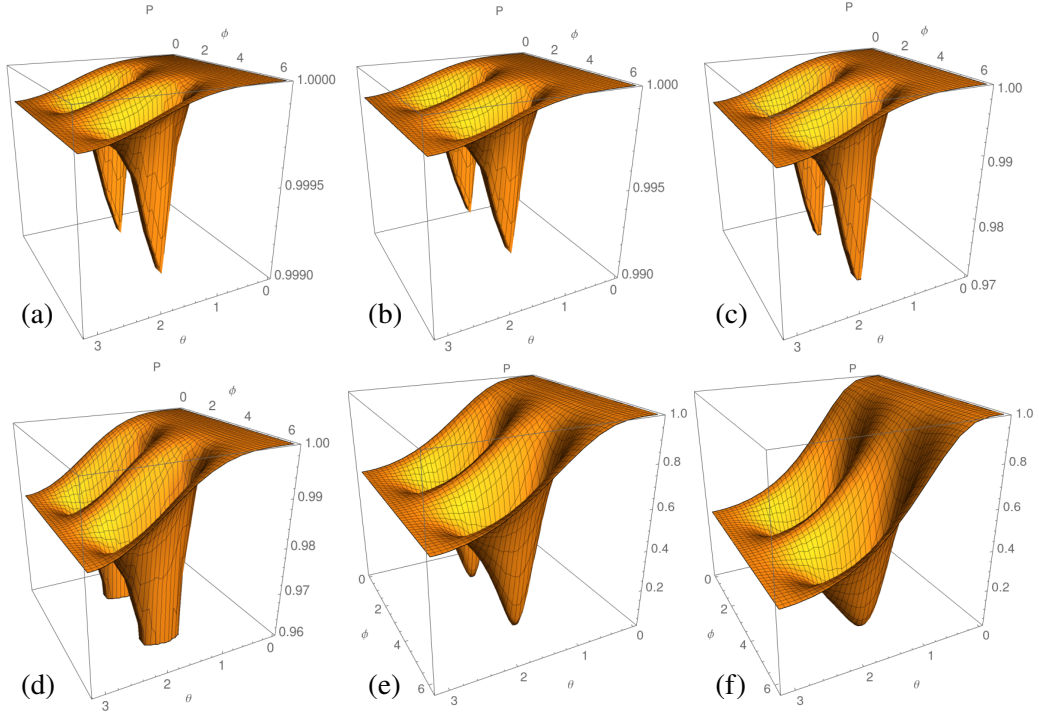
\begin{figure}[t]
\centering
\includegraphics [width=0.90\columnwidth, angle=0]{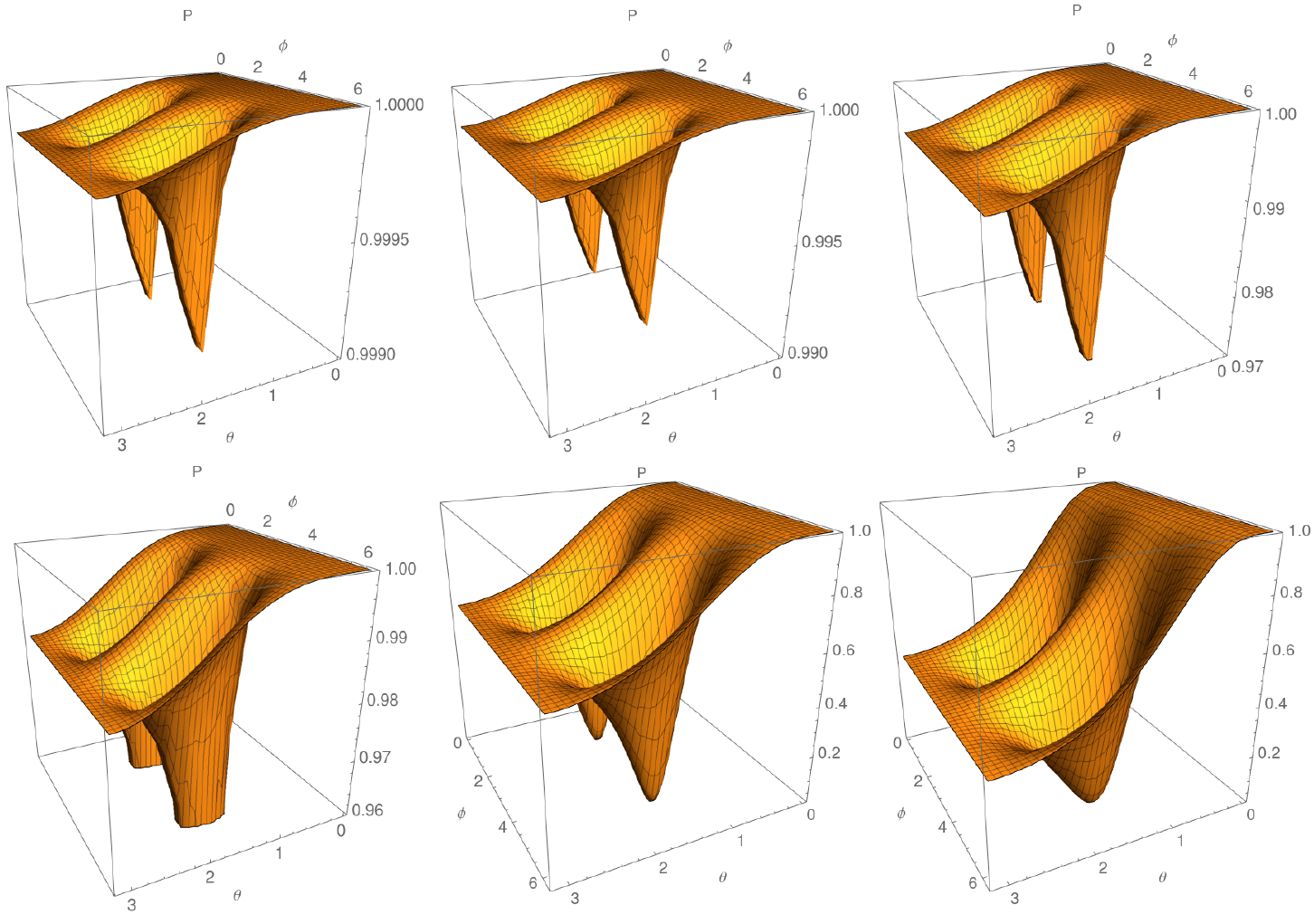}   
\put(-380,150){(a)}
\put(-250,150){(b)}
\put(-120,150){(c)}
\put(-380,10){(d)}
\put(-250,10){(e)}
\put(-120,10){(f)}
\caption {\label{fig_pol_deg_incident_energy} 
Theoretical dependence of the polarization degree with polar scattering angles $\theta$ and $\phi$ for polarized photon incident on unpolarized electron at rest. Incident photon energies: (a)~5~keV, (b)~15~keV, (c)~30~keV, (d)~50~keV, (e)~500~keV, (f)~1000~keV. 
}
\end{figure}

\subsection{Discussion on variation of polarization degree with the incident photon energy}
\label{subsec_PolWithPhotEne}

Let us consider the case of total linear polarization $P_T$~=~1, and thus $T_2$~=~0. In this case, eq.~\eqref{eq_LinearPolDegree_GenCase} can be written as:
\begin{equation}\label{eq_LinearPolDegree_LinPol}
P = \cfrac{2(1 - A \cdot B)}{\cfrac{k_i}{k_f} + \cfrac{k_f}{k_i} - 2 \cdot A \cdot B}.
\end{equation}
Figure \ref{fig_pol_deg_incident_energy} shows the theoretical dependence of the polarization degree with polar scattering angles $\theta$ and $\phi$ for various incident photon energies, as given by the equation above. For example, if we consider the head-on collision of a green laser (532~nm / 2.33~eV) with relativistic electrons of 500~MeV and 1.5~GeV which are the lower and upper energy limits of the NewSUBARU synchrotron, the corresponding photon energy in the electron rest frame will be approximately 4.55~keV and 13.7~keV respectively. These are roughly the cases represented in figures \ref{fig_pol_deg_incident_energy} (a) and (b), which illustrate the theoretical polarization degree distribution for incident photon energies of 5~keV and 15~keV respectively. From these figures it is clear that polarization degree is very high (>~99.5\%) over the entire angular region, with the exception of ($\theta$, $\phi$) pairs of (90$^\circ$, 90$^\circ$) and (90$^\circ$, 270$^\circ$) which are anyhow restricted by the differential Compton cross section. We can also consider the case of green laser photons undergoing head-on collision with 3.5~GeV relativistic electrons from SLEGS facility. In such case, the incident photon will have approximately 31.9 keV energy in the electron rest frame. If we analyze the figure \ref{fig_pol_deg_incident_energy} (c) where the theoretical polarization degree for 30~keV incident photons was represented, we deduce that polarization degree is still very high (>~99\%) over the entire angular range. 

However, if we continue to increase the photon energy, this situation changes drastically. In figures \ref{fig_pol_deg_incident_energy} (d), (e) and (f) are represented the cases for photons with energies of 50~keV, 500~keV and 1~MeV respectively, incident on electrons at rest. By analyzing these figures, it is clear that the polarization degree decreases significantly for back-scattered photons at values ranging from 98\% to less than 40\%, but also for photons scattered at angular regions around (90$^\circ$, 270$^\circ$) pairs. Of course this trend continues with even higher magnitude as the incident photon energy increases in the electron rest frame. Thus, the main conclusion that we draw from this discussion of polarized photons scattering on electrons at rest is that for incident photon energies in the electron rest frame up to tens of keV, the polarization degree of the scattered photons remains very high disregarding the scattering angle, while when increasing the incident energy to hundreds of keV the polarization degree of the scattered photon starts to have a strong dependence on the scattering angles with low values for back-scattered photons.

\section{Vectorial treatment of polarization in Compton scattering}

\subsection{Definition of polarization vector and its gauge invariant transformation}

The polarization of a photon can be described by the polarization quadrivector (4-vector), denoted with $\varepsilon$, which must fulfill the relations
\begin{equation}\label{PolarizationConditions}
\varepsilon \cdot \varepsilon = -1 \quad \mathrm{and} \quad \varepsilon \cdot p = 0\,,
\end{equation}
where $p \equiv (p^0, \boldsymbol p) = (E/c, \boldsymbol p)$ denotes the photon momentum. In the above relations, the $(+---)$ metric was used for the 4-vector scalar product, meaning that first condition can be detailed as
\begin{equation}
\varepsilon \cdot \varepsilon = {\left ( \varepsilon^0 \right )}^2 - \boldsymbol \varepsilon \cdot \boldsymbol \varepsilon,
\end{equation}
with $\varepsilon^0$  denoting the temporal component of the $\varepsilon$ 4-vector, and $\boldsymbol \varepsilon \equiv \left ( \varepsilon^1, \varepsilon^2, \varepsilon^3 \right )$ being the spatial polarization vector.

A polarization 4-vector with real components describes the linear polarization, while the one with complex components describes the circular or, more general, elliptical polarization. 
In the following, we limit the discussion to the case of linear polarization, thus we only consider polarization 4-vectors $\varepsilon \in \mathbb{R}$.

Because the rest mass of photon is zero, it results that
\begin{equation}
p \cdot p = {\left ( p^0 \right )}^2 - {\boldsymbol p}^2 = 0.
\end{equation}
Let us define the 4-vector $n$ associated to the propagation direction of a photon having the momentum $\boldsymbol p = |\boldsymbol p| \cdot \boldsymbol n$, as $n \equiv (1, \boldsymbol n)$. It is obvious that $n \cdot n = 0$. 

It has been demonstrated by Landau in reference~\cite{Landau} (chapter 8) that, if we perform a gauge transformation of polarization 4-vector $\varepsilon$
\begin{equation}\label{GaugeTransformation}
\varepsilon \rightarrow \varepsilon' = \varepsilon + \lambda p
\end{equation}
with $\lambda$ being a real number, this will not have consequences on quantities associated with physical observables. 

The polarization 4-vector $\varepsilon'$ that results after gauge transformation \eqref{GaugeTransformation} fulfills the same conditions from equation~\eqref{PolarizationConditions} that must be satisfied by a polarization 4-vector
\begin{equation}
\varepsilon' \cdot \varepsilon' = (\varepsilon + \lambda p) \cdot (\varepsilon + \lambda p) = -1, \; \; \varepsilon' \cdot p = (\varepsilon + \lambda p) \cdot p = 0
\end{equation}

Thus the quantities having physical meaning, like cross sections for example, will be invariant to a gauge transformation as in equation~\eqref{GaugeTransformation} and consequently we can choose a convenient expression of polarization $\varepsilon$ using such transformation, without affecting the final result. This statement is not true anymore for the quantities that are not direct experimental observable, thus it is not allowed to apply gauge transformation of $\varepsilon$ that enters in the expression of such quantity. 

We can choose to perform a convenient gauge transformation of the polarization 4-vector $\varepsilon$ with the real factors $\lambda$ set as a function of the temporal components of $\varepsilon$ and $p$:
\begin{equation}\label{eq_GaugeTransformationConstants}
\lambda = \cfrac{\varepsilon^0}{p^0}. 
\end{equation}
Thus, after performing the gauge transformation, the polarization 4-vector $\varepsilon$ becomes spatial by losing the temporal 0 component:
\begin{equation}
\varepsilon = (0, \boldsymbol \varepsilon) .
\end{equation}
Because polarization vector $\varepsilon$ must fulfill conditions \eqref{PolarizationConditions}, it results that the spatial vector $\boldsymbol \varepsilon$ is perpendicular on the spatial vector $\boldsymbol p$ and has the modulus equal to 1.

\subsection{Compton scattering cross section of linearly polarized gamma rays}

Let's consider the Compton scattering of a linearly polarized photon characterized initially by a momentum $p_i \equiv (E_i/c, \boldsymbol p_i)$ and polarization $\varepsilon_i$ 4-vectors on a unpolarized electron at rest, while after the scattering the photon is characterized by the momentum $p_f \equiv (E_f/c, \boldsymbol p_f)$ and polarization $\varepsilon_f$ and the scattered electron by the momentum $P_f$. For linearly polarized photons, $\varepsilon_i$ and $\varepsilon_f$ 4-vectors have real components. For linearly polarized photons, $\varepsilon_i$ and $\varepsilon_f$ 4-vectors have real components. In this case, the differential Compton scattering cross-section~\cite{nishina1929} is 
\begin{equation}
\label{eq_ComptonCS_PolarizationVector_Symplified}
\cfrac {\mathrm d \sigma}{\mathrm d \Omega} = \cfrac{1}{2} r_0^2 \Big ( \cfrac{E_f}{E_i} \Big )^2 \left [ \cfrac{1}{2}\left ( \cfrac{E_i}{E_f} + \cfrac{E_f}{E_i} \right ) -1 +2 \cos ^2 {\Theta} \right ]
\end{equation}
where $\Theta$ is the angle between initial and final polarization spatial vectors of photon. 
Eq.~\ref{eq_ComptonCS_PolarizationVector_Symplified} is a particularization of the more general formula given in ref.~\cite{krafft2016}(eq.~53), which considers complex polarization vectors for the case of circularly or elliptically polarized photons.

\begin{figure}[t]
\centering
\includegraphics [width=0.70\columnwidth, angle=0]{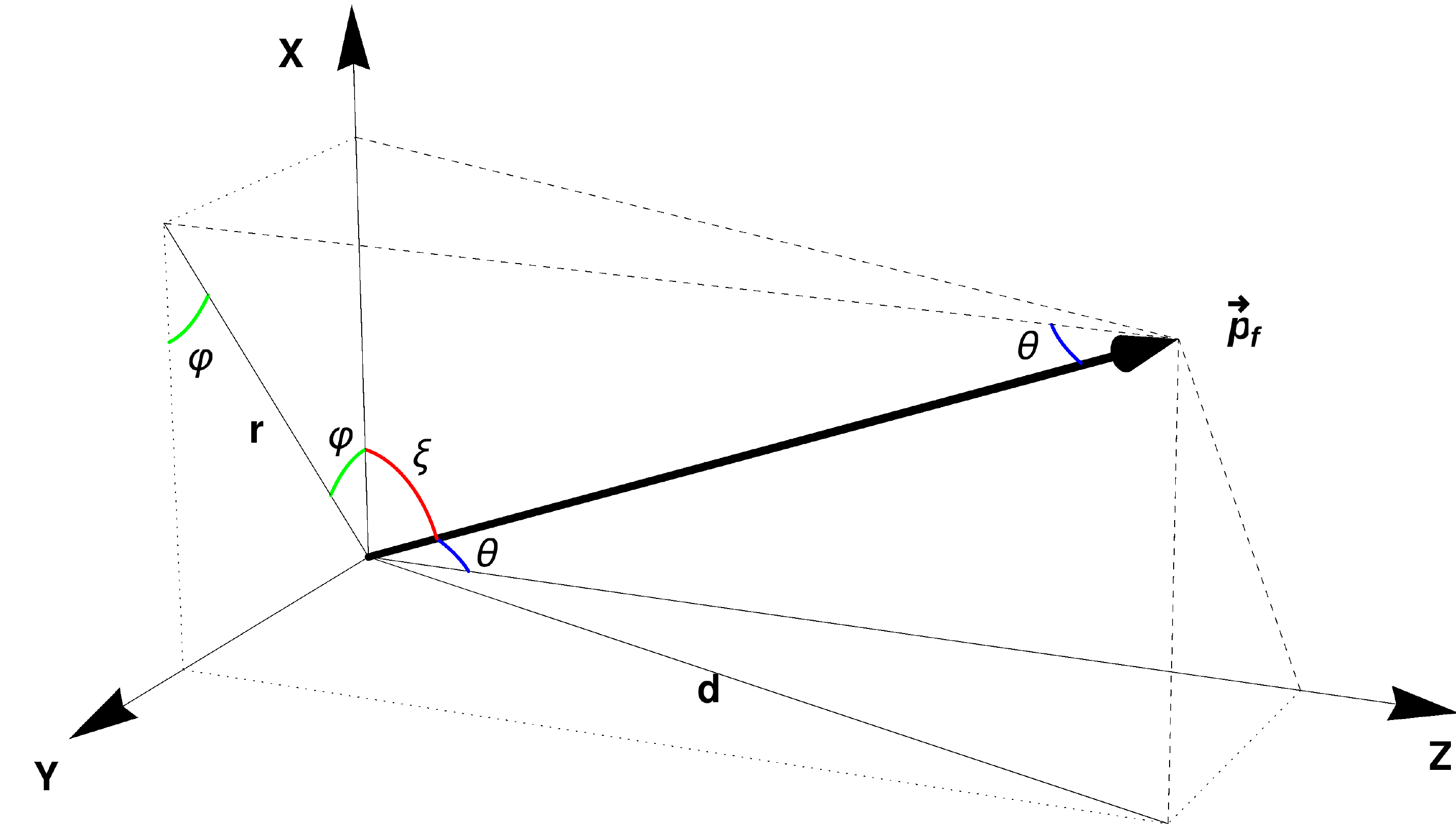} 
\caption {\label{Vectors_Diagram} 
 Vectorial representation of the Compton scattering process and the angles involved. The incident photon momentum $\boldsymbol p_i$ is aligned along $\boldsymbol{OZ}$ axis and has the associated polarization vector $\boldsymbol \varepsilon_i$ aligned along $\boldsymbol{OX}$ axis, heading toward an unpolarized electron which is at rest. Besides the polar $\theta$ and azimuthal $\varphi$ spherical angles, in the diagram there are also depicted the angles $\xi$ (angle between initial polarization vector $\boldsymbol \varepsilon_i$ and photon momentum vector after scattering $\boldsymbol p_f$) and $\alpha$ (angle between projection of $\boldsymbol p_f$ vector on $\boldsymbol{YZ}$ plane and initial photon momentum $\boldsymbol p_i$).
}
\end{figure}

In the following we will consider that the incident photon momentum $\boldsymbol p_i$ is aligned along $\boldsymbol{OZ}$ axis and has the associated polarization vector $\boldsymbol \varepsilon_i$ aligned along $\boldsymbol{OX}$ axis, heading toward an unpolarized electron which is at rest, as mentioned already (see figure~\ref{Vectors_Diagram}). The following useful relation can be deduced from figure~\ref{Vectors_Diagram}:
\begin{equation}\label{eq_1stAngularRelation}
\cos \xi = \cos \varphi \cdot \sin \theta
\end{equation}

When performing an experiment aiming to measure the photon polarization, this always involves an analysis of polarization along two perpendicular axes. Thus, a convenient reference system must be chosen having $\boldsymbol OZ$ axis along photon momentum, and then perform polarization measurement along the other two axes $\boldsymbol OX$ and $\boldsymbol OY$ perpendicular on the photon momentum and perpendicular between them. For example, in the case of polarization treatment using Stokes parameters, as it was shown in the previous section, the convenient reference system is the so called Fano system that has the $\boldsymbol OX$ axis perpendicular on the plane defined by the photon momentum vectors before $\boldsymbol p_i$ and after $\boldsymbol p_f$ the scattering.

In the case of vectorial treatment of polarization, we follow the treatment given by Depaola in ref.~\cite{Depaola} and choose one of the directions of polarization measurement as the perpendicular on the plane defined by the initial polarization vector $\boldsymbol \varepsilon_i$ and the final momentum vector of photon $\boldsymbol p_f$ after scattering. We will denote this direction with $\perp$ symbol. Of course, the other direction will be both perpendicular on propagation direction of the scattered photon defined by the $\boldsymbol p_f$ vector and on the previously defined direction ($\perp$), thus it will be contained in the plane defined by the vectors $\boldsymbol \varepsilon_i$ and $\boldsymbol p_f$. This direction is denoted with $\parallel$ symbol. The plane formed by $\boldsymbol \varepsilon_i$ and $\boldsymbol p_f$ vectors, along with vectorial decomposition of $\boldsymbol \varepsilon_f$ vector and the significant angles are illustrated in figure~\ref{Vectorial_RefSystem}.

From figure~\ref{Vectorial_RefSystem} we can deduce the following relation between angles:
\begin{figure}[t]
\centering
\includegraphics [width=0.50\columnwidth, angle=0]{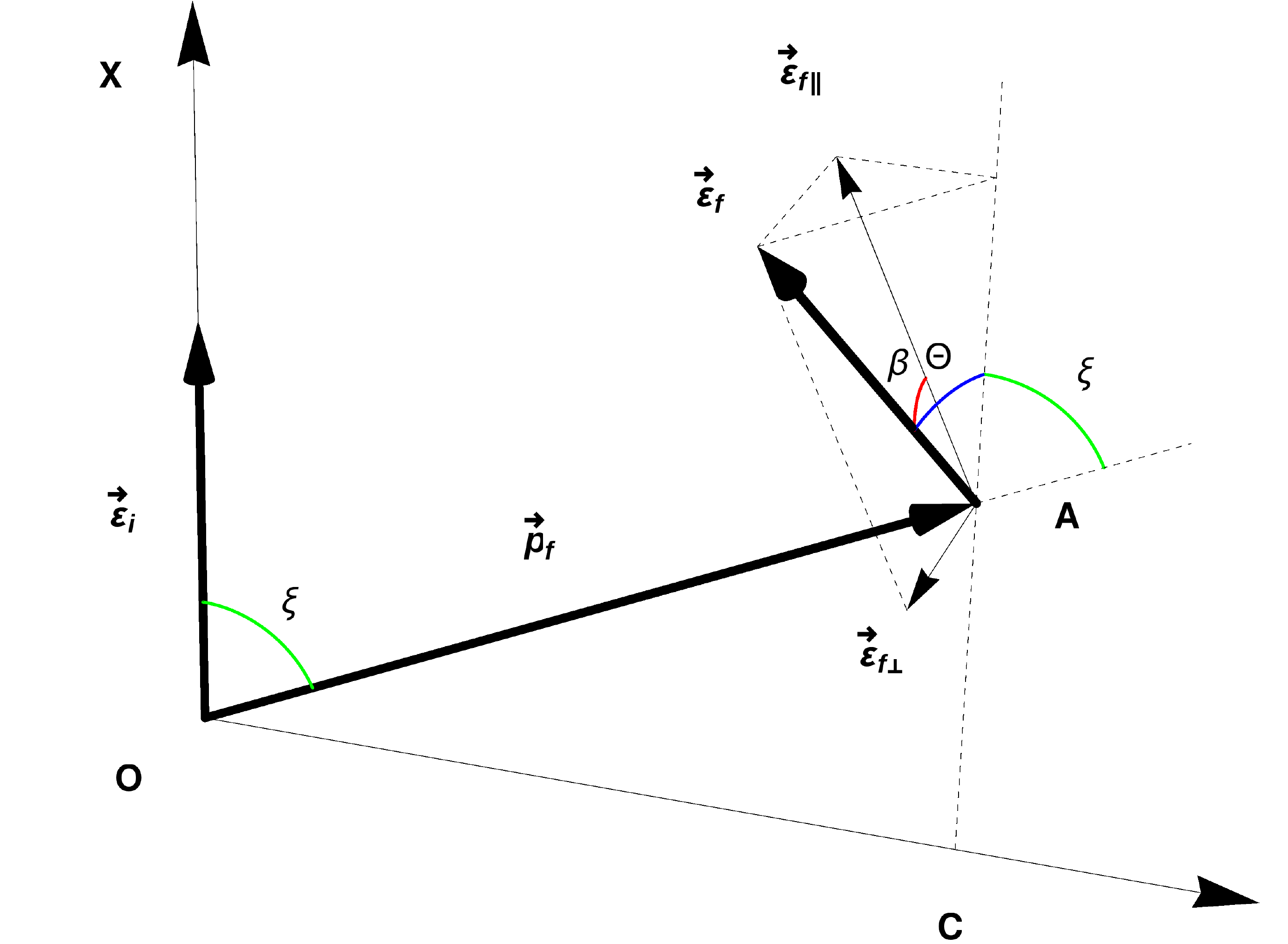} 
\caption {\label{Vectorial_RefSystem} 
 Representation of the plane defined by the vectors $\boldsymbol \varepsilon_i$ and $\boldsymbol p_f$. The polarization vector of the scattered photon $\boldsymbol \varepsilon_f$ makes the angle $\Theta$ with the direction corresponding to initial polarization vector $\boldsymbol \varepsilon_i$ and is decomposed in the $\boldsymbol \varepsilon_{f \perp}$ and $\boldsymbol \varepsilon_{f \parallel}$ components perpendicular and in the $(\boldsymbol \varepsilon_i, \boldsymbol p_f)$ plane respectively. The angle between $\boldsymbol \varepsilon_f$ vector and $(\boldsymbol \varepsilon_i, \boldsymbol p_f)$ plane is denoted with $\beta$.
}
\end{figure}

\begin{equation}\label{eq_3rdAngularRelation}
\cos \Theta = \sin \xi \cdot \cos \beta
\end{equation}

As can be seen from equation~\eqref{eq_ComptonCS_PolarizationVector_Symplified}, the differential Compton scattering cross section depends on the angle between initial and final directions of polarization vector $\cos^2 \Theta = (\boldsymbol \varepsilon_i \; \boldsymbol \varepsilon_f)^2$. Consequently, when performing polarization measurements along perpendicular ($\perp$) and parallel ($\parallel$) directions, we obtain the following cross sections for having the polarization of the scattered photon along $\perp$ or $\parallel$:
\begin{equation}\label{eq_Sig_Perp}
\cfrac {\mathrm d \sigma_{\perp}}{\mathrm d \Omega} = \cfrac{1}{4} r_0^2 \Big ( \cfrac{E_f}{E_i} \Big )^2 \left [ \cfrac{E_i}{E_f} + \cfrac{E_f}{E_i} -2 \right ],
\end{equation}
\begin{equation}
\cfrac {\mathrm d \sigma_{\parallel}}{\mathrm d \Omega} = \cfrac{1}{4} r_0^2 \Big ( \cfrac{E_f}{E_i} \Big )^2 \left [ \cfrac{E_i}{E_f} + \cfrac{E_f}{E_i} -2 +4 \cos ^2 {(\pi /2 - \xi)} \right ],
\end{equation}
and by using \eqref{eq_1stAngularRelation}, we obtain
\begin{equation}\label{eq_Sig_Parallel}
\cfrac {\mathrm d \sigma_{\parallel}}{\mathrm d \Omega} = \cfrac{1}{4} r_0^2 \Big ( \cfrac{E_f}{E_i} \Big )^2 \left [ \cfrac{E_i}{E_f} + \cfrac{E_f}{E_i} -2 +4 (1 - \sin^2 \theta \cos^2 \varphi) \right ]
\end{equation}

The differential Compton scattering cross section disregarding the polarization state of the scattered photon is further obtained by summing the perpendicular~\eqref{eq_Sig_Perp} and parallel~\eqref{eq_Sig_Parallel} components:
\begin{equation}\label{eq_SumCSPerp_and_Par}
\cfrac {\mathrm d \sigma}{\mathrm d \Omega} = \cfrac {\mathrm d \sigma_{\perp}}{\mathrm d \Omega} + \cfrac {\mathrm d \sigma_{\parallel}}{\mathrm d \Omega} = \cfrac{1}{2} r_0^2 \Big ( \cfrac{E_f}{E_i} \Big )^2 \left [ \cfrac{E_i}{E_f} + \cfrac{E_f}{E_i} -2 \sin^2 \theta \cos^2 \varphi \right ].
\end{equation}
The expression above reproduces the results of equation~\eqref{eq_ComptonCS_PolarizedBeam_General_Int_explicit} when taking into account also the relation~\eqref{eq_polar_angle_link} between angles $\phi$ and $\varphi$.

An important aspect that should be discussed is the link between the two types of treatments of polarization, namely the treatment using Stokes parameters formalism and the vectorial treatment. If a polarization vector is given, meaning the photon has total degree linear polarization, then we can express the Stokes parameter $P_1$ in a certain reference system and the Stokes parameter $P_2$ in a reference system rotated with 45$^\circ$ to the right of the previous system as we are looking towards the photon propagation direction. Usually the $\boldsymbol{OZ}$ axis of the reference system relative to which the Stokes $P_1$ parameter is expressed, is chosen along the photon propagation direction. In this case, if we denote with $\varepsilon_1$ and $\varepsilon_2$ the components of $\boldsymbol \varepsilon$ vector on the axes $\boldsymbol{OX}$ and $\boldsymbol{OY}$ respectively, we can write the expressions of $P_1$ and $P_2$ Stokes parameters using equations~\eqref{eq_Stokes1_expression} and \eqref{eq_Stokes2_expression}:
\begin{equation}
P_1 = \varepsilon_1^2 - \varepsilon_2^2
\end{equation}
\begin{equation}
P_2 = 2 \varepsilon_1 \varepsilon_2
\end{equation}

In the cases where the $\boldsymbol {OZ}$ axis is not aligned with the direction of photon propagation, because the component of $\boldsymbol \varepsilon$ polarization vector on $\boldsymbol {OZ}$ axis is non-zero anymore, we have to re-normalize the above two expressions with the value given by equation~\eqref{eq_Stokes0_expression}, in a similar way it was done in equation~\eqref{eq_PolarizationNormalization} in order to re-normalize the experimental Stokes parameters:
\begin{equation}\label{eq_Stokes1FromVector}
P_1 = \cfrac{\varepsilon_1^2 - \varepsilon_2^2}{\varepsilon_1^2 + \varepsilon_2^2}
\end{equation}
\begin{equation}\label{eq_Stokes2FromVector}
P_2 = \cfrac{2 \varepsilon_1 \varepsilon_2}{\varepsilon_1^2 + \varepsilon_2^2}
\end{equation}

\section{Algorithm for Monte Carlo modeling Compton scattering of laser photons on relativistic electrons with the consideration of photon polarization}

Laser Compton scattering on relativistic electrons is used to produce quasi-monochromatic and polarized $\gamma$-ray beams for basic research and applications. Modeling the LCS $\gamma$-ray beam production and transport is therefore necessary for both beam diagnostics and for estimating the beam parameters necessary for specific investigations. 

In order to design LCS sources, fast calculation codes are needed to provide the characteristics of the photon beam and their dependence on important parameters, such as the electron beam emittance and energy resolution, as well as the temporal characteristics of the electron and laser pulses. Thus, codes such as \textsc{iccs}~\cite{ranjan2018} have been developed, that use analytical formulas in the laboratory frame system, numerically integrate the individual electron spectra, and then add up and average the emission from the electron bunch. The \textsc{iccs} code describes the finite pulse effects possible in a real laser pulse given in plane-wave approximation. Another analytical description is given by the \textsc{sense}~\cite{terzic2019} code, which treats the nonlinear effects in the Thomson scattering of high intensity lasers beams on low to medium energy electron beams.  

At the expense of longer computation times, Monte Carlo LCS modeling codes find applications in providing event by event generators, as required by radiation transport Monte Carlo codes, such as \textsc{Geant4}, \textsc{mcnp}~\cite{goorley2012}, \textsc{fluka}~\cite{bohlen2014}, \textsc{egs4}~\cite{dinter1988}, etc. Following the procedure applied in various existing Monte Carlo simulation codes of laser Compton scattering sources, such as \textsc{cain}~\cite{CAIN}, \textsc{mccmpt}~\cite{Sun2011_STAB}, \textsc{cmcc}~\cite{Curatolo2017,Curatolo_PhD_thesis}, \textsc{mclcss}~\cite{Luo2011}, etc., the electron and laser quadrivectors are transformed to obtain a head-on collision in the electron rest frame coordinate system, as has been treated in the sections before.   

Here we describe step-by-step the Monte Carlo algorithm for simulating the laser Compton scattering between laser photons and relativistic electrons with focus on describing the photon polarization. 
Details on the spectral distribution and flux of LCS $\gamma$-ray beams with focus on the realistic modeling of unsynchronized laser and relativistic electron beams will be given in a follow-up paper~\cite{filipescu_NIM_LCS_22}.

\begin{enumerate}[leftmargin=*]   

\subsection{Initialize physical quantities}\label{sec_initialize_phys_quant}

\item Considering the LCS $\gamma$-ray beam produced along a straight path of a circular electron storage~/~synchrotron ring, we define the laboratory frame system relative to the electron accelerator, as following: the $\boldsymbol {OZ}$ axis along the electron beam axis, $\boldsymbol {OX}$ axis in the plane of the accelerator and $\boldsymbol {OY}$ axis perpendicular on the plane of the accelerator.

\item Realistic modeling of the electron and laser beams provides the initial values for the:
\begin{itemize}[leftmargin=*] 
\item[--] laser~--~electron interaction position;
\item[--] initial momentum and energy of the electron;
\item[--] the initial momentum and energy of the photon; 
\item[--] initial orientation of photon polarization vector in a plane perpendicular to the direction of photon propagation. 
The polarization vector is defined by the unit vectorial product between the direction of photon propagation and the rotated $\boldsymbol {OY}$ axis. We specify that the $\boldsymbol {OY}$ axis is rotated around the $\boldsymbol {OZ}$ axis with the linear polarization angle $\tau$ of the incident laser photon. At this stage, the 4-vector associated to the laser photon polarization is of spatial type, namely it has the temporal 0-component equal with 0, $\varepsilon_i = (0, \boldsymbol \varepsilon_i)$.
\end{itemize}
Having initialized the above quantities, the algorithm proceeds with the simulation of the interaction process.

\subsection{Transformation to electron rest frame system}\label{sec_transformation_el_rest}

\item \label{it_ElectronPolarAng} We rotate the electron momentum, photon momentum and photon polarization vectors in order to align the electron momentum vector along the $\boldsymbol {OZ}$ axis of the laboratory frame system. 

\item \label{it_RelBoost} We perform a boost transformation of the electron momentum, photon momentum and photon polarization 4-vectors with relativistic ratio $-\beta$ of the electron along the $\boldsymbol {OZ}$ axis, in order to reach the electron rest frame;
 
\item \label{it_LaserPolarAng} In the electron rest frame, the photon momentum and polarization vectors are rotated in order to align the photon momentum along the $\boldsymbol {OZ}$ axis towards positive values. 
At this stage, the 4-vector associated to the photon polarization most probable presents a non-zero temporal component. We perform the gauge transformation as described in the equations~\eqref{GaugeTransformation} and \eqref{eq_GaugeTransformationConstants}, generating in this way spatial polarization vector perpendicular on the photon momentum vector, which imply that the polarization vector has non-zero components only in the $\boldsymbol{XOY}$ plane.

\item \label{it_TauRotation} We perform a final rotation (with an angle denoted with $\tau$) of the polarization vector around $\boldsymbol {OZ}$ axis in order to align the polarization vector with the $\boldsymbol{OX}$ axis.

\subsection{Random generation of scattered photon and electron 4-momentum}

\item We generate the photon energy after Compton scattering. For this, we generate a uniformly distributed random number $k$ in the interval given by \eqref{eq_k_interval}.
Then, we generate a second uniform random number between $0$ and the maximum value specified by the equation~\eqref{eq_ComptonCS_Maximum} and apply the acception-rejection method: if the later random number is lower than the cross section given by \eqref{eq_ComptonCS_General_DiffEne}, we accept the $k$ generated value or otherwise we reject it and repeat the current step. Having so generated the photon energy after the Compton scattering, we use equation~\eqref{eq_compton_int_K_1} to compute the photon scattering polar angle $\theta$.

\item \label{it_PhiGeneration} We generate the angle  $\phi$ between the normal on the plane given by the initial and final momentum of the photon $\boldsymbol e_{\perp} \propto \boldsymbol p_0 \times \boldsymbol p$ and the initial polarization vector $\boldsymbol \varepsilon_0$ (see equation~\eqref{eq_Rotate_PartialLinear_Stokes} and figure~\eqref{Fano_ref_system}). To do this, we use equation~\eqref{eq_ComptonCS_PolarizedBeam_General_DiffEne} and apply again the acception-rejection method. Then, we apply \eqref{eq_polar_angle_link} to obtain the azimuthal angle $\varphi$. Knowing the photon energy after scattering and its direction given by the polar angles $\theta$ and $\varphi$, we can compute the energy and direction of the electron after the Compton scattering.

\subsection{Photon polarization in vectorial treatment and Stokes parameters formalism}

\item \label{it_PolarizationGenetation} \textbf{Vectorial treatment.} We decide if the polarization of the photon after scattering is along $\boldsymbol \varepsilon_{f \parallel}$ or along $\boldsymbol \varepsilon_{f \perp}$. We use equations~\eqref{eq_Sig_Perp}, \eqref{eq_Sig_Parallel} and \eqref{eq_SumCSPerp_and_Par} and choose between the two variants based on the probabilities:
\begin{equation}
\mathcal{P}_{\perp} = \left. \cfrac {\mathrm d \sigma_{\perp}}{\mathrm d \Omega} \middle/ \cfrac {\mathrm d \sigma}{\mathrm d \Omega} \right. \;\; \mathrm{and} \;\; \mathcal{P}_{\parallel} = \left. \cfrac {\mathrm d \sigma_{\parallel}}{\mathrm d \Omega} \middle/ \cfrac {\mathrm d \sigma}{\mathrm d \Omega} \right.
\end{equation}
The method follows the procedure recommended in reference~\cite{danxu_2005} and adopted in \textsc{Geant4}~\cite{geant_agostinelli_2003,geant_allison_2006,geant_allison_2016} in the \verb|G4LivermorePolarizedComptonModel| class. In reference~\cite{Depaola} of Depaola, it is suggested to generate the orientation of the polarization vector $\boldsymbol \varepsilon_f$ according to equation~\eqref{eq_ComptonCS_PolarizationVector_Symplified}, and this option has also been implemented in the present code. However, we find that it tends to excessively randomize the polarization vector orientation. We note that, in the process of generating the $\theta$ and $\varphi$ polar angles described in the previous steps, we already took into account the correlation of the angular distribution of the scattered photon with the initial direction of photon's linear polarization vector. In addition, as already stated above, when performing an experiment aiming to measure polarization of a given photon scattered in the direction defined by the $\theta$ and $\varphi$ polar angles, this involves an analysis of polarization along two perpendicular axes, fulfilling flux conservation condition~\eqref{eq_SumCSPerp_and_Par}.

\item \label{it_LandauStokesTransformation} \textbf{Stokes parameters formalism.} In parallel with the vectorial treatment of step~\ref{it_PolarizationGenetation}, we model the polarization also with Stokes parameters formalism. In order to do this, we compute the initial Stokes parameters in Fano reference system, before Compton scattering on electron at rest. According to equation~\eqref{eq_Rotate_PartialLinear_Stokes}, the expression for the initial Stokes parameters are:
\begin{equation}\label{eq_rotation_pol_degree}
P_1 = P_T \cos{2\phi}, \;\; P_2 = -P_T \sin{2\phi}, \;\; P_3 = P_C
\end{equation}
where $P_T$ and $P_C$ are the degree of linear and circular polarization and are provided at input. $P_T$ and $P_C$ parameters have to fulfill the condition:
\begin{equation}
P^2 = P_T^2 + P_C^2 \leq 1
\end{equation}
As demonstrated by Landau in reference~\cite{Landau} (chapter 8), quantities $P_T=\sqrt{P_1^2 + P_2^2}$ and $P_C=P_3$ are Lorentz invariant and remain unchanged to boost transformations. In addition, the initial values $P_T$ and $P_C$ of the laser photon are not changed during various rotation operations. As shown in expression~\eqref{Rotation_Matrix} of the rotation operator matrix for the Stokes parameters, indeed the share between Stokes parameters $P_1$ and $P_2$ changes with rotation angle $\phi$, but in such a way that $\sqrt{P_1^2 + P_2^2}$ and $P_3$ remain constant. Thus, after all operations performed in section~\ref{sec_transformation_el_rest}, the initial laser photon $P_T$ and $P_C$ remain unchanged.

Angle $\phi$ from equations~\eqref{eq_rotation_pol_degree} is the one generated at step~\ref{it_PhiGeneration} and represented in figure~\ref{Fano_ref_system}.

Having the initial Stokes parameter before Compton Scattering, we apply equations~\eqref{eq_Stokes_Compton_transformation} in order to obtain the Stokes parameters of the scattered photon.

At this stage, we are able to compute the degree of linear polarization after Compton scattering using the formula:
\begin{equation}\label{eq_LinearPolDegree}
{P'_T}^2 = \sqrt{{P'_1}^2 + {P'_2}^2}
\end{equation}

For crosscheck purposes, we compute also the intensity ratio $\left. {\mathrm d \sigma_{\perp}(\Omega)} \middle/ {\mathrm d \sigma_{\parallel}(\Omega)} \right.$ using equation~\eqref{eq_Intensity_ratio_GenCase}.

\item Rotate back with angle $-\tau$ from step \ref{it_TauRotation} the photon momentum after scattering $\boldsymbol p_f$ and its polarization vector $\boldsymbol \varepsilon_f$ and the electron momentum after scattering $\boldsymbol P_f$.

\item \label{it_StokesInFanoFromVector} \textbf{Intervalidation of vectorial and Stokes parameters treatments.} At this point, we have the option to obtain the photon Stokes parameters after Compton scattering not only from Landau's transformation applied at step~\ref{it_LandauStokesTransformation}, but also directly from the polarization vector $\boldsymbol \varepsilon_f$. The Stokes parameters must be determined in the Fano reference system shown in figure~\ref{Fano_ref_system}. 
To do this, we compute the versors $\boldsymbol e_{\perp} \propto \boldsymbol p_i \times \boldsymbol p_f$ and $\boldsymbol e'_{\parallel} \propto \boldsymbol p_f \times \boldsymbol e_{\perp}$ and decompose the polarization vector $\boldsymbol \varepsilon_f$ generated at step~\ref{it_PolarizationGenetation} along these directions in order to obtain the components $\boldsymbol {\varepsilon'}_1$ and $\boldsymbol {\varepsilon'}_2$. Having the components $\boldsymbol {\varepsilon'}_1$ and $\boldsymbol {\varepsilon'}_2$ of photon polarization vector $\boldsymbol \varepsilon_f$, we use equations~\eqref{eq_Stokes1FromVector} and \eqref{eq_Stokes2FromVector} to compute the Stokes parameters of the photon, after Compton scattering, in the Fano reference system.

This vectorial formalism assumes a priori that the incident photon has full linear polarization degree. For treating partially polarized beams, the degree of linear polarization after Compton scattering given by equation~\eqref{eq_LinearPolDegree_GenCase} is used to re-normalize the Stokes parameters obtained as mentioned before. The procedure reproduces the average Stokes values obtained in the exclusive Stokes formalism give at step~\ref{it_LandauStokesTransformation}, which is the recommended option in the present algorithm. 

As in the case of Stokes formalism of step~\ref{it_LandauStokesTransformation}, the intensity ratio $\left. {\mathrm d \sigma_{\perp}(\Omega)} \middle/ {\mathrm d \sigma_{\parallel}(\Omega)} \right.$ can also be obtained. Having computed the vector components $\boldsymbol {\varepsilon'}_1$ and $\boldsymbol {\varepsilon'}_2$ of photon polarization vector $\boldsymbol \varepsilon_f$ in Fano reference system along $\boldsymbol e_{\perp}$ and $\boldsymbol e'_{\parallel}$ directions respectively, we replace successively the $\cos \Theta$ term from equation~\eqref{eq_ComptonCS_PolarizationVector_Symplified} with $\boldsymbol \varepsilon_i \cdot \boldsymbol {\varepsilon'}_1$ and $\boldsymbol \varepsilon_i \cdot \boldsymbol {\varepsilon'}_2$ scalar products in order to compute $ \mathrm d \sigma_{\perp}/{\mathrm d \Omega}$ and $\mathrm d \sigma_{\parallel}/\mathrm d \Omega$.

\subsection{Transformation to laboratory system and to \textsc{Geant4} \emph{"Particle Reference Frame"}}

\item \label{it_TransformBack} In this step, we apply a set of transformations to the following 4-vectors: electron momentum $P_f$, photon momentum $p_f$, photon linear polarization vector $\varepsilon_f$, versors of the Fano reference system $e_{\perp} = (0,\boldsymbol e_{\perp})$ and $e'_{\parallel} = (0,\boldsymbol e'_{\parallel})$. First we rotate back all vectors with spherical angles used at step~\ref{it_LaserPolarAng} in order to obtain the incident photon momentum aligned along the $\boldsymbol {OZ}$ axis towards positive values. Then we perform back the relativistic boost of the 4-vectors in order to reach the laboratory system, using the relativistic ratio $\beta$ from step~\ref{it_RelBoost}. Finally, we rotate back all vectors with the spherical angles from step~\ref{it_ElectronPolarAng} in order to align the electron momentum vector along the $\boldsymbol {OZ}$ axis of the laboratory system.

\item \label{it_GaugeTransformation} Following transformations performed in the previous step, the polarization type 4-vectors $\varepsilon_f$, $e_{\perp}$ and $e'_{\parallel}$ have non-zero temporal component. Consequently, we perform the gauge transformation as described in equations~\eqref{GaugeTransformation} and \eqref{eq_GaugeTransformationConstants}, generating spatial polarization vectors in the laboratory reference system, orthogonal on the photon momentum vector $\boldsymbol p_f$. We note that, after gauge transformation, the $\boldsymbol e_{\perp}$ and $\boldsymbol e'_{\parallel}$ vectors are also mutually perpendicular. 

\item \label{it_StokesInPRF} The present algorithm has been implemented in the \textsc{Geant4} framework to perform simulation of radiation transport. As specified in \textsc{Geant4} \emph{Physics Reference Manual} \cite{GEANTPhysRef}, if the Stokes parameters of the photon are provided, they have to be provided in the so-called \emph{"Particle reference frame"} (PRF). According to the \textsc{Geant4} manual (\cite{GEANTPhysRef} Chapter 17.1.3, page 258), the PRF $z$-direction coincides with the particle's direction of motion and the $x$-axis lies in the $\boldsymbol{X-Y}$ plane of world frame (laboratory frame in our case). We denote with $\boldsymbol e_x$, $\boldsymbol e_y$ and $\boldsymbol e_z$ the versors of the \emph{"Particle reference frame"}. Versor $\boldsymbol e_x$ is obviously perpendicular on $\boldsymbol e_z$, and, as it is contained in the $\boldsymbol{X-Y}$ plane, it must be perpendicular also on the $\boldsymbol{OZ}$ axis of the laboratory reference system: 
\begin{equation}\label{eq_eX_versor_expression}
\boldsymbol e_x = \cfrac{\boldsymbol e_Z \times \boldsymbol e_z}{|\boldsymbol e_Z \times \boldsymbol e_z|}; \;\; \boldsymbol e_Z = (0,0,1).
\end{equation}
In order to provide the photon Stokes parameters in this \emph{"Particle reference frame"}, we recall that, when performing the transformation from the rest electron frame, where the Stokes parameters were obtained at step~\ref{it_LandauStokesTransformation}, to laboratory frame, we perform only vector rotations and relativistic boost transformations, operations for which the linear $P'_T$ and circular $P'_C$ degree of polarization are invariant. This means that the Stokes parameters $P^{(PRF)}_1$ and $P^{(PRF)}_2$ in the \emph{"Particle reference frame"} can be obtained from the Stokes parameters $P'_1$ and $P'_2$ obtained at step~\ref{it_LandauStokesTransformation} by performing a simple rotation operation as shown in equation~\eqref{eq_StokesRotationTransformation}.

We notice that the orthogonal versors $\boldsymbol e_{\perp}$ and $\boldsymbol e'_{\parallel}$ obtained in the previous step are contained in a plane perpendicular on the photon propagation direction $\boldsymbol p_f$, in a similar way with the orthogonal system composed of $\boldsymbol e_x$ and $\boldsymbol e_y$ versors. Thus, the orthogonal system consisting from $\boldsymbol e_x$ and $\boldsymbol e_y$ versors can be generated by a rotation with a certain angle $\phi_0$ of the orthogonal versors $\boldsymbol e_{\perp}$ and $\boldsymbol e'_{\parallel}$. Consequently, the rotation angle used in \eqref{eq_StokesRotationTransformation} to transform the Stokes parameters in order to provide them into the \emph{"Particle reference frame"} is exactly the angle $\phi_0$ expressed as the angle between the versors $\boldsymbol e_x$ and $\boldsymbol e_{\perp}$.

\item \label{it_StokesFromVectorInPRF} The Stokes parameters in the \emph{"Particle reference frame"} can be deduced also from the polarization vector $\boldsymbol \varepsilon_f$ generated at step \ref{it_GaugeTransformation}. For this, we decompose $\boldsymbol \varepsilon_f$ vector in the \emph{"Particle reference frame"} and use its components $\boldsymbol {\varepsilon'}_1$ and $\boldsymbol {\varepsilon'}_2$ along $\boldsymbol e_x$ and $\boldsymbol e_y$ versors defined at the previous step, in equations~\eqref{eq_Stokes1FromVector} and \eqref{eq_Stokes2FromVector} to compute the Stokes parameters.
\end{enumerate}

\subsection{Remarks on algorithm implementation in \textsc{Geant4}}

\subsubsection{Stokes parameters representation in laboratory frame system}

Because the Stokes parameters do not look very intuitive when they are represented in the \emph{"Particle reference frame"}, we can represent the Stokes parameters also in the laboratory frame system.

We have to remark that due to the high relativistic boost the photon gains in the Compton back-scattering process on high energy electron, the gamma photons are very forward focused, and in fact they are directed almost along the electron beam, meaning we can consider that the propagation direction of the photon in the laboratory frame system is along $\boldsymbol{OZ}$ axis.

So, for the purpose of determining the Stokes parameters in the laboratory system, we determine the angle between $\boldsymbol e_X = (1,0,0)$ and $\boldsymbol e_{\perp}$ versors, because both of them can be considered approximately perpendicular on the photon propagation direction according to the above reasoning,  and perform a rotation of the Stokes vector obtained at the step~\ref{it_LandauStokesTransformation}, using this angle.

We can produce the same result in the vectorial treatment, using the projections of $\boldsymbol \varepsilon_f$ polarization vector on the $\boldsymbol{OX}$ and $\boldsymbol{OY}$ axes of the laboratory frame, in equations~\eqref{eq_Stokes1FromVector} and \eqref{eq_Stokes2FromVector} (we neglect the very small component of $\boldsymbol \varepsilon_f$ on $\boldsymbol{OZ}$ axis due to the same reasoning explained above. We denote these Stokes parameters with $P^{(LAB)}_1$ and $P^{(LAB)}_2$.

\subsubsection{Different input polarization vectors requested by \texttt{G4PolarizedCompton} \\ and \texttt{G4LivermorePolarizedComptonModel} classes}

The prescription in \textsc{Geant4} manual is to provide as polarization property of the photon a three dimensional vector containing the Stokes parameters in the \emph{"Particle reference frame"} for further use of this property by the physics processes classes involved in the particle transport simulation procedure. Thus, the standard classes (see \verb|G4PolarizedCompton| which treats the Compton scattering of photons on polarized electrons, in which the circular polarization of photon, namely the Stokes parameter $P_3$ plays an important role) require to provide a three dimension vector $(P_1,P_2,P_3)$ whose components represent the Stokes parameters. 

However, some physics classes require that the components of the same three-vector to have a different significance. For example, the \verb|G4LivermorePolarizedComptonModel| requires the components of the polarization \verb|G4ThreeVector| to be the three components of the photon polarization vector. Thus, based on the employed physics classes, the user must select the information to be put in the polarization \verb|G4ThreeVector|.  

\paragraph{Stokes parameters required. } The user can either apply the:
\begin{enumerate}[leftmargin=*]
\item Stokes formalism, and provide directly the parameters computed at step \ref{it_StokesInPRF} in the \emph{"Particle reference frame"}, or
\item vectorial treatment, and deduce the Stokes parameters from the polarization vector, as shown at step \ref{it_StokesFromVectorInPRF}.
\end{enumerate}

\paragraph{Photon polarization vector required. } The user has two options implemented in the algorithm:
\begin{enumerate}[leftmargin=*]   
\item The polarization vector components obtained in step~\ref{it_GaugeTransformation} can be directly provided; 
\item One can obtain the components of the linear polarization vector from the components $P^{(PRF)}_1$ and $P^{(PRF)}_2$ of the Stokes vector in the \emph{"Particle reference frame"}. To do this we use again Equation \ref{eq_StokesRotationTransformation}, and the fact that the photon propagates along the $\boldsymbol{oz}$ axis of the \emph{"Particle reference frame"}. If we consider that the polarization vector is aligned along the $\boldsymbol{OX}$ axis of a particular reference system having the same $\boldsymbol{oz}$ axis as \emph{"Particle reference frame"} (means $P_1 = +1$), if we rotate this particular reference system around $\boldsymbol{oz}$ axis with angle $\phi$, the Stokes parameters in the rotated system can be expressed as:
\begin{equation}
P^{(PRF)}_1 = \cos 2\phi, \;\; P^{(PRF)}_2 = -\sin 2\phi
\end{equation}
Thus we can extract the angle of polarization vector in the $\boldsymbol{xy}$ plane:
\begin{equation}
\phi = \cfrac{1}{2} \arctan \left ( -\cfrac{P^{(PRF)}_2}{P^{(PRF)}_1} \right )
\end{equation}
We also have to take into account the fact that the $\arctan$ function is only defined over $[-\pi/2,\pi/2]$ interval. In order to fully resolve this uncertainty introduced by the inverse trigonometric function, we have to look into an additional information provided by sign of the first Stokes parameter $P^{(PRF)}_1$. More exactly, if sign of $P^{(PRF)}_1$ is different from the sign of $\cos 2\phi$, we further add to, or subtract half $\pi$ from $\phi$ angle depending on its sign, as following:
\begin{align}
\phi = \phi + \pi/2 \quad\text{if}\quad \phi<0 \\
\nonumber \phi = \phi - \pi/2 \quad\text{if}\quad \phi>0
\end{align}
Having so obtained the $\phi$ angle, we can generate the photon polarization vector $\boldsymbol \varepsilon_f$ in the $\boldsymbol{xy}$ plane of the \emph{"Particle reference frame"} by rotating the $\boldsymbol e_x$ versor from equation~\eqref{eq_eX_versor_expression} around $\boldsymbol p_f$ photon final propagation direction with $-\phi$ angle. Having expressed the polarization vector in the \emph{"Particle reference frame"}, we can extract the polar angle of  the polarization vector $\boldsymbol \varepsilon_f$ in the laboratory frame system.
\end{enumerate}

\section{Results for head-on laser--relativistic electron collisions}\label{sec_results_head_on}

If the incident laser is characterized by a full degree ($100\%$) of linear polarization, than both treatments, using Stokes parameter formalism or the polarization vector formalism will provide identical results. If we consider a partial polarization for the incident laser, we can simulate this in the frame of the polarization vector treatment by considering that a certain amount of the laser photons, proportional with the non-polarization degree, presents a random orientation of the polarization vector. Although both the vectorial treatment and Stokes formalism produce identical average results, the recommended option in the algorithm is the Stokes formalism due to its more general treatment of partial polarization.  

\begin{figure}[t]
\centering
\includegraphics [width=0.98\columnwidth, angle=0]{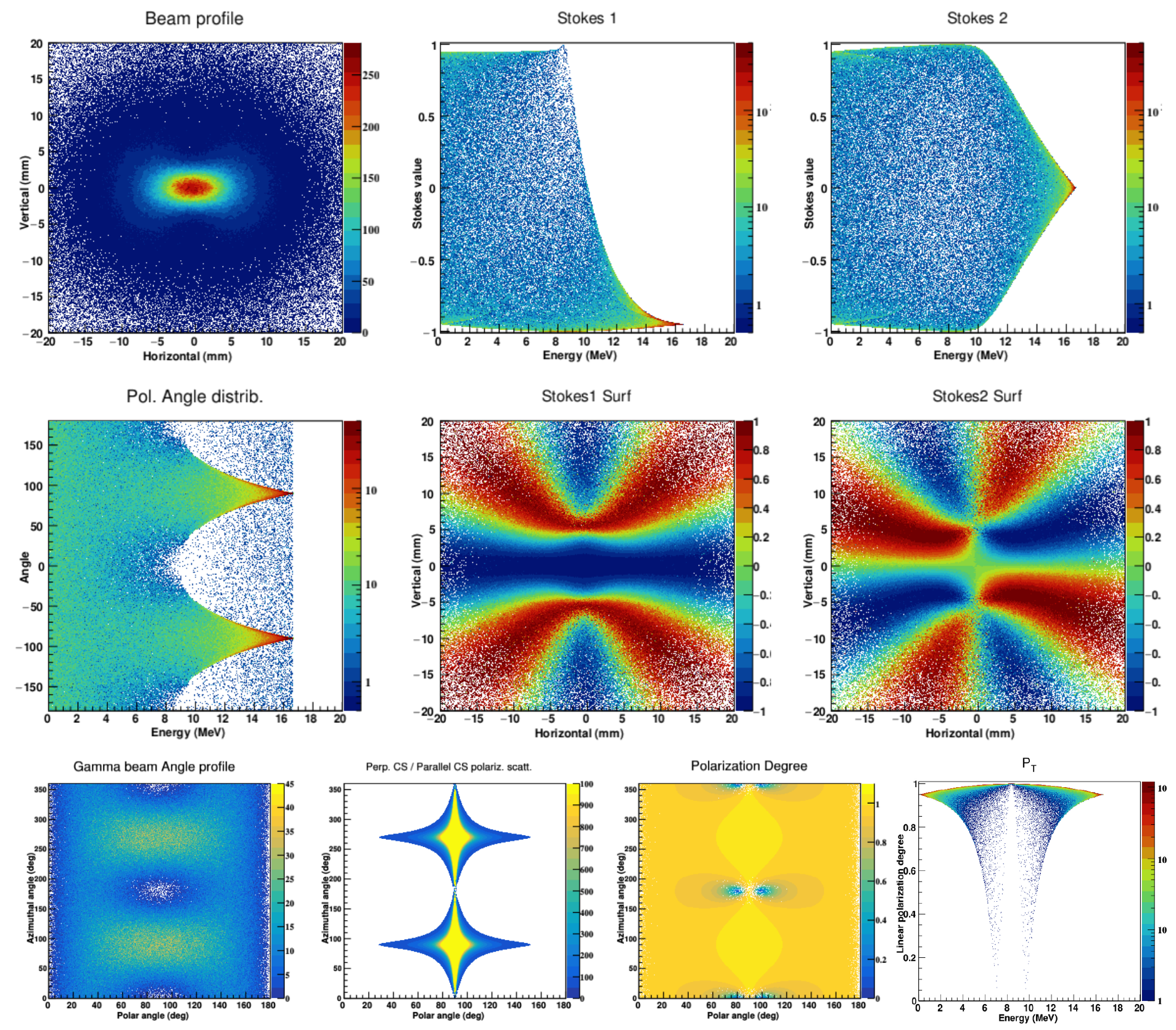} 
\put(-405,357){{(a)}}
\put(-265,357){{(b)}}
\put(-125,357){{(c)}}
\put(-405,221){{(d)}}
\put(-265,221){{(e)}}
\put(-125,221){{(f)}}
\put(-410,90){{(g)}}
\put(-305,90){{(h)}}
\put(-195,90){{(i)}}
\put(-95,90){{(j)}}
\caption{\label{General_95Plin_090Log}
Simulations for Compton scattering of a Gaussian 1064~nm laser beam with $95\%$ linear polarization on realistic 974~MeV electron beam. 
\emph{Laboratory frame system representations:} 
(a) $\gamma$-ray beam intensity spatial distribution at collimator position;  
(b) $P^{(LAB)}_1$ and (c) $P^{(LAB)}_2$ Stokes parameters distributions as function of $\gamma$-ray energy; 
(d) $\gamma$-ray energy distributions of polarization vector azimuthal angle; 
(e) $P^{(LAB)}_1$ and (f) $P^{(LAB)}_2$ spatial distribution;
\emph{Electron rest frame system representations:}
(g) distribution of $\theta$\&$\varphi$ spherical angles defined in fig.~\ref{Fano_ref_system};
(h) distribution of ratio $p = \left. \mathrm d \sigma_{\perp}(\Omega) \middle/ \mathrm d \sigma_{\parallel}(\Omega) \right.$ between perpendicular and parallel polarization cross sections relative to the Compton scattering plane (Fano reference system) defined in~\eqref{eq_Intensity_ratio} as function of spherical angles; 
(i) distribution of the scattered photon polarization degree $P_T$ defined in~\eqref{Total_polarization} as function of spherical angles. 
(j) \emph{Laboratory frame system representation} for 2D distribution of the linear polarization degree $P_T$ as function of the $\gamma$-ray energy.
}
\end{figure}
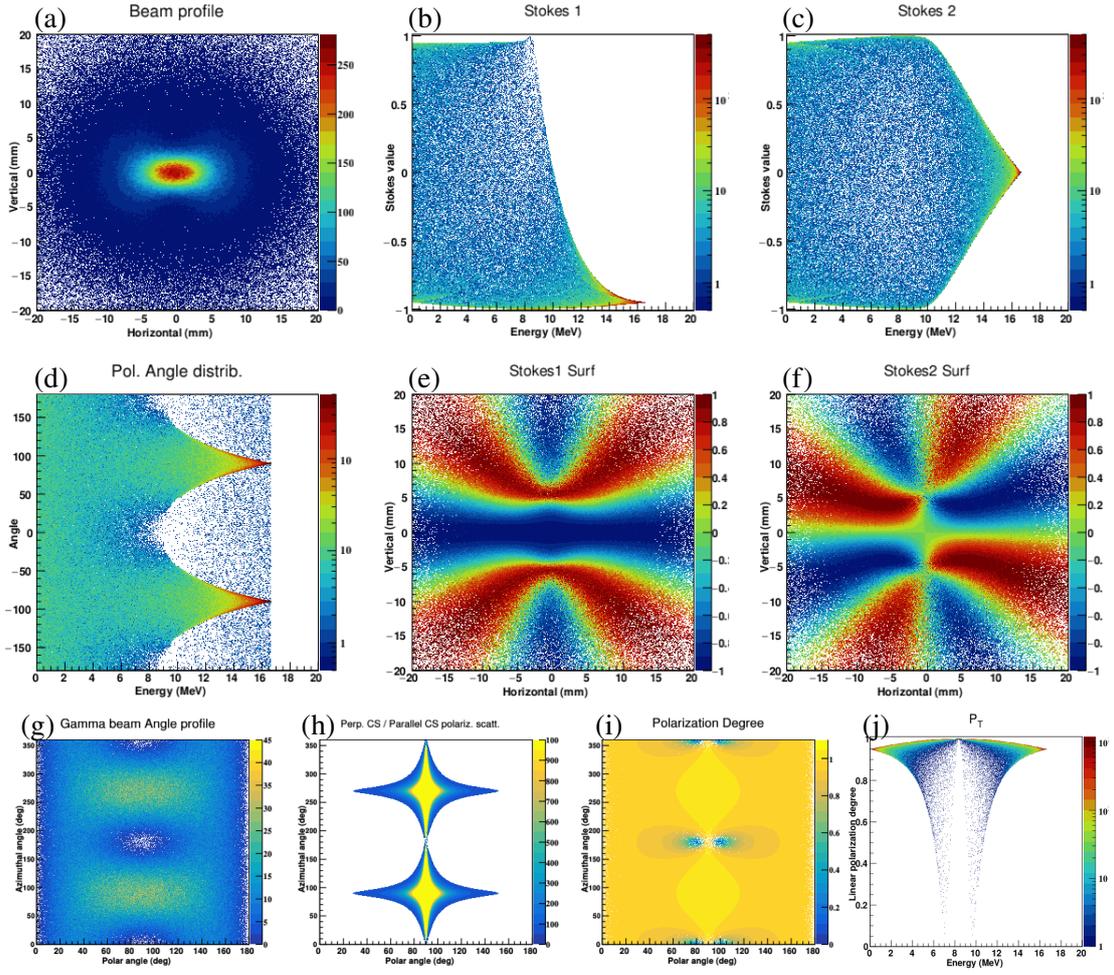

\subsection{Realistic electron and partial linear polarized laser beams}

Here we have considered a realistic laser-Compton scattering (LCS) process between relativistic electron beam having 974 MeV (this corresponds to the top-up energy of the NewSUBARU electron synchrotron) and a 1064~nm wavelength laser (this corresponds to the INAZUMA laser installed on the BL01 beamline at NewSUBARU). Electron beam phase space distribution was described using Twiss parameters formalism, while the laser beam was described using Gaussian beam formalism. The laser was considered to be $95\%$ linearly polarized and the azimuthal angle orientation of the polarization vector $\tau = 90^\circ$. We remind that, for head-on collisions, the azimuthal angle is defined with regard to the $\boldsymbol{OX}$ axis of the laboratory frame system. Thus, an azimuthal angle $\tau = 0^\circ$ defines a laser polarization in the plane of the synchrotron accelerator, while $\tau = 90^\circ$ defines a polarization perpendicular on the accelerator plane. 

\subsubsection{Polarization properties in laboratory frame system}

Figure~\ref{General_95Plin_090Log} shows intensity spatial distributions on an imaginary imaging plate and energy distributions. The imaginary 40$\times$40 mm$^2$ imaging plate was placed at 12~m downstream of the interaction region center, in the position of the first (C1) collimator used at the NewSUBARU BL01 beamline. 

Figure~\ref{General_95Plin_090Log}(a) shows the $\gamma$-ray beam intensity spatial distribution at C1 collimator position. We note the clear correlation between laser polarization vector orientation and the $\gamma$-ray beam spatial distribution. As the laser polarization vector is perpendicular on the accelerator plane, the photons are mostly scattered in the accelerator plane.

Figures~\ref{General_95Plin_090Log}(b,~c) show the $P^{(LAB)}_1$ and $P^{(LAB)}_2$ Stokes parameters distributions as function of $\gamma$-ray energy, in the laboratory frame system. We note that the high energy $\sim16.5$~MeV $\gamma$-ray photons, obtained by Compton back scattering at $\sim180^\circ$, posses the same polarization as the initial laser photon ($P_1 \approx -0.95$ and $P_2 \approx 0$).

Figure~\ref{General_95Plin_090Log}(d) shows the $\gamma$-ray energy distribution for the azimuthal angle of the polarization vector in the laboratory frame system. We note the behavior present also in figures~\ref{General_95Plin_090Log}(b,~c): the polarization vector of the maximum energy back-scattered photons is distributed along the $-90^\circ$ to $+90^\circ$ direction. The $5\%$ unpolarized component can be noticed as a random background in the azimuthal angle distribution of the polarization vector.

Figures~\ref{General_95Plin_090Log}(e,~f) show the $P^{(LAB)}_1$,~$P^{(LAB)}_2$ Stokes parameters spatial distribution on the imaginary imaging plate placed at C1 collimator position. The spatial distributions were obtained by incrementing the histograms with the Stokes parameter value for each event and normalizing the result to the $\gamma$-ray beam intensity spatial distribution shown in figure~\ref{General_95Plin_090Log}(a). 

\subsubsection{Polarization properties in electron rest frame system} 

It is worth to represent and analyze the same process represented in figure~\ref{General_95Plin_090Log} in the electron rest frame system. Figure~\ref{General_95Plin_090Log}(g) shows the distribution of the polar $\theta$ and azimuthal $\varphi$ spherical angles of the scattered photon, as defined in figure~\ref{Fano_ref_system}. We have to remind that in order to reach the electron rest frame system, we perform rotations and Lorentz transformations in such a way that we finally place the polarization vector of the incident photon along the $\boldsymbol{OX}$ axis, and its momentum along the $\boldsymbol{OZ}$ axis. Having this in mind, we note that most of the photons are scattered on the azimuthal direction defined by the $\boldsymbol{OY}$ axis (meaning $\varphi = 90^\circ \; \div \; 270^\circ$), or in other words - perpendicular to the initial polarization direction. This dependence is the most notable for polar scattering angle $\theta = 90^\circ$ and can be easily explained using equation~\eqref{eq_ComptonCS_PolarizedBeam_General_Int_explicit}.

Figure~\ref{General_95Plin_090Log}(h) shows the distribution of ratio $p = \left. {\mathrm d \sigma_{\perp}(\Omega)} \middle/ {\mathrm d \sigma_{\parallel}(\Omega)} \right.$ between perpendicular and parallel polarization cross section relative to the Compton scattering plane (Fano reference system) as a function of polar angles and normalized to the previous distribution of polar angles given in figure~\ref{General_95Plin_090Log}(g). This quantity is defined in \eqref{eq_Intensity_ratio} and can be computed, depending on the desired polarization model, with equation~\eqref{eq_Intensity_ratio_GenCase} in the treatment using Stokes parameters or using equation~\eqref{eq_ComptonCS_PolarizationVector_Symplified} as described at step~\ref{it_StokesInFanoFromVector} in the vectorial treatment. 

One can observe in figure~\ref{General_95Plin_090Log}(h) that the cross section ratio distribution is highly peaked for the angle pairs $(\theta,\varphi) = (90^\circ,90^\circ) \;\&\; (90^\circ,270^\circ)$. We note that, for the visualization of the small cross section distributions around the two peaks, the linear 2D color code of the distribution was cut at a value significantly lower than the peak maximum. We conclude that, for the two peaks, the photon polarization is very high - the probability to detect a photon having the polarization vector perpendicular on the scattering plane being close to $100\%$, or in other words, the ratio $p\rightarrow +\infty$.

The linear polarization degree $P_T$ distribution, computed with equations~\eqref{eq_LinearPolDegree_GenCase} or \eqref{eq_LinearPolDegree}, is represented in figure~\ref{General_95Plin_090Log}(i) as function of polar angles, normalized to the distribution of polar angles from figure~\ref{General_95Plin_090Log}(g). One can extract the same conclusion drawn from the cross section ratio distribution, namely that linear polarization degree around $(\theta,\varphi) = (90^\circ,90^\circ) \;\&\; (90^\circ,270^\circ)$ peaks is $\approx100\%$.

\subsubsection{Energy dependence of total polarization degree} 

In figure~\ref{General_95Plin_090Log}(j) we represent the linear polarization degree as function of the $\gamma$-ray energy in the laboratory frame. In fact, as it was already stated at the step~\ref{it_LandauStokesTransformation}, it is demonstrated by Landau in reference~\cite{Landau} (chapter 8) that the degree of linear polarization $P_T$ is Lorentz invariant, thus once computed in the electron rest frame does not change by rotations and Lorentz boost performed in order to reach the laboratory frame. Analyzing the distribution from figure~\ref{General_95Plin_090Log}(j) we draw again the same conclusion: $\gamma$-rays produced by Compton back-scattering at $\approx180^\circ$ retain the incident laser linear polarization degree ($P_T\approx0.95\%$).

Another interesting fact that can be noticed from figure~\ref{General_95Plin_090Log}(j) is that at $\gamma$-ray energies around $\approx8.5$ MeV, which corresponds to the scattering of photons at $\theta\approx90^\circ$ polar angle in the electron rest frame (the two peaks mentioned in the distributions from figures\ref{General_95Plin_090Log}(h)~\&~(i) of cross section ratio and polarization degree $P_T$ respectively), the total linear polarization degree becomes $P_T\approx100\%$. This is a general feature of LCS $\gamma$-ray beams. Nevertheless, due to the different azimuthal angle orientation of these photons around central axis defined by the electron beam, the complete polarization of these particular photons is hindered in the laboratory system frame as can be seen in figure~\ref{General_95Plin_090Log}(b) $\&$ (c) (the polarization is shared between $P^{(LAB)}_1$ and $P^{(LAB)}_2$ Stokes parameters depending on the azimuthal angle $\varphi$).

\begin{figure}[t]
\centering
\begin{tabular}{cccc}
\includegraphics[width=0.49\textwidth, angle=0]{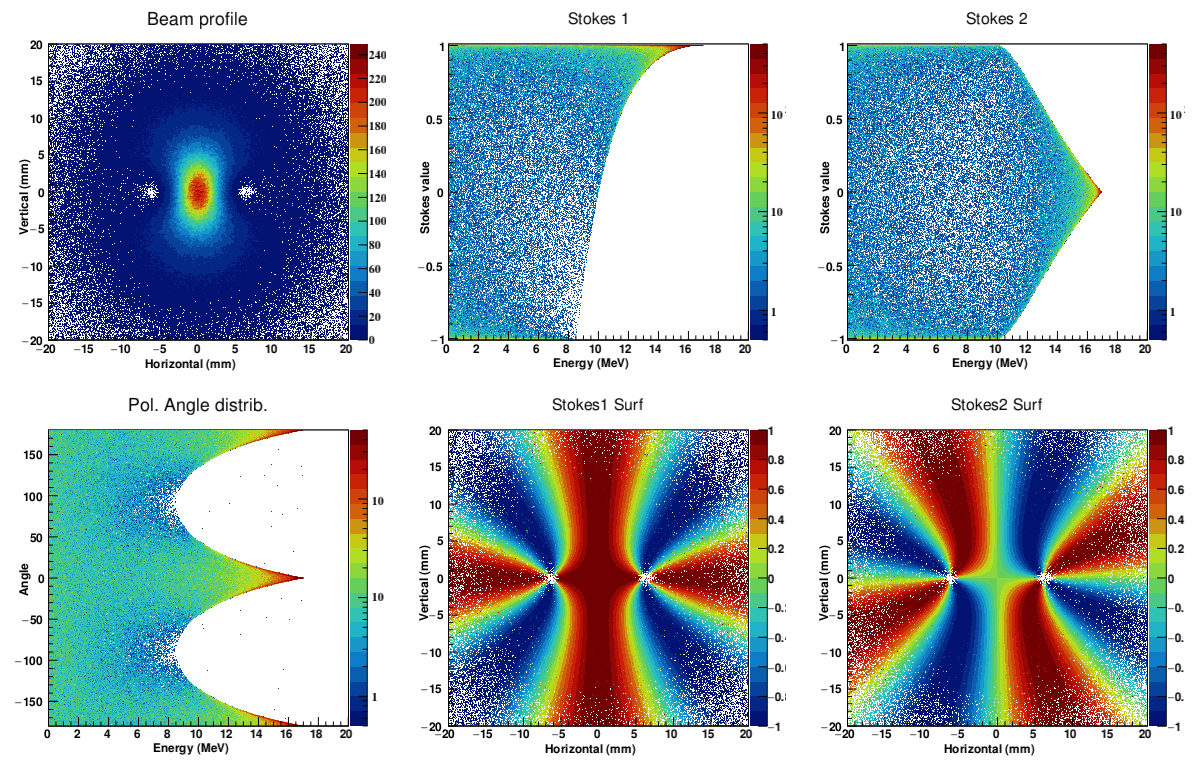} & 
\includegraphics[width=0.49\textwidth, angle=0]{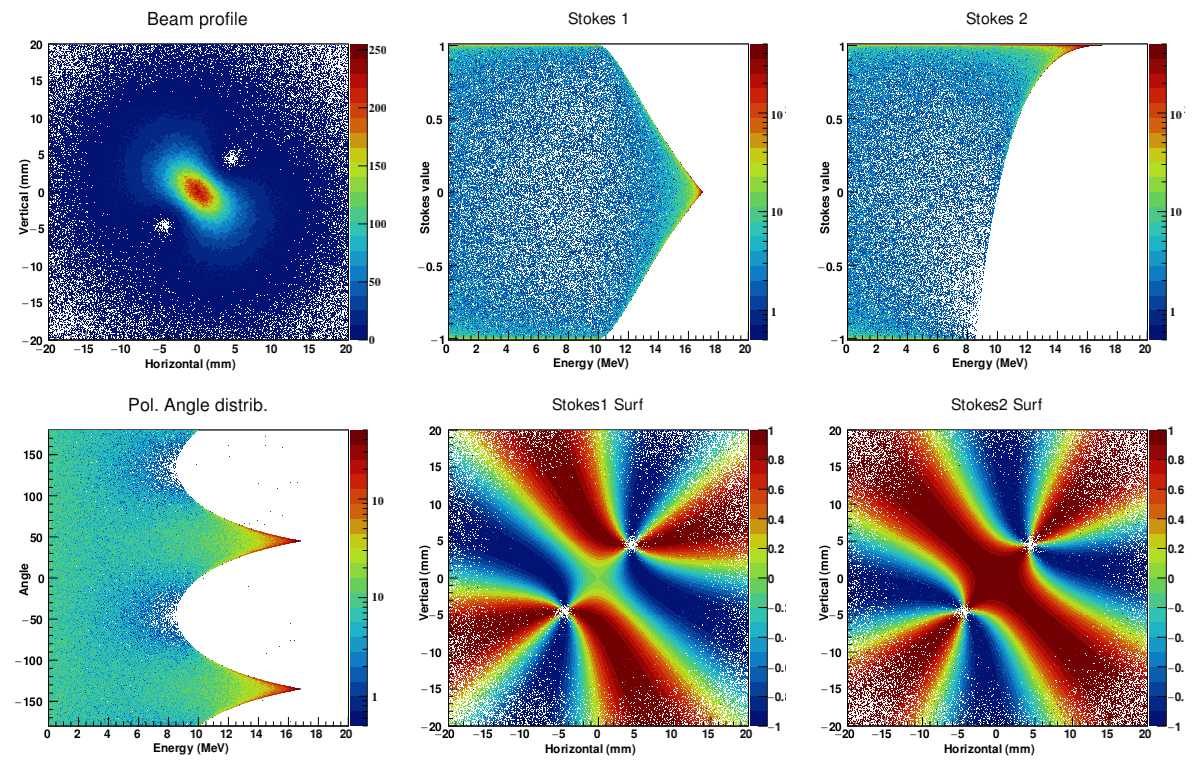} \\ 
{(a)} $\tau = 0^\circ \;\; \& \;\; 180^\circ$ & {(b)} $\tau = 45^\circ \;\; \& \;\; 225^\circ$ \\[6pt]
\end{tabular}
\begin{tabular}{cccc}
\includegraphics[width=0.49\textwidth, angle=0]{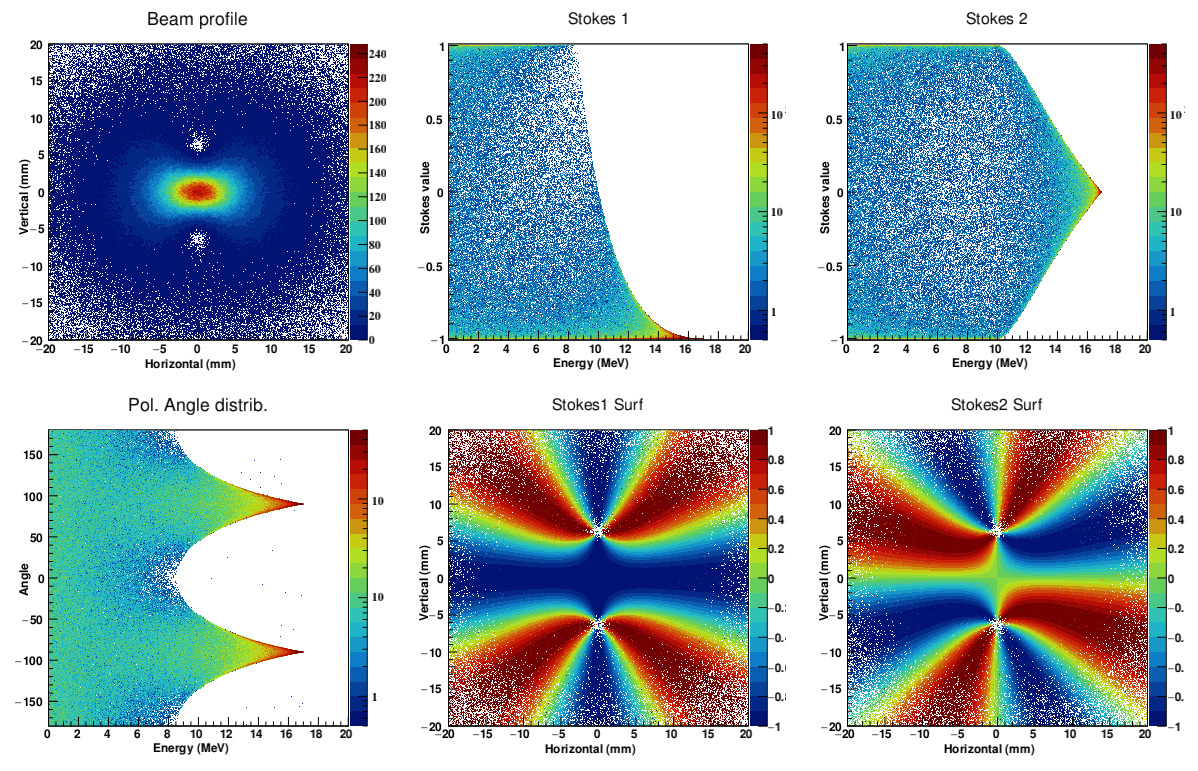} & 
\includegraphics[width=0.49\textwidth, angle=0]{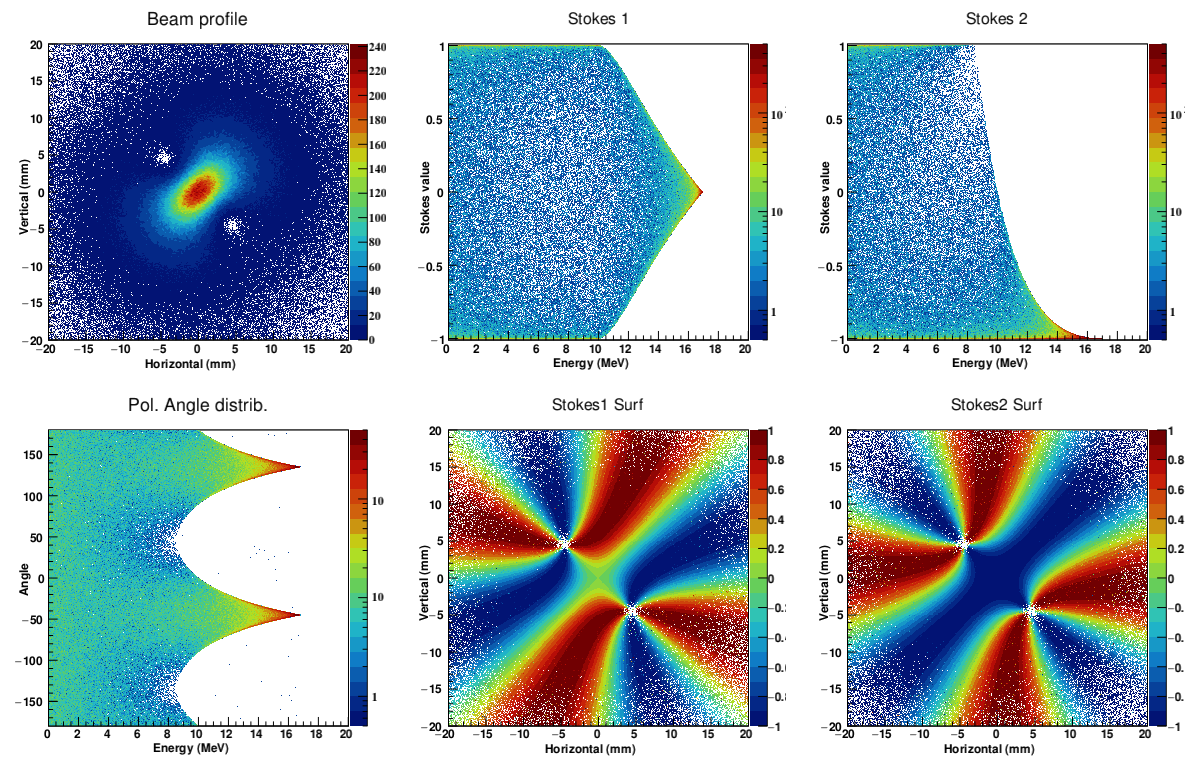} \\ 
{(c)} $\tau = 90^\circ \;\; \& \;\; 270^\circ$ & {(d)} $\tau = 135^\circ \;\; \& \;\; 315^\circ$ \\[6pt]
\end{tabular}
\caption{
Laboratory frame system representation of correlation between polarization of $\gamma$-ray beam produced by LCS and the orientation of the linear polarization vector of the incident laser. Electron and laser beams are considered to be ideal pencil-like. A $100\%$ linear polarization is considered for the laser beam. The significance of each sub-figure is the same as in figure~\ref{General_95Plin_090Log}. The orientation angle $\tau$ pairs of the laser linear polarization vector are: (a) $\tau = 0^\circ \;\; \& \;\; 180^\circ$, (b) $\tau = 45^\circ \;\; \& \;\; 225^\circ$, (c) $\tau = 90^\circ \;\; \& \;\; 270^\circ$, (d) $\tau = 135^\circ \;\; \& \;\; 315^\circ$
}
\label{Ideal_Log_Rotation}       
\end{figure}

\subsection{Ideal electron and laser beams}

Here, simulations were performed for ideal pencil-like electron and laser beams. The spatial distributions are again shown on an imaginary 40$\times$40 mm$^2$ imaging plate placed at 12~m downstream of the center of the interaction region, in the position corresponding to the NewSUBARU C1 collimator.

\subsubsection{Total linear polarized laser}

In order to show the correlation between the polarization of the LCS $\gamma$-ray beam and the initial linear polarization vector of the incident laser beam, the laser polarization orientation angle $\tau$ was varied in $45^\circ$ step. Here $\tau=0^\circ$ corresponds to polarization vector in the accelerator plane and $\tau=90^\circ$ to polarization vector perpendicular on the accelerator plane. The laser beam was considered to be completely linear polarized. 

Figure~\ref{Ideal_Log_Rotation} shows the resulting intensity spatial distributions and energy distributions for the (a)~$\tau$~=~0$^\circ$\&180$^\circ$, (b)~$\tau$~=~45$^\circ$\&225$^\circ$, (c)~$\tau$~=~90$^\circ$\&270$^\circ$ and (d)~$\tau$~=~135$^\circ$\&315$^\circ$ pairs of the incident laser beam linear polarization vector orientation. For each $\tau$ angle pair,
\begin{itemize}
\item upper left corner figure shows the well-known $\gamma$-ray beam intensity spatial distribution~\cite{Sun2011_STAB,petrillo2015,zhijun_chi2020,zhijun_chi2022,Hajima2021}, where one can notice that the photons are mostly scattered in a plane perpendicular to the laser polarization vector. 
\item upper middle and right corner figures show the $P^{(LAB)}_1$ and $P^{(LAB)}_2$ Stokes parameters distributions as a function of $\gamma$-ray energy, in the laboratory frame system. Again, the high-energy back-scattered photons retain the polarization of the initial laser photon. 
\item bottom left corner figure shows the $\gamma$-ray energy distribution for the azimuthal angle of the polarization vector in the laboratory frame system. 
The sharp angle distribution of the back-scattered photons show the well defined direction of the linear polarization vector. 
\item bottom middle and right corner figures show the $P^{(LAB)}_1$ $\&$ $P^{(LAB)}_2$ Stokes parameters spatial distribution, which reproduce the analytical results of Petrillo~\cite{petrillo2015} and of Zhijun~Chi in the first harmonic~\cite{zhijun_chi2020,zhijun_chi2022}. For all treated polarization vector orientations, the Stokes parameters distribution is well defined in the collimation area from the center of the imaging plate. Although the distributions shown here are constructed for ideal electron and laser beams, we note that the ones shown in figures~\ref{General_95Plin_090Log}(e,~f) for partially linear polarized laser and realistic electron and laser beams are also well defined in the collimation region, with only small smearing effects introduced by the non-ideal beam conditions. 
\end{itemize}

From both spatial and energy distributions of Stokes parameters represented in figure~\ref{Ideal_Log_Rotation}, one can notice how the linear polarization of the scattered $\gamma$-ray beam transfers between $P^{(LAB)}_1$ and $P^{(LAB)}_2$ Stokes parameters when rotating the polarization vector of the incident laser photon beam.

\begin{figure}[t]
\centering
\includegraphics [width=0.98\columnwidth, angle=0]{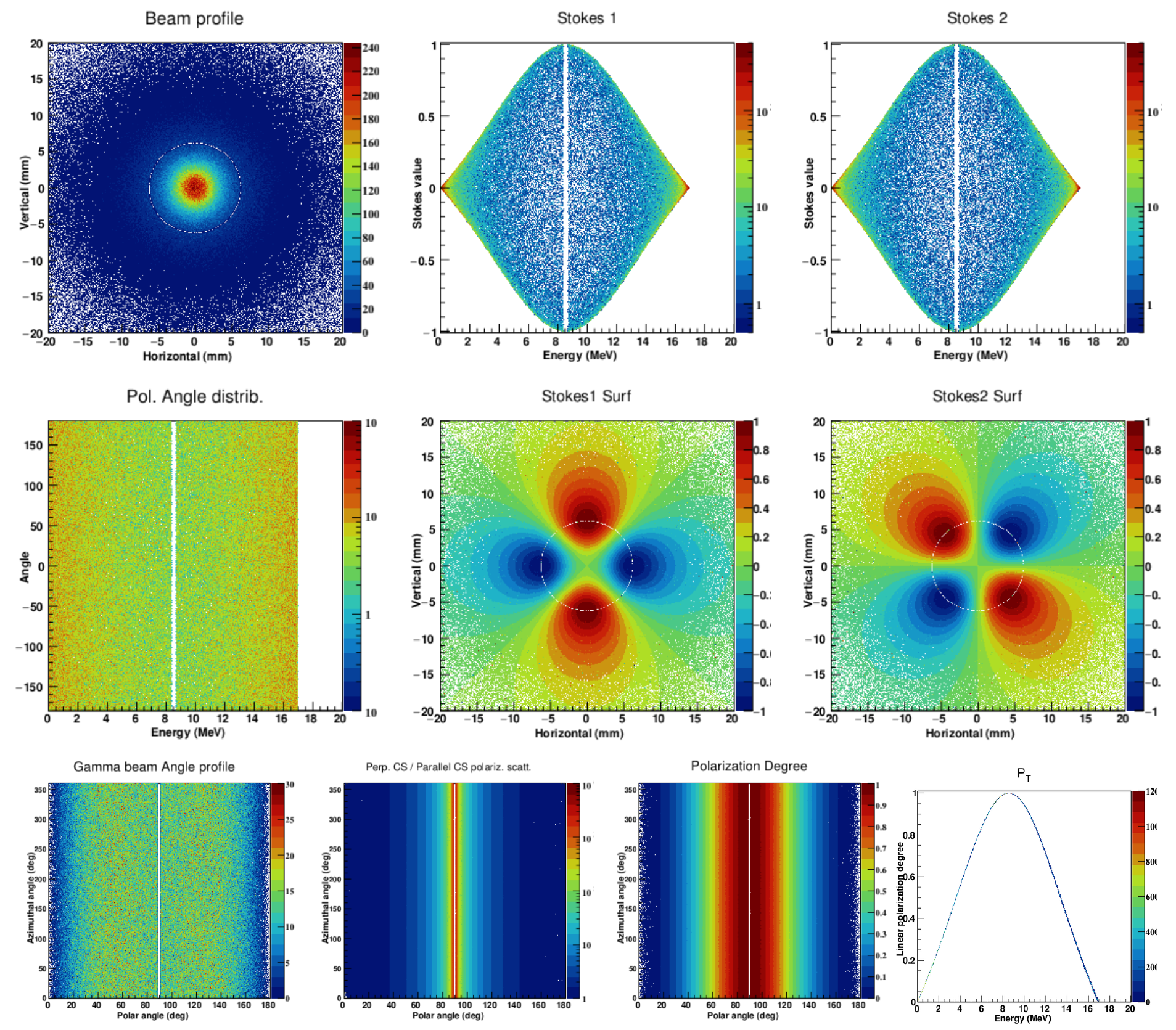} 
\put(-405,357){{(a)}}
\put(-265,357){{(b)}}
\put(-125,357){{(c)}}
\put(-405,221){{(d)}}
\put(-265,221){{(e)}}
\put(-125,221){{(f)}}
\put(-410,90){{(g)}}
\put(-305,90){{(h)}}
\put(-195,90){{(i)}}
\put(-95,90){{(j)}}
\caption{\label{FIG_UNPOL}
Simulations for Compton scattering of ideal, fully unpolarized 1064 nm laser beam on ideal 974 MeV electron beam. 
\emph{Laboratory frame system representations:} 
(a) $\gamma$-ray beam intensity spatial distribution at collimator position;  
(b)~$P^{(LAB)}_1$ and (c)~$P^{(LAB)}_2$ Stokes parameters distributions as function of $\gamma$-ray energy; (d) $\gamma$-ray energy distributions of polarization vector azimuthal angle; 
(e)~$P^{(LAB)}_1$~and (f)~$P^{(LAB)}_2$ spatial distribution. 
\emph{Electron rest frame system representations:}
(g) distribution of $\theta$\&$\varphi$ spherical angles defined in fig.~\ref{Fano_ref_system}; 
(h) distribution of ratio $p = \left. \mathrm d \sigma_{\perp}(\Omega) \middle/ \mathrm d \sigma_{\parallel}(\Omega) \right.$ defined in~\eqref{eq_Intensity_ratio} as function of spherical angles; 
(i) distribution of the scattered photon linear polarization degree $P_T$ defined in~\eqref{Total_polarization} as function of spherical angles. 
Histograms shown in (h,~i) were normalized to the distribution of spherical angles given in (g). 
(j) \emph{Laboratory frame system representation} for 2D distribution of linear polarization degree $P_T$ as function of $\gamma$-ray energy. 
For visualizing the contribution of photons scattered at $\theta$~=~90$^\circ$ in the electron rest frame, we eliminated in the simulation process the emission in the polar angle range 89$^\circ$~--~91$^\circ$.
}
\end{figure}
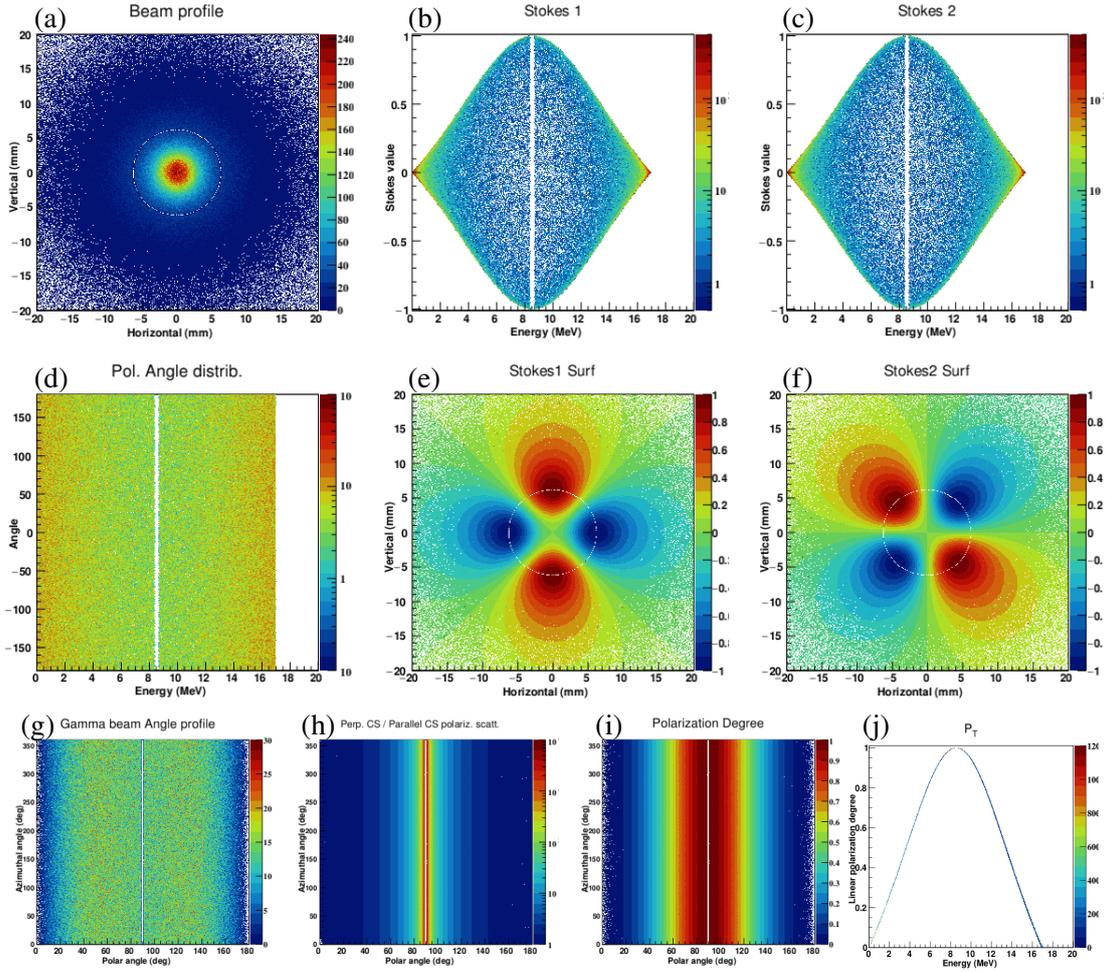

\subsubsection{Totally unpolarized laser}

Another interesting case is to consider a totally unpolarized laser beam. The simulation results for ideal pencil-like electron beam and ideal, fully unpolarized laser beam conditions are shown in the laboratory frame system and electron rest frame system representations in figure~\ref{FIG_UNPOL}, where the sub-figures significance is consistent with the one in figure~\ref{General_95Plin_090Log}.  

Figure~\ref{FIG_UNPOL}(a) shows that, for incident unpolarized laser beam, the azimuthal dependence of Compton scattered $\gamma$-ray beam spatial distribution is isotropic around the axis defined by the electron beam direction. Although the experimental measurement with an unpolarized laser of ref.~\cite{amano09} shows a slight spatial anisotropy, it can be shown that it is generated by the spatial anisotropy of the electron beam, characteristic for electron synchrotrons. The isotropic azimuthal dependence is again noticed in the distribution of $\theta$ and $\varphi$ spherical angles in electron rest frame system representation shown in figure~\ref{FIG_UNPOL}(g). 

Also, the Compton back-scattered photons that generate the maximum energy part of the $\gamma$-ray spectrum are completely unpolarized, as can be seen in figure~\ref{FIG_UNPOL}(b,~c) depicting $P^{(LAB)}_1$ and $P^{(LAB)}_2$ Stokes parameters in the laboratory frame. Consequently, the corresponding polarization vector shown in figure~\ref{FIG_UNPOL}(d) presents a random orientation. Nevertheless, if we look at the linear polarization degree represented in figure~\ref{FIG_UNPOL}(h,~i), we again note that for polar angle $\theta=90^\circ$ in the electron rest frame, the linear polarization of the scattered photons becomes 100~\%. This is again visible in figure~\ref{FIG_UNPOL}(j), where we represent the linear polarization degree $P_T$ incremented in the laboratory frame system as a function of the $\gamma$-ray energy.

For visualizing the contribution of photons scattered at polar angle $\theta$~=~90$^\circ$ in the electron rest frame, we eliminated in the simulation process the emission in the polar angle range 89$^\circ$~--~91$^\circ$. This can be noticed by the presence of white areas in figure~\ref{FIG_UNPOL}. The spatial intensity distributions at C1 collimator position given in figures~\ref{FIG_UNPOL}(a,~e,~f) show that the photons scattered at $\theta$~=~90$^\circ$ in the electron rest frame are localized in a ring centered on the electron beam axis, namely these photons are emitted in a narrow cone. The fact that these photons are fully linearly polarized can be observed in figures~\ref{FIG_UNPOL}(e,~f), where the ($P^{(LAB)}_1$,~$P^{(LAB)}_2$) Stokes parameters pairs on the 90$^\circ$ emission ring are ($-$1,~0) and (1,~0) on OX and OY axes, while in the system rotated by 45$^\circ$ they are (0,~$-$1) and (0,~1) on the rotated OX and OY axes. Thus, in any point located on the 90$^\circ$ emission ring, the linear polarization degree $P_T$~=~1.  

\begin{figure}[t]
\centering
\includegraphics [width=0.98\columnwidth, angle=0]{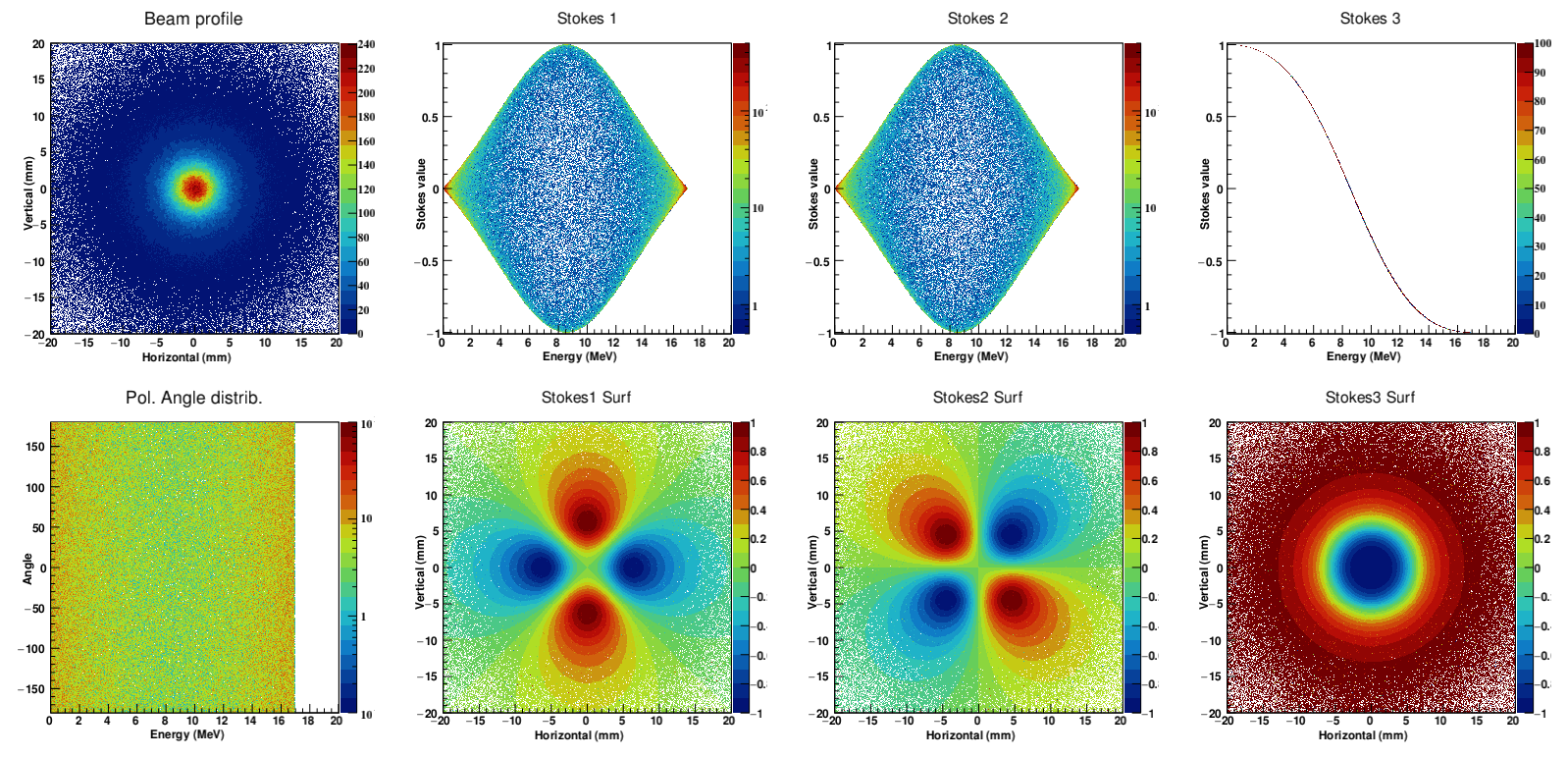}  
\put(-408,195){{(a)}}
\put(-303,195){{(b)}}
\put(-198,195){{(c)}}
\put(-93,195){{(d)}}
\put(-408,93){{(e)}}
\put(-303,93){{(f)}}
\put(-198,93){{(g)}}
\put(-93,93){{(h)}}
\caption {\label{Ideal_100Pcirc_000Log}
Laboratory frame system representation of simulations for Compton scattering of ideal, $100\%$ right circular polarized 1064 nm laser beam on ideal 974 MeV electron beam. 
(a) $\gamma$-ray beam intensity spatial distribution; 
(b)~$P^{(LAB)}_1$, (c)~$P^{(LAB)}_2$ and (d)~$P^{(LAB)}_3$ Stokes parameters distributions as function of $\gamma$-ray energy; 
(e) $\gamma$-ray energy distributions of polarization vector azimuthal angle; 
(f)~$P^{(LAB)}_1$, (g)~$P^{(LAB)}_2$ and (h)~$P^{(LAB)}_3$ spatial distribution.
}
\end{figure}
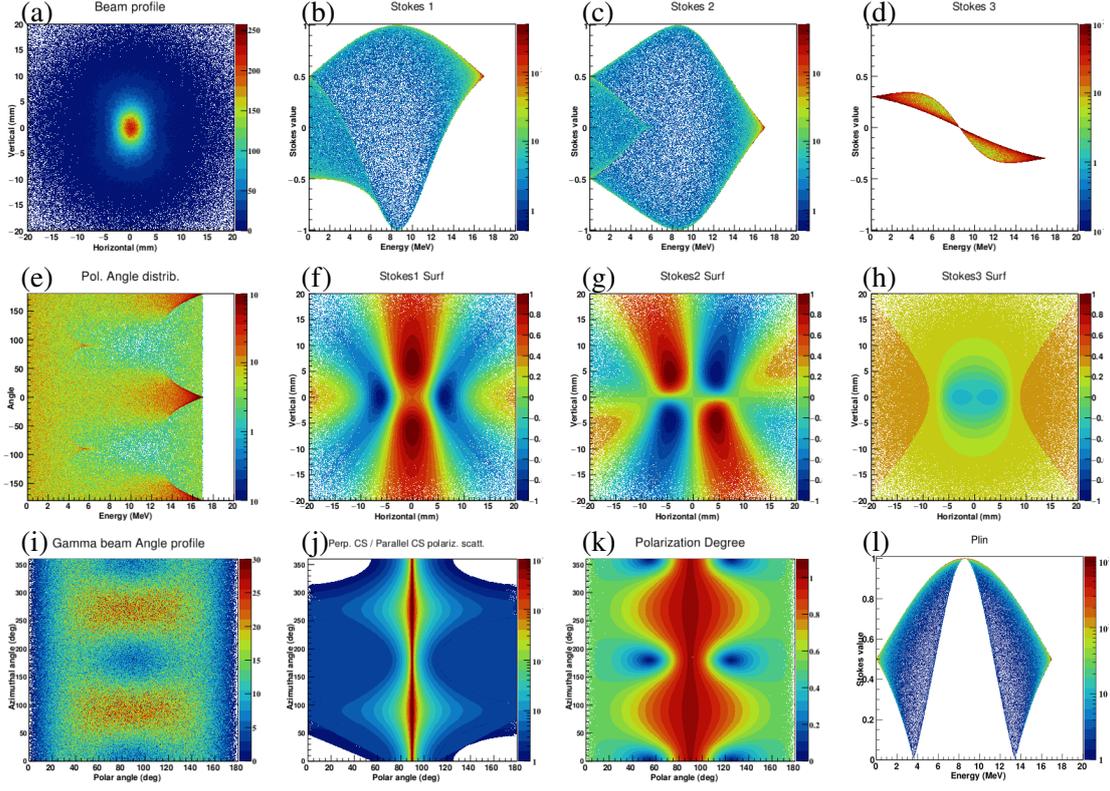
\begin{figure}[t]
\centering
\includegraphics [width=0.98\columnwidth, angle=0]{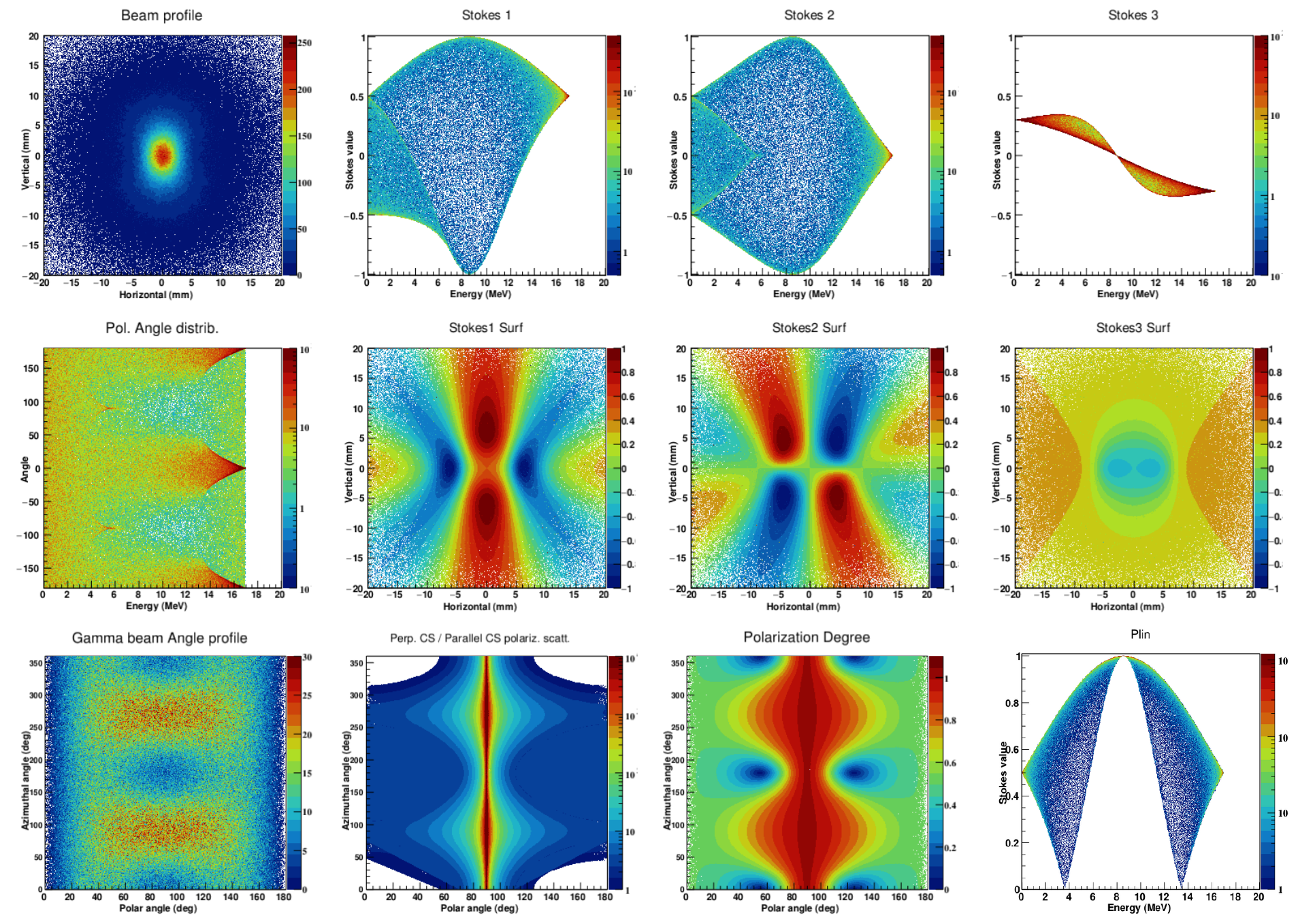} 
\put(-410,290){{(a)}}
\put(-305,290){{(b)}}
\put(-200,290){{(c)}}
\put(-95,290){{(d)}}
\put(-410,190){{(e)}}
\put(-305,190){{(f)}}
\put(-200,190){{(g)}}
\put(-95,190){{(h)}}
\put(-410,90){{(i)}}
\put(-305,90){{(j)}}
\put(-200,90){{(k)}}
\put(-95,90){{(l)}}
\caption {\label{Ideal_05Plin_03Pcirc_000Log}
Simulations for Compton scattering of ideal 1064 nm laser beam presenting mixed polarization of $50\%$ linear degree (polarization vector in the accelerator plane) and $30\%$ right circular degree on an ideal 974 MeV electron beam. 
\emph{Laboratory frame system representations:} 
(a) $\gamma$-ray beam intensity spatial distribution at collimator position; 
(b)~$P^{(LAB)}_1$, (c)~$P^{(LAB)}_2$ and (d)~$P^{(LAB)}_3$ Stokes parameters distributions as functions of $\gamma$-ray energy; 
(e) $\gamma$-ray energy distribution of polarization vector azimuthal angle;  
(f)~$P^{(LAB)}_1$, (g)~$P^{(LAB)}_2$ and (h)~$P^{(LAB)}_3$ spatial distribution.
\emph{Electron rest frame system representations:} 
(i) distribution of $\theta$\&$\varphi$ spherical angles defined in fig.~\ref{Fano_ref_system}; 
(j) distribution of ratio $p = \left. \mathrm d \sigma_{\perp}(\Omega) \middle/ \mathrm d \sigma_{\parallel}(\Omega) \right.$ defined in~\eqref{eq_Intensity_ratio} as function of spherical angles;
(k) distribution of the scattered photon linear polarization degree $P_T$ defined in~\eqref{Total_polarization} as function of spherical angles. 
Histograms shown in (j,~k) were normalized to the distribution of spherical angles given in (i).
(l) \emph{Laboratory frame system representation} for 2D distribution of linear polarization degree $P_T$ as function of $\gamma$-ray energy.
}
\end{figure}

\subsubsection{Circular polarized laser}

Let us consider the Compton scattering of fully circular polarized laser photons on relativistic electrons.  Figure~\ref{Ideal_100Pcirc_000Log} shows the laboratory frame system representation for simulations of Compton scattering of an ideal, $100\%$ right circular polarized ($P_3=+1$) laser beam of 1064~nm wavelength on ideal pencil-like relativistic electron beam of 974~MeV energy. 

Figures~\ref{Ideal_100Pcirc_000Log} shows results similar to the unpolarized laser beam case for the (a) $\gamma$-ray beam intensity spatial distribution, $P^{(LAB)}_1$ $\&$ $P^{(LAB)}_2$ (b,~c) spatial and (f,~g) energy distributions as well as for the (e) energy distribution of the polarization vector azimuthal angle. The azimuthal dependence of Compton scattered $\gamma$-ray beam spatial distribution is isotropic around the axis defined by the electron beam direction, reproducing the experimental and analytical results given in references~\cite{Sun2011_STAB,petrillo2015,zhijun_chi2020}.  

\begin{figure}[t]
\centering
\includegraphics [width=0.9\columnwidth, angle=0]{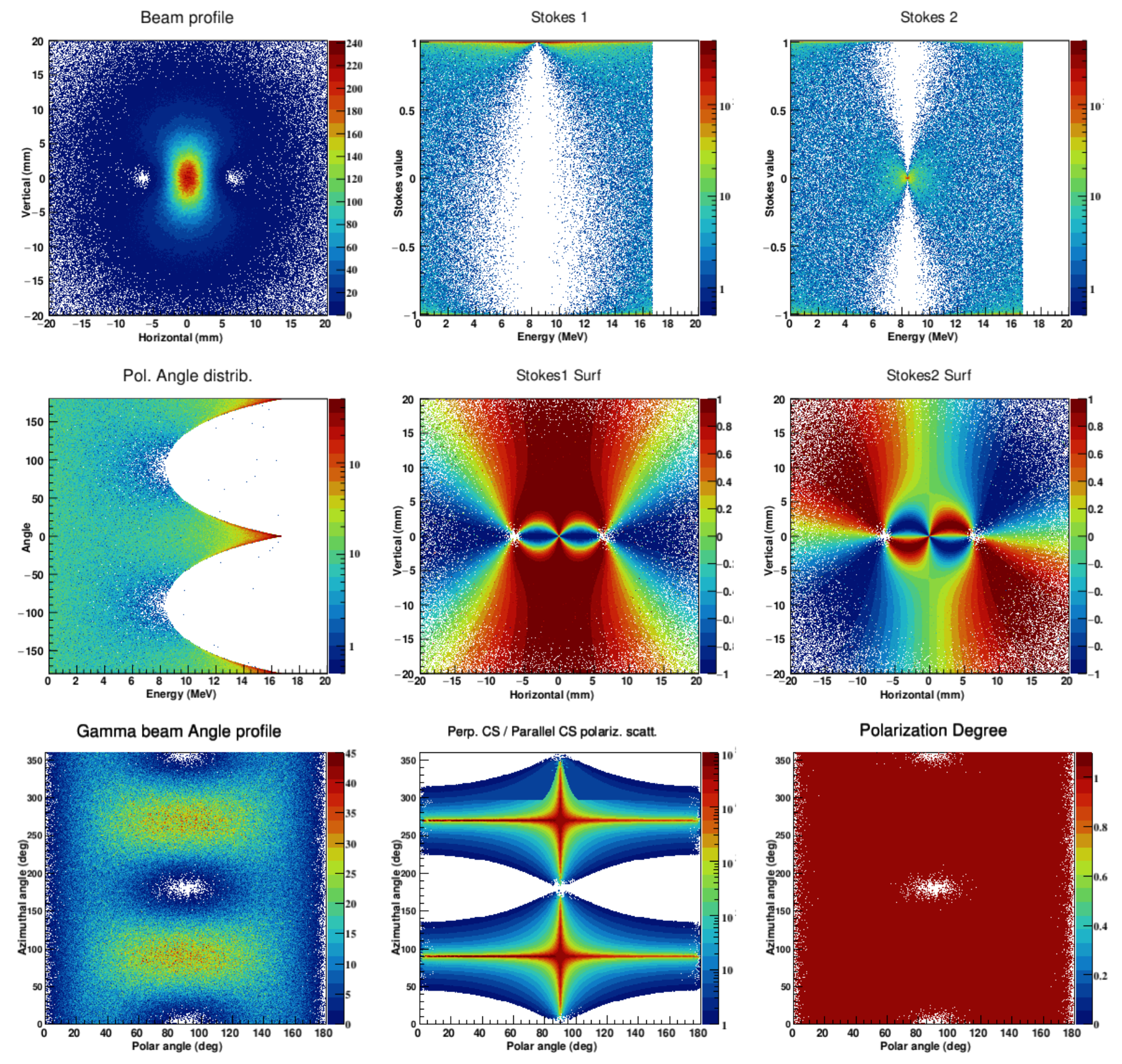} 
\put(-375,358){{(a)}}
\put(-248,358){{(b)}}
\put(-118,358){{(c)}}
\put(-375,232){{(d)}}
\put(-248,232){{(e)}}
\put(-118,232){{(f)}}
\put(-375,110){{(g)}}
\put(-248,110){{(h)}}
\put(-118,110){{(i)}}
\caption {\label{Ideal_000SCMLog}
Simulations for Compton scattering of ideal, 100~\% linear polarized (in accelerator plane) 1064 nm laser beam on ideal 974 MeV electron beam. 
\textsc{Geant4} \emph{"Particle reference frame" (PRF) representations:} 
(a) $\gamma$-ray beam intensity spatial distribution at collimator position;  
(b)~$P^{(PRF)}_1$ and (c)~$P^{(PRF)}_2$ Stokes parameters distributions as function of $\gamma$-ray energy; 
(d) $\gamma$-ray energy distributions of polarization vector azimuthal angle; 
(e)~$P^{(PRF)}_1$ and (f)~$P^{(PRF)}_2$ spatial distribution. 
\emph{Electron rest frame system representations:}
(g) distribution of $\theta$\&$\varphi$ spherical angles defined in fig.~\ref{Fano_ref_system}; 
(h) distribution of ratio $p = \left. \mathrm d \sigma_{\perp}(\Omega) \middle/ \mathrm d \sigma_{\parallel}(\Omega) \right.$ defined in~\eqref{eq_Intensity_ratio} as function of spherical angles; 
(i) distribution of the scattered photon linear polarization degree $P_T$ defined in~\eqref{Total_polarization} as function of spherical angles. 
Histograms shown in (h,~i) were normalized to the distribution of spherical angles given in (g). 
}
\end{figure}

The well known helicity-flip effect~\cite{Angelo2000,Sun2011_STAB,petrillo2015,zhijun_chi2020} can be observed in figures~\ref{Ideal_100Pcirc_000Log} (d,~h), where the $\gamma$-rays produced by Compton back-scattering at $\approx180^\circ$ in the high energy region of the spectrum have $\approx100\%$ left circular polarization ($P^{(LAB)}_3=-1$). Thus, the initial laser polarization changes from right to left. Going to lower energies determined by increasing the polar angle $\theta$ values, the third Stokes parameter $P^{(LAB)}_3$ varies from $-1$ to $+1$, passing through 0 for polar scattering angle $\theta$~=~90$^\circ$ in the electron rest frame, which corresponds to $\approx8.5$ MeV scattered $\gamma$-ray energy. At this energy, the total linear polarization reaches $100\%$ (because the total linear polarization degree in this case have the same behavior as in figure~\ref{FIG_UNPOL}(j), it is not represented here any more).

\subsubsection{Mixed linear - circular polarization}

The last case presented here is the one of an incident laser presenting mixed types of polarization. Figure~\ref{Ideal_05Plin_03Pcirc_000Log} shows the laboratory frame system and electron rest frame system representations for simulations of Compton interaction between an ideal 1064 nm laser having $50\%$ linear polarization degree and $30\%$ circular polarization degree with an ideal electron beam of 974 MeV. 

We notice in figure~\ref{Ideal_05Plin_03Pcirc_000Log}(a) the azimuthal asymmetry of the $\gamma$-ray beam spatial distribution induced by the linear polarization, but also the induced circular polarization of the $\gamma$-ray beam in figures~\ref{Ideal_05Plin_03Pcirc_000Log}(d,~h) due to the partial circular polarization degree of the incident laser.

The histograms incremented in the electron rest frame system shown in figures~\ref{Ideal_05Plin_03Pcirc_000Log}(i~--~k) are mostly defined by and correlated with the linear polarization degree of the laser beam.

For the same simulation, figure~\ref{Ideal_05Plin_03Pcirc_000Log}(l) shows the linear polarization degree of the $\gamma$ beam as function of the $\gamma$-ray beam energy in the laboratory frame system. 
We notice again that the high-energy back-scattered photons retain the initial laser photon linear polarization, here of 50\% degree. 

For the mixed linear-circular polarization case treated here, we notice in figure~\ref{Ideal_05Plin_03Pcirc_000Log}(d) that the circular polarization $P^{(LAB)}_3$ Stokes parameter values for a given photon energy have a continuous  distribution, whereas, for the circular-only case shown in figure~\ref{Ideal_100Pcirc_000Log}(d), the $P^{(LAB)}_3$ has a fixed value for a given scattered photon energy.  

Figure~\ref{Ideal_100Pcirc_000Log}(d) shows that, at scattered photon energies around 12~MeV, the circular Stokes parameter modulus $|P^{(LAB)}_3|$ can reach values higher than the initial circular polarization degree of the laser. Thus, we can deduce that, at these scattered photon energies, there is a transfer from linear polarization to circular polarization. This can also be observed in figure~\ref{Ideal_100Pcirc_000Log}(h) by the presence of the two regions (dark blue) with absolute $|P^{(LAB)}_3|$ Stokes values higher than the initial circular polarization degree of the laser retained by the back-scattered photons, here in the center of the spatial Stokes distribution (cyan). The positions of the dark-blue regions are correlated with the orientation of the linear polarization vector. The $P^{(LAB)}_3$ Stokes parameters spatial distribution shown in figure~\ref{Ideal_100Pcirc_000Log}(h) reproduces the results obtained by classical electrodynamics treatment for the case of Thomson scattered laser photon with elliptical polarization, and shown in figure~6(a,b) of reference~\cite{zhijun_chi2020}. 

\subsection{Stokes parameters in \textsc{Geant4} \emph{"Particle Reference Frame"} }

Finally, figures~\ref{Ideal_000SCMLog}(a~--~f) show the Stokes parameters in \emph{"Particle reference frame"} as they are requested by some \textsc{Geant4} physics classes. A LCS between an ideal 974 MeV electron beam and an ideal fully linearly polarized (in accelerator plane) 1064 nm was considered in order to produce this simulation and some of the quantities describing this interaction in the electron rest frame are represented in figures~\ref{Ideal_000SCMLog}(g~--~i).

The $P^{(PRF)}_1$ and $P^{(PRF)}_2$ Stokes parameters distributions as function of $\gamma$-ray energy shown in figures~\ref{Ideal_000SCMLog}(b) and respectively (c), maintain their shapes regardless of the orientation of the laser linear polarization vector. Thus, the graphical representation of the Stokes parameters distributions in the \emph{"Particle reference frame"} is not as intuitive as the laboratory frame system representation, which has been shown up to now in the present work. 

We also note that, comparing the distributions (g) and (i) as function of spherical angles $\theta$ and $\phi$ in electron rest frame system shown here in figure~\ref{Ideal_000SCMLog} and the ones in figure~\ref{General_95Plin_090Log}, we notice a clear smearing effect introduced by the realistic laser and electron beam modeling as well as by the partial linear polarization degree of 95\%. 

\section{Results for arbitrary angle collisions and variable incident photon energy}

The previous section treated only the case of head-on collision of laser photons on relativistic electrons with its impact on polarization properties. This was done because the present algorithm and the associated computer simulation code were developed especially to provide a tool to be used in the analysis process of the experimental data acquired at the NewSUBARU facility. We recall that a head-on collision geometry is employed at NewSUBARU, where the LCS $\gamma$-ray beam energy variation is based on the continuously variable electron beam energy. However, the recently developed SLEGS LCS $\gamma$-ray beam facility~\cite{wang_fan_2022,hao_fan_2022} makes use of a fixed 3.5~GeV energy electron beam, where the $\gamma$-ray beam energy variation relies on the variation of the laser incident angle, in a so-called \emph{slant-scattering} mode~\cite{xu_fan_2022}. Also, recent works~\cite{zhijun_chi2020} give analytical treatments on the interaction between a photon beam and an electron beam at a non-head-on collision angle. Consequently, these aspects triggered the test of the present algorithm for the case of interaction of a photon beam with an electron beam at an arbitrary angle of incidence. 

\subsection{Arbitrary angle collisions of laser photons on relativistic electrons}\label{sec_results_laser_ele_arbitrary_angle}

The following convention is made: the incidence angle $\theta_i$ is considered to be 0$^\circ$ for head-on collision between the photon beam and the electron beam, and is considered to increase up to $\lesssim 180^\circ$ for the case when the photon beam tends to become parallel with the electron beam and both propagate in the same direction. For an incidence angle variation from 0$^\circ$ to 180$^\circ$, the photon energy in the rest electron frame decreases from the maximum value:
\begin{equation}
E^\mathrm{max}(\theta_i = 0^\circ) = E_0 \gamma ( 1 + \beta ) 
\end{equation}
to the minimum value:
\begin{equation}
E^\mathrm{min}(\theta_i = 180^\circ) = \cfrac{E_0} {\gamma ( 1 + \beta )} 
\end{equation}


Thus, for the case of a green laser (532 nm, 2.33 eV) scattering on a 500 MeV electron beam, which is the lowest energy limit of the NewSUBARU synchrotron, the photon energy in the electron rest frame varies from 4.55~keV in the case of head-on collision ($\theta_i=0^\circ$), down to 1.19~meV in the case when laser is send in the same direction with the electron beam ($\theta_i=180^\circ$). Going to the upper energy limit of the NewSUBARU electron beam of 1.5~GeV, we find that the photon energy in the electron rest frame varies from 13.7~keV at $\theta_i=0^\circ$ to 0.39~meV at $\theta_i=180^\circ$. Recalling the discussion and remarks made in subsection \ref{subsec_PolWithPhotEne} for the scattering of polarized photons on unpolarized electrons, we note that at all incidence angles, the scattered photon polarization degree remains very high (>99.5$\%$) over the entire scattering angle range in the rest electron frame. Consequently, the polarization properties obtained for \emph{any} laser--electron collision angle are identical to the ones obtained for head-on collision and shown in the previous section~\ref{sec_results_head_on}. This can be observed in figure~5 of the supplementary material~\cite{supplementary_material}, where simulation results for polarization related properties are shown for the case of green laser (532~nm) incident on 500~MeV relativistic electron beam at 10 incident collision angles in the 0$^\circ$ to 162$^\circ$ range. Our results are in agreement with the investigation of reference~\cite{zhijun_chi2020}, where it is demonstrated by classical electrodynamics treatment that the laser polarization degree is completely transferred to the Thomson scattered photon on unpolarized relativistic electron.  

The endpoint energy of the $\gamma$-ray beam spectrum is the only quantity which varies significantly with the incidence angle. Otherwise, the $\gamma$-ray beam direction is still very focused and given by the electron beam direction, the $\gamma$-ray beam spot shape is correlated with the orientation of the linear polarization plane and the $\gamma$-ray beam retains the polarization degree of the laser in the direction along the electron beam axis, disregarding the angle of incidence.

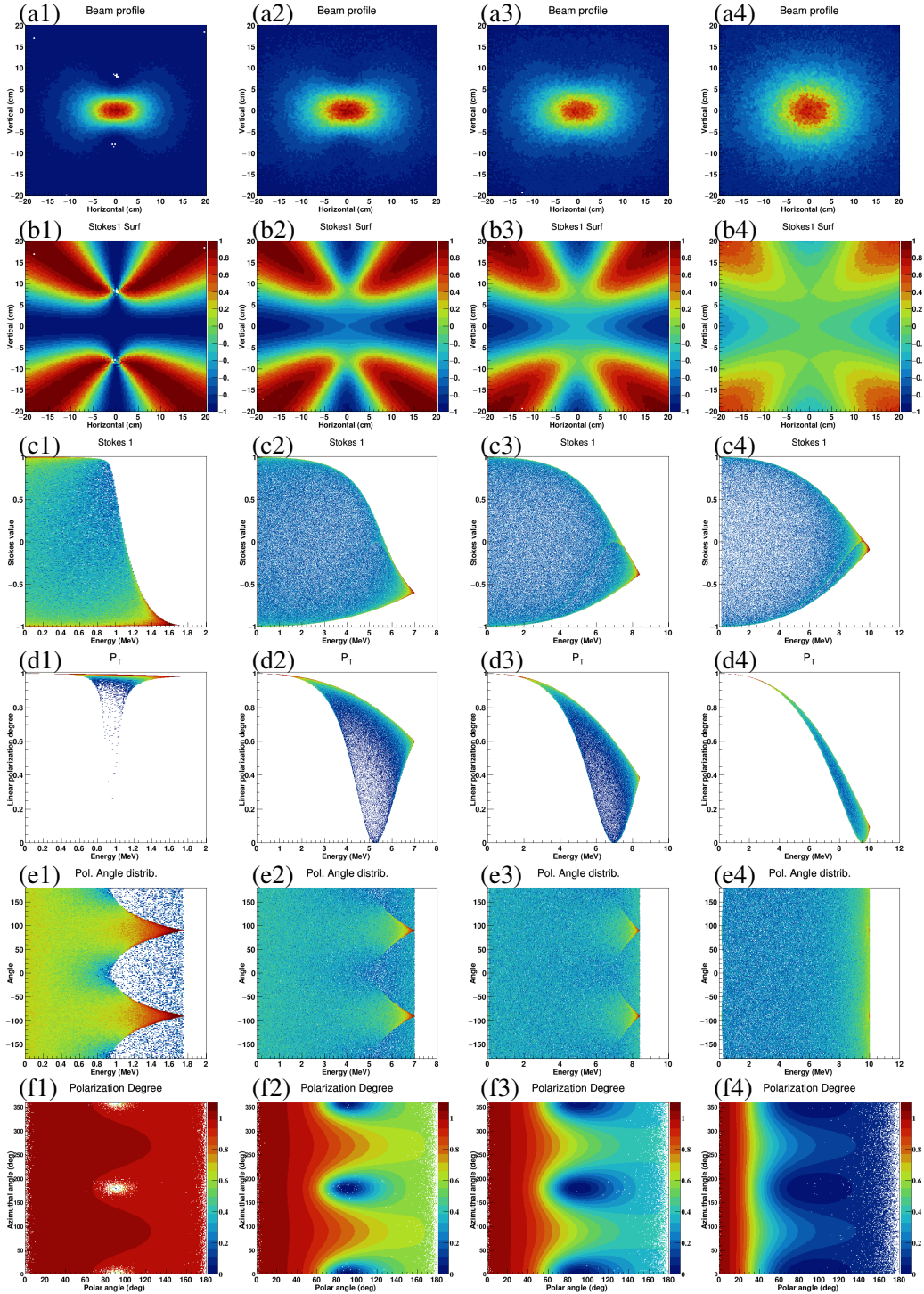
\begin{figure}[t]
\centering
\includegraphics [width=0.9\columnwidth, angle=0]{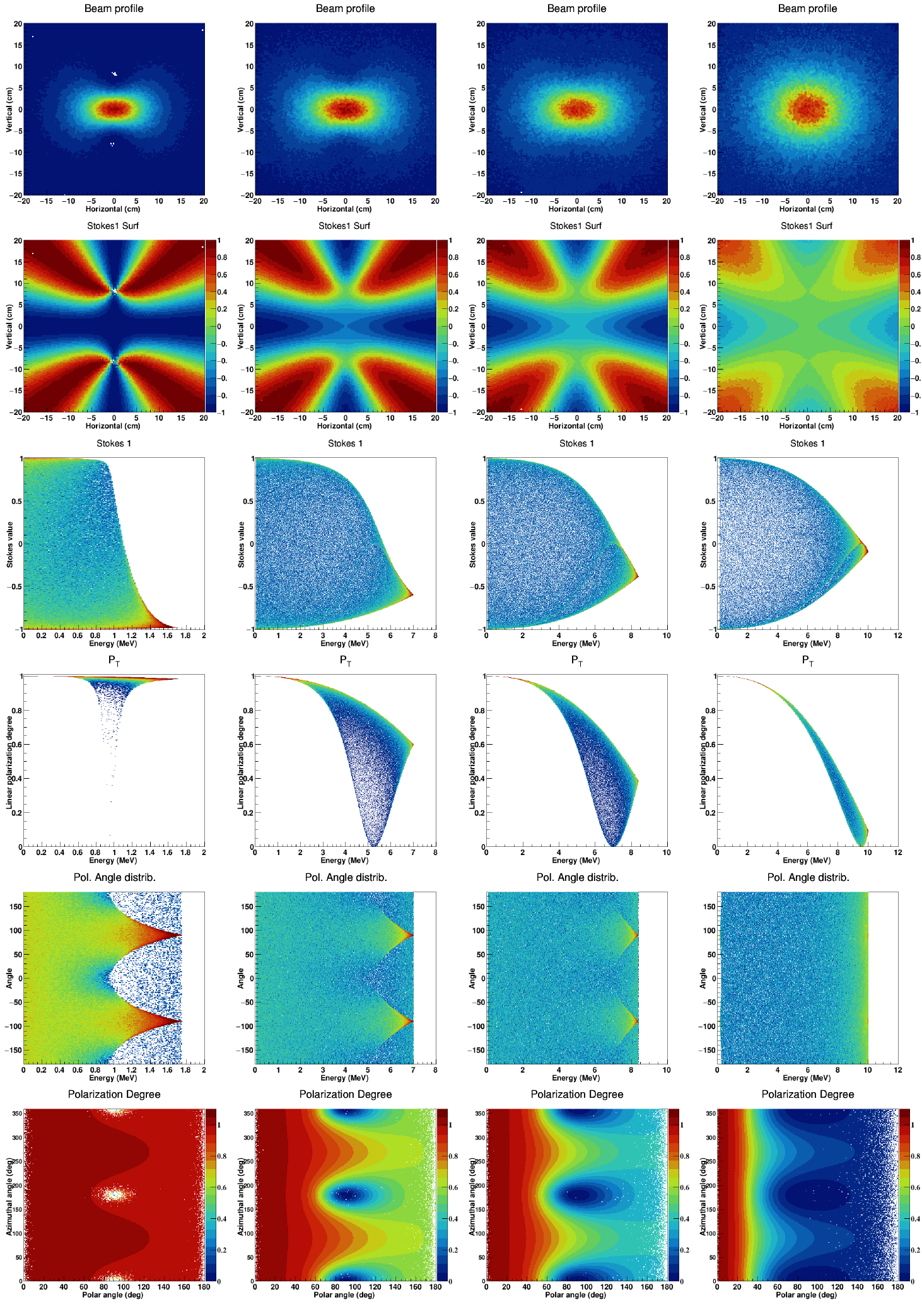} 
\put(-380,545){(a1)}
\put(-282,545){(a2)}
\put(-186,545){(a3)}
\put( -88,545){(a4)}
\put(-380,453){(b1)}
\put(-282,453){(b2)}
\put(-186,453){(b3)}
\put( -88,453){(b4)}
\put(-380,362){(c1)}
\put(-282,362){(c2)}
\put(-186,362){(c3)}
\put( -88,362){(c4)}
\put(-380,271){(d1)}
\put(-282,271){(d2)}
\put(-186,271){(d3)}
\put( -88,271){(d4)}
\put(-380,180){(e1)}
\put(-282,180){(e2)}
\put(-186,180){(e3)}
\put( -88,180){(e4)}
\put(-380, 90){(f1)}
\put(-282, 90){(f2)}
\put(-186, 90){(f3)}
\put( -88, 90){(f4)}
\caption{\label{fig_ene}
Compton scattering of 100~\% linear polarized ($\tau=90^\circ$) 1~nm~(1), 100~pm~(2), 50~pm~(3) and 10~pm~(4) wavelength X-rays on 10~MeV electrons. 
(a) $\gamma$-ray beam intensity spatial distribution;
$P^{(LAB)}_1$ distribution (b) spatial and (c) as function of $\gamma$-ray energy;
(d) $P_T$ distribution as function of $\gamma$-ray energy.
(e) $\gamma$-ray energy distribution of polarization vector azimuthal angle; 
(f) $P_T$ distribution as function of $\theta\&\phi$. 
}
\end{figure}

Let us also consider the scattering a green laser (532 nm, 2.33 eV) on a 3.5~GeV electron beam from SLEGS facility, in which case we note that the photon energy in the electron rest frame varies from 31.9 keV at $\theta_i=0^\circ$ down to 0.17~meV at $\theta_i=180^\circ$. Simulated polarization properties obtained for various incident photon--electron angles are shown in figure~6 of the supplementary material~\cite{supplementary_material}, where no obvious variation can be observed with the angle of incidence. For all incident angles, the scattered photons are highly polarized, with polarization degree better than 99$\%$. However, in the angle distribution histograms shown in columns 5 and 6 of figure~6 in ref.~\cite{supplementary_material}, a slight depolarization effect starts to be visible at head-on and small collision angles, when compared with the previous case of lower, 500~MeV electron energy. 

A simulation of the EuroGammaS laser recirculator~\cite{Dupraz2014,Ndiaye2019} is also shown in figure~7 of the supplementary material~\cite{supplementary_material}. In the simulation, a 515~nm laser bunch is sent 32 times against a 750~MeV electron beam with 7.5$^\circ$ incidence angle (see figure~7(a) of ref.\cite{supplementary_material}). At each cycle, the laser momentum azimuthal angle is shifted with a fixed step, in order to preserve the integrity of the two confocal parabolic mirrors of the optical resonator. Thus, the laser polarization vector at the interaction point depicts a specific pattern in the polar angles space, as described in ref.~\cite{Dupraz2014} and also shown in figure~7(b) of ref.\cite{supplementary_material}. The present algorithm reproduced the $\gamma$-ray polarization phasor angular distribution given in fig.~19(a) of ref.~\cite{Dupraz2014} (see figure~7(c) of ref.\cite{supplementary_material}). We show in figure~7(d) of ref.\cite{supplementary_material} that the $\gamma$-ray linear polarization degree is higher than 99.95$\%$ for photons back-scattered along the electron beam axis. We find that the 7.5$^\circ$ incident angle given by the recirculator system and the small laser polarization vector wiggle do not affect the polarization degree of the collimated LCS $\gamma$-ray beam.  

\subsection{Arbitrary angle collisions of X-rays on relativistic electrons}

\begin{table}[t]
\centering
\caption{\label{table_PhotonEnergy} Main properties for the four cases of X-ray scattering on 10~MeV electrons: incident photon energy in laboratory system, incident energy in electron rest frame system for for $\theta_i=0^\circ$ head-on and $\theta_i=162^\circ$ collision angle, integrated Compton scattering cross section for head-on collision:
}
\smallskip
\begin{tabular}{|c|c|c|c|}
\hline
X-ray energy  & \multicolumn{2}{c|}{Photon energy in electron}     & Integrated Compton \\ 
laboratory    & \multicolumn{2}{c|}{rest frame system [keV]}       & cross section for     \\ \cline{2-3}
system [keV]  & \quad\quad$\theta_i=0^\circ$ \quad\quad & \quad\quad$\theta_i=162^\circ$\quad\quad & head-on collision [mb]\\  \hline
1.24  & 50.9  & 1.27 & 560\\
12.4  & 509.9 & 12.77 & 287 \\
24.8  & 1020. & 25.52 & 209\\
124.  & 5098. & 127. & 82  \\
\hline
\end{tabular}
\end{table}

As concluded in section~\ref{subsec_PolWithPhotEne}, in order to observe a clear variation of polarization variables with the angle of incidence between interacting photon and electron beams, one must start from hundreds of keV photon energies in the electron rest frame for head-on collision ($\theta_i=0^\circ$). To test the present algorithm in such cases, some extreme hypothetical cases have to be imagined. We will consider the interaction between a 10~MeV relativistic electron beam with a hard X-ray beam having the wavelengths/energies of: 1~nm/1.24~keV, 100~pm/12.4~keV, 50~pm/24.8~keV and 10~pm/124~keV. The lower energy of the electron beam was chosen in order highlight some interesting kinematic effects on the final $\gamma$-ray beam. The angle of incidence was varied with 18$^\circ$ steps from head-on $\theta_i=0^\circ$ up to $\theta_i=162^\circ$. The corresponding photon energies of the four cases of X-ray beams in the electron rest frame for incidence angles $\theta_i=0^\circ\,\&\,162^\circ$ are given in table~\ref{table_PhotonEnergy}.

We note that for three of the four cases, the condition imposed in concluding remarks of the subsection \ref{subsec_PolWithPhotEne} is met, namely while increasing the incidence angle between photon and electron beams from 0$^\circ$ to 162$^\circ$, the photon energy in the electron rest frame drops from hundreds of keV to tens of keV and consequently we expect a large variation of polarization related variables, although, as shown in table~\ref{table_PhotonEnergy}, the integrated Compton cross section for head-on collision is visibly decreasing with the incident energy increase. 

In order to perform the simulations, we considered ideal electron and X-ray beams. Due to the low energy of the electron beam, the imaging plate used as beam monitor was placed at 1.5~m downstream of the interaction point, along the electron beam line. Also, the monitor screen dimension was increased because the wider $\gamma$-ray emission cone. For all cases, a linear polarized X-ray beam was considered, with the plane of polarization situated in plane ($\tau=0^\circ$) and perpendicular ($\tau=90^\circ$) to the plane defined by the incident X-ray and electron beams. 

Complete simulation results for all four cases are given in the supplementary material~\cite{supplementary_material}. For each laser--electron configuration and for each incident angle, figures~1~--~4 of ref.~\cite{supplementary_material} show the $\gamma$-ray beam intensity spatial distribution, the $P^{(LAB)}_1$ Stokes parameter distribution as function of $\gamma$-ray energy and its spatial distribution, and the $\gamma$-ray energy distribution of polarization vector azimuthal angle. Selected results will displayed in the present paper in the following sections.  

\subsubsection{Dependence with the incident photon energy in the electron rest frame system}

Figure~\ref{fig_ene} shows the beam spatial distribution and polarization properties results for the head-on collision of 1~nm~(column~1), 100~pm~(column~2), 50~pm~(column~3) and 10~pm~(column~4) wavelength X-ray photons on 10~MeV electron beam. The polarization plane of the X-ray beam was considered perpendicular ($\tau=90^\circ$) to the plane defined by the incident X-ray and electron beams. Figures~\ref{fig_ene}(a1-a4) show the $\gamma$-ray beam intensity spatial distribution. One can observe that the $\gamma$-ray beamspot increases with the increase of initial X-ray energy and becomes progressively symmetric around central axis defined by the electron beam. Thus, as the Compton generated $\gamma$-ray beam polarization decreases, the $\gamma$-ray beam intensity spatial distribution is increasingly more uncorrelated with the initial orientation of the X-ray polarization plane. 

The decrease of the $\gamma$-ray beam polarization degree with the increase of the incident X-ray energy can be seen also when looking on the spatial distribution on monitor screen of Stokes~$P^{(LAB)}_1$ parameter in figure~\ref{fig_ene}(b1-4), on Stokes~$P^{(LAB)}_1$ parameter (figure~\ref{fig_ene}(c1-4)), linear polarization degree (figure~\ref{fig_ene}(d1-4)) and polarization vector angle orientation (figure~\ref{fig_ene}(e1-4)) dependence with $\gamma$-ray energy. In figures~\ref{fig_ene}(f1-4), it can be observed that in the electron rest frame system, the photons tend to be scattered and polarized more in the forward direction as the X-ray incident energy increases.

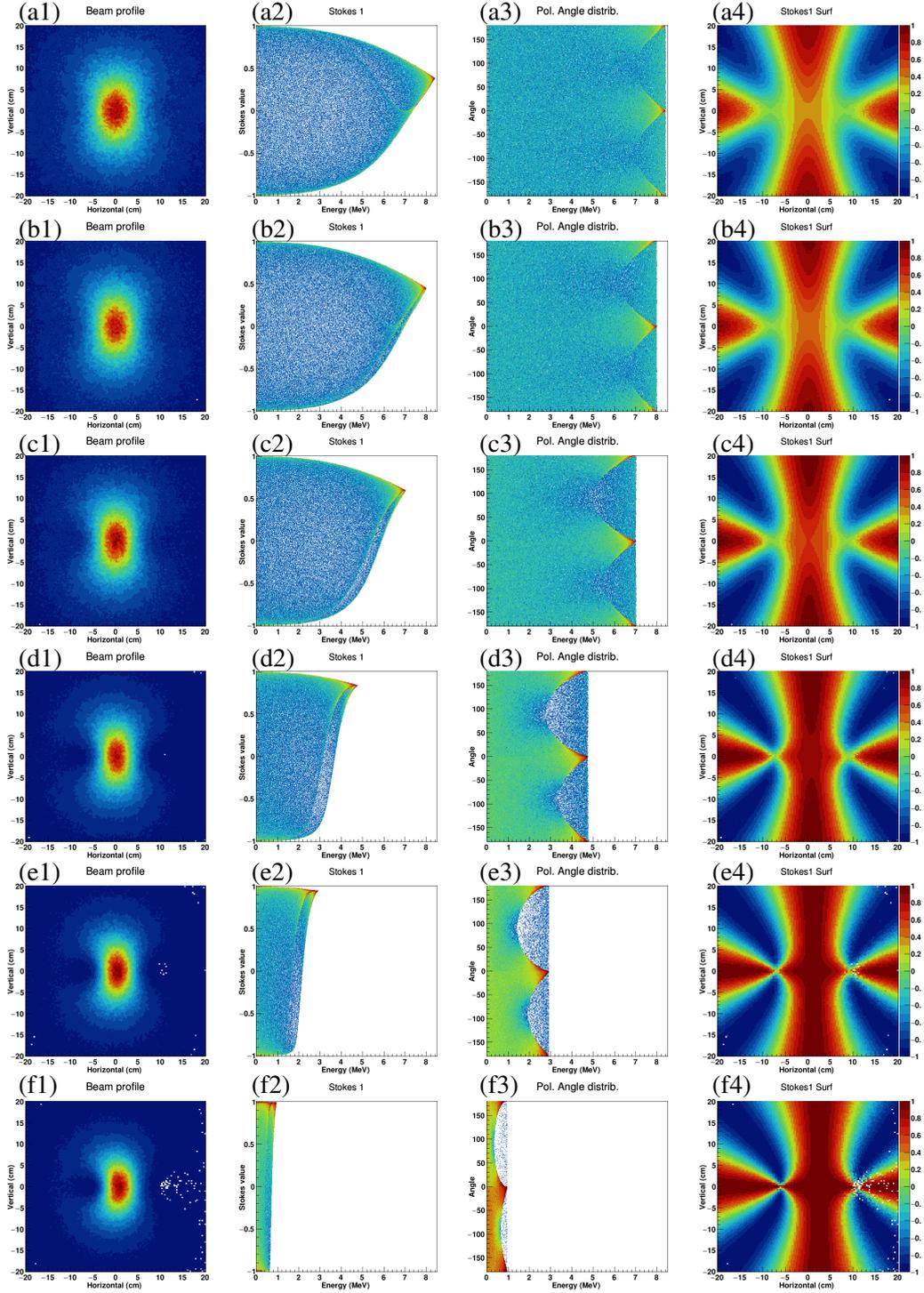
\begin{figure}[t]
\centering
\includegraphics [width=0.9\columnwidth, angle=0]{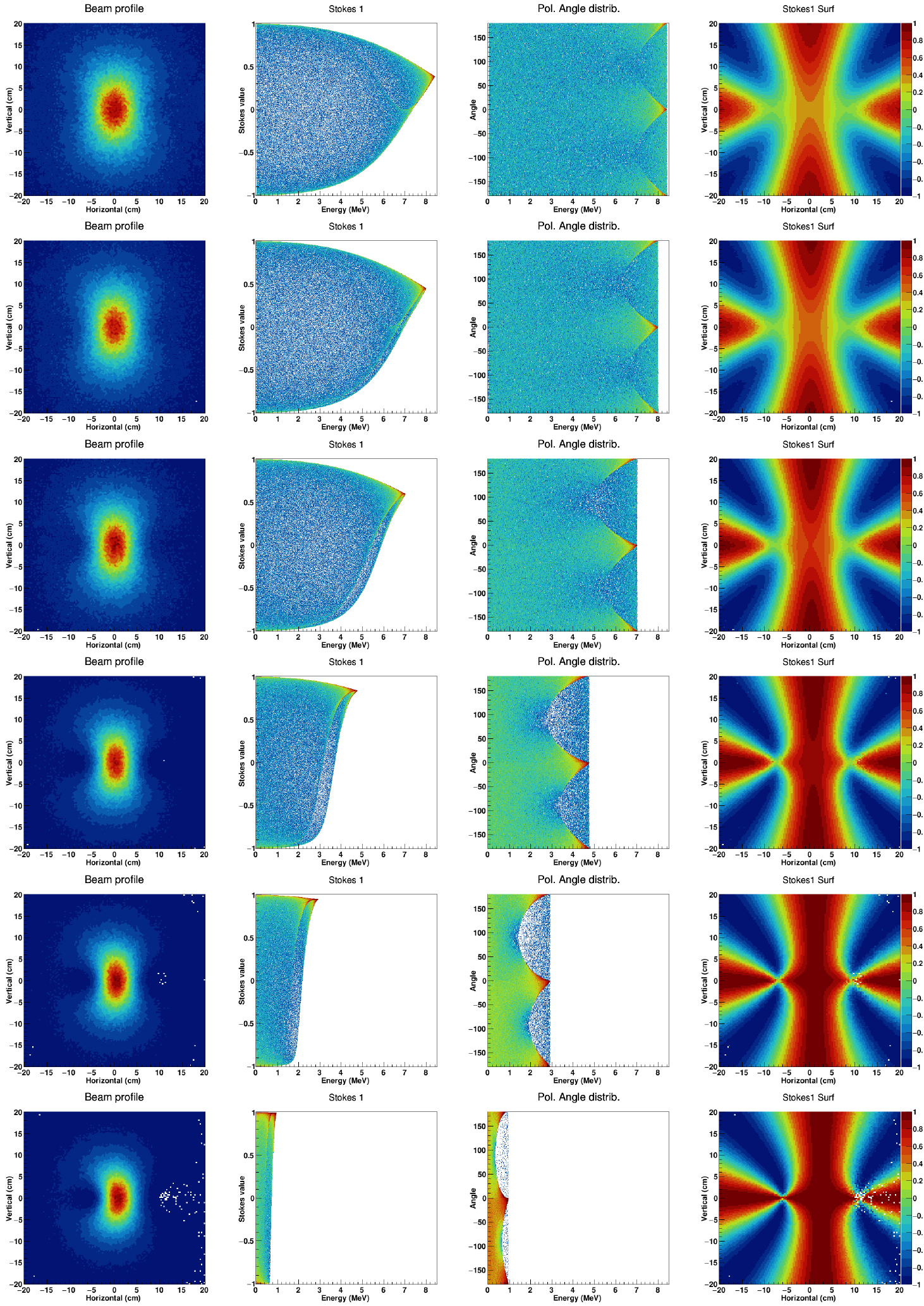} 
\put(-380,540){(a1)}
\put(-282,540){(a2)}
\put(-186,540){(a3)}
\put( -88,540){(a4)}
\put(-380,448){(b1)}
\put(-282,448){(b2)}
\put(-186,448){(b3)}
\put( -88,448){(b4)}
\put(-380,357){(c1)}
\put(-282,357){(c2)}
\put(-186,357){(c3)}
\put( -88,357){(c4)}
\put(-380,266){(d1)}
\put(-282,266){(d2)}
\put(-186,266){(d3)}
\put( -88,266){(d4)}
\put(-380,175){(e1)}
\put(-282,175){(e2)}
\put(-186,175){(e3)}
\put( -88,175){(e4)}
\put(-380, 85){(f1)}
\put(-282, 85){(f2)}
\put(-186, 85){(f3)}
\put( -88, 85){(f4)}
\caption{\label{fig_ang}
Compton scattering of 100~\% linear polarized ($\tau=0^\circ$) 50~pm~wavelength X-rays on 10~MeV electrons, with collision angles: (a)~0$^\circ$, (b)~54$^\circ$, (c)~90$^\circ$, (d)~126$^\circ$, (e)~144$^\circ$, (f)~162$^\circ$.
(col.1) $\gamma$-ray beam intensity spatial distribution;
(col.2) $P^{(LAB)}_1$ distribution as function of $\gamma$-ray energy; 
(col.3) $\gamma$-ray energy distribution of polarization vector azimuthal angle; 
(col.4) $P^{(LAB)}_1$ spatial distribution.
}
\end{figure}

\subsubsection{Dependence with the collision angle}

Simulation results on the polarization properties dependence with the collision angles for the four X-ray Compton scattering hypothetical scenarios listed in table~\ref{table_PhotonEnergy} are given in figures~1~--~4 of the supplementary material~\cite{supplementary_material}. Figure~\ref{fig_ang} shows results for selected collision angles in the case of 24.8~keV X-ray fully linearly polarized beam with the polarization in the plane defined by the X-ray and the 10~MeV electron beam axes, a case which we find conclusive for the present study. As given in table~\ref{table_PhotonEnergy}, in this case, the photon energy in the electron rest frame system varies between 1020.~keV at head-on collision to 25.52~keV for $\theta_i$~=~162$^\circ$. Following the discussion given in section~\ref{subsec_PolWithPhotEne}, in this scenario, the polarization properties of the scattered photon change from partial polarization degree at head-on to almost completely polarized at high incident angles. The integral Compton cross section for head-on collision for this case, given in table~\ref{table_PhotonEnergy}, is sufficiently high to be observed experimentally in moderate photon flux conditions.  

Figures~\ref{fig_ang}(a1-f1) show the $\gamma$-ray beam intensity spatial distribution on an imaging plate placed at 1.5~meters from the interaction point. The $\gamma$-ray beam-spot size decreases with the increase in the collision angle and the shape looses the vertical symmetry for large collision angles. This is a kinematic effect, which shows the side momentum impinged by the initial X-ray beam.   

Figures~\ref{fig_ang}(a2-f2) show the $P^{(LAB)}_1$ Stokes parameter distribution as function of $\gamma$-ray energy. We can notice that the polarization degree increases with the increase in the collision angle, in this case from 40$\%$ at head-on to $\sim$100$\%$ at $\theta_i$~=~162$^\circ$. The increase in the polarization degree with the collision angle is also noticed in figures~\ref{fig_ang}(a3-f3) showing the $\gamma$-ray energy distribution of polarization vector azimuthal angle. Here, the azimuthal asymmetry is again visible for large scattering angles. Figures~\ref{fig_ang}(a4-f4) show the $P^{(LAB)}_1$ spatial distribution on the imaging plate. The increase of the polarization degree in the central region is again visible, but one can also observe a slight shift on the horizontal axis corresponding to the plane defined by the X-ray and electron beam axes. 

For a better visualization, figure~\ref{fig_ang_proje}(a) shows the projection on the horizontal axis of the $\gamma$-ray beam intensity spatial distributions given in figures~\ref{fig_ang}(a1-f1). One can notice that the distribution becomes increasingly asymmetric with the increase in the collision angle. Figure~\ref{fig_ang_proje}(b) shows the projections for the case of polarization perpendicular on the plane defined by the X-ray and electron beam axes. Although the projected distribution is wider due to the 90$^\circ$ beam-spot rotation, the distribution preserves the symmetry for all incident angles. As a conclusion, whenever the employed experiment doesn't require a particular polarization orientation, the laser linear polarization should be oriented perpendicular on the plane defined by the incident photon~--~electron beam axes to avoid the beam-spot distortion. 

Figures~\ref{fig_ang_proje}(c,~d) show the profile distributions of the $P^{(LAB)}_1$ Stokes parameter and for the $P_T$ linear polarization degree, respectively, as function of $\gamma$-ray energy for all considered collision angles. One can notice that the minimum value of the $P_T$ is increasing with the increase in the collision angle. From the $P^{(LAB)}_1$ dependence with $\gamma$-ray energy, the polarization of the $\gamma$-rays scattered along the electron beam axis, which are usually selected by collimation, is increasing with the interaction angle. 

\subsection{Energy variable LCS $\gamma$-ray source with fixed energy electron beam -- discussion on polarization properties}

\begin{figure}[t]
\centering
\includegraphics [width=0.9\columnwidth, angle=0]{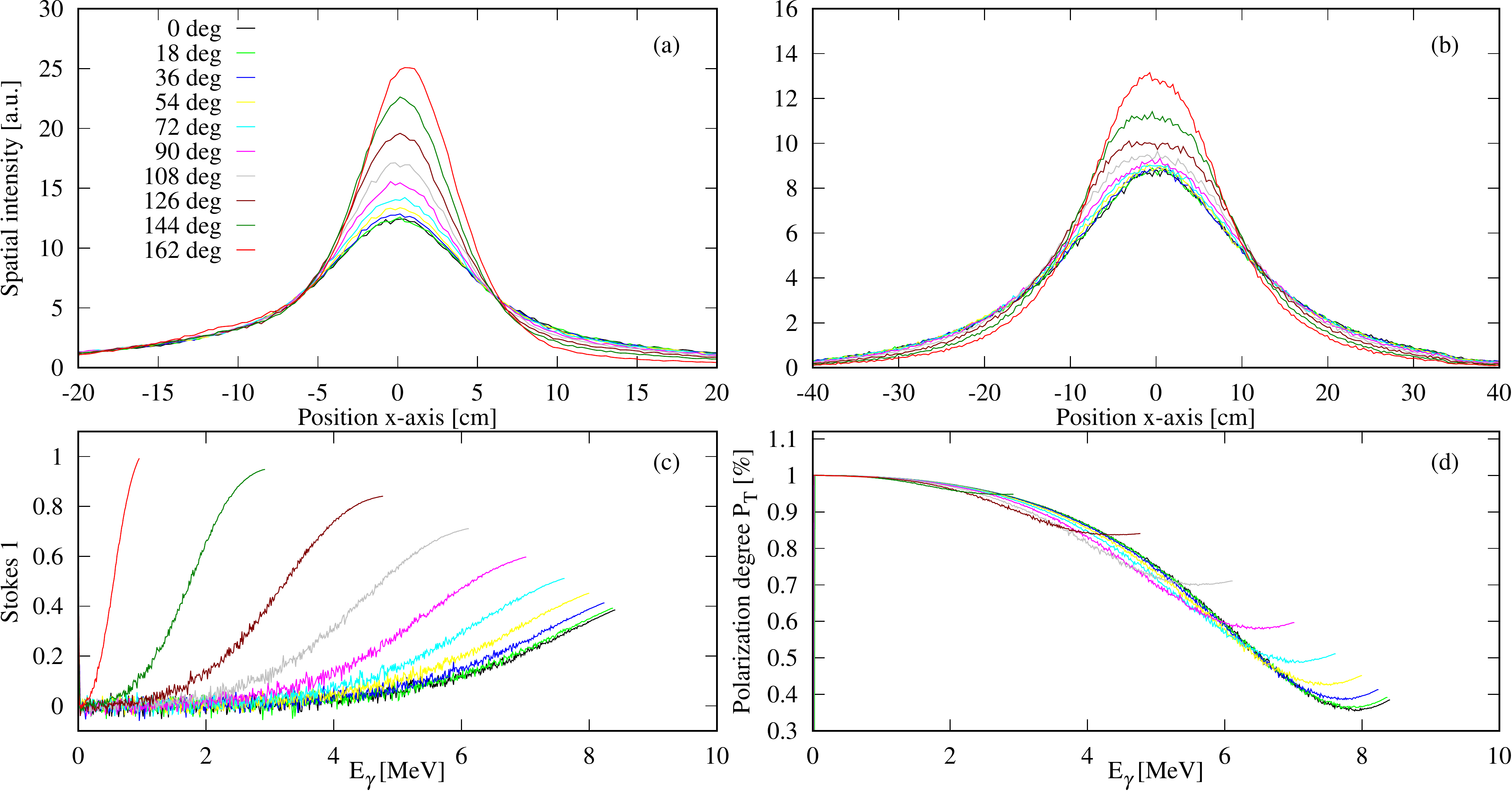} 
\caption{\label{fig_ang_proje}
Simulations for Compton scattering of 24.8~keV X-ray fully linearly polarized beam on 10~MeV unpolarized electron beam. Incident collision angles between 0$^\circ$ (head-on) and 162$^\circ$ have been considered. Projection on the horizontal axis of the $\gamma$-ray beam intensity spatial distributions for linear polarization (a)~\emph{in} and (b)~\emph{perpendicular~on} the plane defined by the X-ray and the electron beam axes; profile distributions for the (c) $P^{(LAB)}_1$ Stokes parameter and (d) $P_T$ linear polarization degree as function of $\gamma$-ray energy. 
}
\end{figure}

In reference~\cite{ohgaki_2007}, two experimental techniques are investigated in order to produce energy tunable LCS $\gamma$-ray beams using fixed energy electron beams. The first investigated method involves head-on laser--electron collisions and a sliding collimator~--~absorber pair in order to select the annular sector of quasi-monoenergetic scattered $\gamma$-rays. As shown in figure~\ref{fig_ang_proje}(c), the annular sector selects a given slice of the Stokes parameter energy profile. The Stokes parameter absolute value decreases with the decrease in the selected energy slice, although figure~\ref{fig_ang_proje}(d) shows that the polarization degree increases with the energy decrease. This apparent contradiction can be explained as following: although along a given annular sector the photons have a polarization degree higher than the one of the back-scattered photons, the different azimuthal orientations of the polarization vector around the selected ring smears the total polarization degree of the given annular sector. This is in agreement with the statement made in ref.~\cite{ohgaki_2007},  that the polarization of the $\gamma$-ray beam severely suffers from a large scattering angle. 

The second method proposed in reference~\cite{ohgaki_2007} and currently used at SLEGS is the \emph{slant-scattering} operating mode, in which the collision angle is varied in order vary the $\gamma$-ray beam energy. In section~\ref{sec_results_laser_ele_arbitrary_angle} and in the supplementary material~\cite{supplementary_material}, it has been shown that for Compton scattering of laser photons on relativistic electrons, the polarization properties of the scattered $\gamma$-ray are essentially unaffected by the variation of the collision angle. Further more, in the case of X-ray scattering on relativistic electrons, figure~\ref{fig_ang_proje}(c) and (d) show that the polarization degree of the photons scattered along the electron beam axis increases with the increase in the collision angle. Thus, the present results disagree with the conclusion given in reference~\cite{ohgaki_2007}, that the polarization of the scattered photons degrades with the increase in the collision angle. 

\section{Conclusions}

Following a series of photoneutron experimental studies at the laser Compton scattering $\gamma$-ray beam line of the NewSUBARU synchrotron radiation facility at SPring8, Japan, a dedicated Monte Carlo algorithm for modeling the energy spectra, flux and polarization of the incident $\gamma$-ray beams has been developed. The \texttt{eliLaBr} code is implemented using the \textsc{Geant4} package and is available on the github repository~\cite{eliLaBr_github}. The present work describes step-by-step the algorithm for modeling the laser Compton scattering between laser photons and relativistic electrons with focus on modeling the polarization properties of the scattered photon.  

The polarization is treated independently both in the Stokes parameters and in the polarization vector formalisms. Both methods provide identical results when dealing with full 100$\%$ linear polarization. When dealing with partial polarized laser or mixed linear/circular polarization, although both treatments produce identical average results, the recommended option in the algorithm is the Stokes formalism due to its more general treatment of partial polarization. In the future, we plan to include the circular and elliptical photon polarization cases also in the vectorial treatment, following the Klein-Nishina formulation given in ref.~\cite{krafft2016} for general complex polarization vectors. User recommendations are also given based on polarization state description requirements of different \textsc{Geant4} physics classes. 

The intensity spatial distributions and energy distributions for the scattered $\gamma$-ray beam and its Stokes parameters are obtained for head-on collisions of laser photons and unpolarized relativistic electrons, where several incident laser polarization states were considered: linear, unpolarized, circular  and mixed linear and circular polarization. 
The Monte Carlo algorithm described in the present work reproduces the analytical results presented in references~\cite{Sun2011_STAB,petrillo2015,zhijun_chi2020} and the fundamental harmonic results from reference~\cite{zhijun_chi2022}.  

The polarization properties of the Compton scattered photon for arbitrary angle collisions were also investigated. The present work shows that the degree of polarization transfer from the incident photon to the scattered photon increases with the collision angle. However, the polarization transfer is mainly determined by the energy of the incident photon in the electron rest frame system. Thus, for $\gamma$-ray sources based on Compton scattering of low-energy laser photons on relativistic electrons, the present work shows that the polarization degree of the incident photon is almost completely transferred to the scattered photon for any incident angle, where realistic simulation test-cases specific to the NewSUBARU, SLEGS and EuroGammaS LCS $\gamma$-ray sources have been considered. This is in agreement with the recent investigation of reference~\cite{zhijun_chi2020} on the polarization transfer for Thomson scattering at arbitrary angles.  

\acknowledgments

The author gratefully acknowledges Hiroaki Utsunomiya for useful discussions and continuous support. The author thanks Shuji Miyamoto for providing technical information on the NewSUBARU synchrotron and lasers employed at BL01. The author thanks Madalina Boca for useful discussions on gauge invariance. Work supported by the Romanian Ministry of Research, Innovation and Digitalization/Institute of Atomic Physics from the National Research - Development and Innovation Plan III for 2015 -2020/Programme 5/Subprogramme 5.1 ELI-RO, project GANT-Photofiss No 14/16.10.2020.


\begin{thebibliography}{99}

\bibitem{Kawano2020} T. Kawano et al., \emph{IAEA Photonuclear Data Library 2019}, \emph{Nucl. Data Sheets} {\bf 163} (2020) 109.
\bibitem{amano09}  S. Amano et al., \emph{Several-MeV $\gamma$-ray generation  at NewSUBARU by laser Compton backscattering}, \emph{Nucl. Instrum. Methods Phys. Res. A} \textbf{602} (2009) 337.
\bibitem{Horikawa2010} Ken~Horikawa et al., \emph{Measurements for the energy and flux of Compton scattering $\gamma$-ray photons generated in an electron storage ring: NewSUBARU}, \emph{Nucl. Instrum. Methods Phys. Res. A} \textbf{618} (2010) 209~--~215. 
\bibitem{utsunomiyaNimDNM} Hiroaki~Utsunomiya, Ioana Gheorghe et al., \emph{Direct neutron-multiplicity sorting with a flat-efficiency detector, Nucl. Instrum. Methods Phys. Res. A} {\bf 871} (2017) 135--141.
\bibitem{IGheorghe_2021_MF} Ioana~Gheorghe, Hiroaki~Utsunomiya et al., \emph{Updated neutron-multiplicity sorting method for producing photoneutron average energies and resolving multiple firing events, Nucl. Instrum. Methods Phys. Res. A} {\bf 1019} (2021) 165867.
\bibitem{Gheorghe2017} I.~Gheorghe, H.~Utsunomiya et al., \emph{Photoneutron cross-section measurements in the $^{209}$Bi($\gamma,\,xn$) reaction with a new method of direct neutron-multiplicity sorting}, \emph{Phys. Rev. C} \textbf{96} (2017) 044604.
\bibitem{filipescu22ND}  D.~Filipescu et al., \emph{Photofission and photoneutron cross sections for $^{238}$U and $^{232}$Th, Proceedings for the 15$^{th}$ International Conference on Nuclear Data for Science and Technology (ND202)}, submitted (2022).
\bibitem{geant_agostinelli_2003} S.~Agostinelli et al., \emph{Geant4--a simulation toolkit}, \emph{Nucl. Instrum. Methods Phys. Res. A}, {\bf 506} (2003) 250.   
\bibitem{geant_allison_2006} J.~Allison et al., \emph{Geant4 developments and applications}, \emph{IEEE Trans. Nucl. Sc.} {\bf 53} (2006) 270. 
\bibitem{geant_allison_2016} J.~Allison et al., \emph{Recent developments in Geant4}, \emph{Nucl. Instrum. Methods Phys. Res. A}, {\bf 835} (2016) 186.   
\bibitem{eliLaBr_github} D.~Filipescu and I.~Gheorghe, \emph{eliLaBr -- GEANT4 simulation code for LCS gamma-ray sources and flat efficiency moderated He-3 counters array dedicated to photoneutron reaction studies}, \href{https://github.com/dan-mihai-filipescu/eliLaBr}{https://github.com/dan-mihai-filipescu/eliLaBr} (2022).
\bibitem{filipescu_NIM_LCS_22} D.~Filipescu et al., \emph{Spectral distribution and flux of $\gamma$-ray beams produced through Compton scattering of unsynchronized laser and electron beams}, \emph{Nucl. Instrum. Methods Phys. Res. A}, submitted (2022).   
\bibitem{Pietralla_2002_NIM} N.~Pietralla et al., \emph{Parity measurements of nuclear levels using a free-electron-laser generated $\gamma$-ray beam}, \emph{Phys. Rev. Lett.} {\bf 88} (2001) 012502.
\bibitem{Beller_2015} J.~Beller et al., \emph{Separation of the 1+/1- parity doublet in 20Ne}, \emph{Phys. Lett. B} {\bf 741} (2015) 128.
\bibitem{Kondo_2012} T.~Kondo, H.~Utsunomiya et al., \emph{Total and partial photoneutron cross sections for Pb isotopes}, \emph{Phys. Rev. C} {\bf 86} (2012) 014316.
\bibitem{Angelo2000} A.~D'Angelo et al., \emph{Generation of Compton backscattering $\gamma$-ray beams}, \emph{Nucl. Instrum. Methods Phys. Res. A} {\bf 455} (2000) 1. 
\bibitem{litvinenko_1997} V.N.~Litvinenko et al., \emph{Gamma-ray production in a storage ring free-electron laser}, \emph{Phys. Rev. Lett.} {\bf 78} (1997) 4569--4572.  
\bibitem{Sun2011_STAB} C.~Sun and Y.K.~Wu, \emph{Theoretical and simulation studies of characteristics of a Compton light source}, \emph{Phys. Rev. ST Accel. Beams} {\bf 14} (2011)  044701.   
\bibitem{petrillo2015} V.~Petrillo et al., \emph{Polarization of x-gamma radiation produced by a Thomson and Compton inverse scattering}, \emph{Phys. Rev. ST Accel. Beams} {\bf 18} (2015) 110701. 
\bibitem{zhijun_chi2020} Zhijun~Chi, \emph{Polarization transfer from a laser to x rays via Thomson scattering with relativistic electrons: A dipole radiation perspective}, \emph{J. Appl. Phys.}, {\bf 128} (2020) 244904.
\bibitem{ohgaki_2007} Hideaki Ohgaki et al., \emph{Study on Energy Variable Laser-Compton Gamma-ray with a Fixed Energy Electron Beam}, \emph{J. Nucl. Sci. Tech.}, {\bf 44} (2007) 698--702.   
\bibitem{xu_fan_2022} H.H.~Xu, G.T.~Fan et al., \emph{Interaction chamber for laser Compton slant-scattering in SLEGS beamline at Shanghai Light Source}, \emph{Nucl. Instrum. Methods Phys. Res. A}, {\bf 1033} (2022) 166742.   
\bibitem{wang_fan_2022} Hong-Wei~Wang, Gong-Tao~Fan et al., \emph{Commissioning of laser electron gamma beamline SLEGS at SSRF}, \emph{Nucl. Sci. Tech.}, {\bf 33} (2022) 87.   
\bibitem{hao_fan_2022} Z.R.~Hao, G.T.~Fan et al., \emph{Collimator system of SLEGS beamline at Shanghai Light Source}, \emph{Nucl. Instrum. Methods Phys. Res. A}, {\bf 1013} (2021) 165638.   
\bibitem{zhijun_chi2022} Zhijun~Chi, \emph{X-ray polarization characteristics in the nonlinear Thomson scattering of a laser with relativistic electrons}, \emph{Nucl. Instrum. Methods Phys. Res. A}, {\bf 1033} (2022) 166681.
\bibitem{McMaster1} William H. McMaster, \emph{Polarization and the Stokes Parameters}, \emph{American Journal of Physics} {\bf 22} (1954) 351--362.
\bibitem{McMaster2} William H. McMaster, \emph{Matrix Representation of Polarization}, \emph{Reviews of Modern Physics} {\bf 33} (1961) 8--28.
\bibitem{Landau} L. Landau, E. Lifchitz, V. Berestetski, L. Pitayevski, \emph{Th\'eorie Quantique Relativiste - premi\'ere partie}, \'Editions MIR, Moscou (1972).
\bibitem{Depaola} G.O. Depaola, \emph{New Monte Carlo method for Compton and Rayleigh scattering by polarized gamma rays}, \emph{Nucl. Instrum. Methods Phys. Res. A} {\bf 512} (2003) 619--630.
\bibitem{nishina1929} Y.~Nishina, \emph{Die Polarisation der Comptonstreuung nach der Diracschen Theorie des Elektrons}, \emph{Zeits. f. Physik.} {\bf 52} (1929) 869. 
\bibitem{wightman1948} A.~Wightman, \emph{Note on Polarization Effects in Compton Scattering}, \emph{Physical Review} {\bf 74} (1948) 1813. 
\bibitem{fano1949} U.~Fano, \emph{Remarks of the Classical and Quantum-Mechanical Treatment of Partial Polarization}, \emph{Journal of the Optical Society of America} {\bf 39} (1949) 859. 
\bibitem{krafft2016} G. A. Krafft et al., \emph{Laser pulsing in linear Compton scattering}, \emph{Phys. Rev. Accel. Beams} {\bf 19} (2016) 121302.
\bibitem{ranjan2018} N. Ranjan et al., \emph{Simulation of inverse Compton scattering and its implications on the scattered linewidth}, \emph{Phys. Rev. Accel. Beams} {\bf 21} (2018) 030701.
\bibitem{terzic2019} B. Terzi{\'{c}} et al., \emph{Improving performance of inverse Compton sources through laser chirping}, \emph{Europhysics Letters} {\bf 126} (2019) 12003.
\bibitem{goorley2012} J. T. Goorley, et al., \emph{Initial MCNP6 Release Overview}, \emph{Nuclear Technology} {\bf 180} (2012) 298-315.
\bibitem{bohlen2014} T.T. B\"ohlen et al., \emph{The FLUKA Code: Developments and Challenges for High Energy and Medical Applications}, \emph{Nuclear Data Sheets} {\bf 120} (2014) 211-214.  
\bibitem{dinter1988} H. Dinter et al., \emph{Calculations of Doses Due to Electron-Photon Stray Radiation from a High Energy Electron Beam Behind Lateral Shielding}, \emph{Radiation Protection Dosimetry} {\bf 25} (1988) 107-116.
\bibitem{CAIN} CAIN user manual. 
\bibitem{Curatolo2017} C.~Curatolo et al., \emph{Analytical description of photon beam phase space in inverse Compton scattering sources}, \emph{Phys. Rev. ST Accel. Beams} {\bf 20} (2017) 080701. 
\bibitem{Curatolo_PhD_thesis} C. Curatolo, Ph.D. thesis, Universit\`{a} degli Studi di Milano, 2016, https://air.unimi.it/handle/2434/358227.
\bibitem{Luo2011} W.~Luo et al., \emph{A 4D Monte Carlo laser-compton scattering simulation code for the characterization of the future energy-tunable SLEGS}, \emph{Nucl. Instrum. Methods Phys. Res. A} {\bf 660} (2011) 108--115. 
\bibitem{danxu_2005} Dan~Xu, Zhong~He and Feng~Zhang, \emph{Detection of Gamma Ray Polarization Using a 3-D Position-Sensitive CdZnTe Detector}, \emph{IEEE Trans. Nucl. Sc.} {\bf 52} (2005) 1060. 
\bibitem{GEANTPhysRef} Geant4 Collaboration, \emph{"Physics Reference Manual - Release 11.0"}, Rev.6.0: December (2021)
\bibitem{Hajima2021} Ryoichi~Hajima, \emph{Bandwidth of a Compton radiation source with an electron beam of asymmetric emittance}, \emph{Nucl. Instrum. Methods Phys. Res. A} {\bf 985} (2021) 164655. 
\bibitem{supplementary_material} See supplementary material at \href{https://raw.githubusercontent.com/dan-mihai-filipescu/eliLaBr/main/doc/DFilipescu-SpplMat-POL-LCS.pdf}{https://raw.githubusercontent.com/dan-mihai-filipescu/eliLaBr/main/doc/DFilipescu-SpplMat-POL-LCS.pdf} for detailed simulation results related to the polarization properties of the Compton scattering cases listed in Table~1.  
\bibitem{Dupraz2014} K.~Dupraz et al., \emph{Design and optimization of a highly efficient optical multipass system for $\gamma$-ray beam production from electron laser beam Compton scattering}, \emph{Phys. Rev. ST Accel. Beams} {\bf 17} (2014) 033501.
\bibitem{Ndiaye2019} Cheikh Fall Ndiaye et al., \emph{Low power commissioning of an innovative laser beam circulator for inverse Compton scattering $\gamma$-ray source}, \emph{Phys. Rev. Accel. Beams} {\bf 22} (2019) 093501.

\end{thebibliography}
\end{document}